\documentclass{aa}  

\usepackage{graphicx,txfonts,lscape,natbib}
\bibpunct{(}{)}{;}{a}{}{,} 

\begin{document}
\newcommand{\APer}{$\alpha$\,Per}
\newcommand{\gbp}{{G_\mathrm{BP}}}
\newcommand{\grp}{{G_\mathrm{RP}}}

   \title{A 5D view of the \APer{}, Pleiades, and Praesepe clusters}


   \author{N.\ Lodieu \inst{1,2}
       \and
       A.\ P\'erez-Garrido \inst{3}
        \and
       R.\ L.\ Smart \inst{4}
        \and
       R.\ Silvotti \inst{4}
        }

   \institute{Instituto de Astrof\'isica de Canarias (IAC), Calle V\'ia L\'actea s/n, E-38200 La Laguna, Tenerife, Spain \\
       \email{nlodieu@iac.es}
       \and
       Departamento de Astrof\'isica, Universidad de La Laguna (ULL), E-38206 La Laguna, Tenerife, Spain
       \and
       Dpto.\ F\'\i sica Aplicada, Universidad Polit\'ecnica de Cartagena, E-30202  Cartagena, Murcia, Spain
       \and
       Istituto Nazionale di Astrofisica, Osservatorio Astrofisico di Torino, Strada Osservatorio 20, I-10025 Pino Torinese, Italy
       }

   \date{Received \today{}; accepted (date)}

 
  \abstract
   {}
   {Our scientific goal is to provide revised membership lists of the \APer{}, Pleiades, 
   and Praesepe clusters exploiting the second data release of $Gaia$ and produce five-dimensional
   maps ($\alpha$, $\delta$, $\pi$, $\mu_{\alpha}cos\delta$, $\mu_{\delta}$) of these clusters.
    }
   {We implemented the kinematic method combined with the statistical
    treatment of parallaxes and proper motions to identify astrometric 
    member candidates of three of the most nearby and best studied open clusters in the sky.
   }
   {We cross-correlated the $Gaia$ catalogue with large-scale public surveys to complement the
   astrometry of $Gaia$ with multi-band photometry from the optical to the mid-infrared.
   We identified 517, 1248, and 721 bona-fide astrometric member candidates inside the tidal radius of 
   the \APer{}, the Pleiades, and Praesepe, respectively. We cross-matched our final samples with 
   catalogues from previous surveys to address the level of completeness. We update the main physical
   properties of the clusters, including mean distance and velocity as well as core, half-mass, and
   tidal radii. We infer updated ages from the white dwarf members of the Pleiades and Praesepe.
   We derive the luminosity and mass functions of the three clusters and compare them
   to the field mass function. We compute the positions in space of all member candidates in the 
   three regions to investigate their distribution in space.
   }
   {We provide updated distances and kinematics for the three clusters. We identify a list of members 
    in the \APer{}, Pleiades, and Praesepe clusters from the most massive stars all the way down into 
    the hydrogen-burning limit with a higher confidence and better astrometry than previous studies. 
    We produce complete 5D maps 
    of stellar and substellar bona-fide members in these three regions. The photometric sequences derived 
    in several colour-magnitude diagrams represent benchmark cluster sequences at ages from 90 to 600\,Myr.
    We note the presence of a stream around the Pleiades cluster extending up to 40 pc from the cluster centre.
   }  


  \authorrunning{Lodieu et al.}
  \titlerunning{A 5D view of the \APer{}, Pleiades, and Praesepe clusters}

   \maketitle
%

%
%
\section{Introduction}
\label{clustersGaia:intro}

Open clusters are small entities of a few hundreds to thousands stars loosely bound in a small
region of the sky \citep{lada03}. They represent ideal laboratories to study stellar evolution 
because their members share the same mean distance, age, and metallicity.
The wide range of masses present in these clusters is key to investigate the present-day
mass function \citep{salpeter55,miller79,scalo86,kroupa01b,chabrier03}
and study processes responsible for the formation of low-mass stars and brown dwarfs 
\citep[e.g.][]{chabrier00a}. More recently, the Kepler K2 mission \citep{borucki10,howell14}
has rejuvenized the study of open clusters providing thousands of light curves with exquisite 
photometric precision \citep{nardiello15a,rebull16a,nardiello16a,stauffer16a,douglas16a,rebull17,douglas17a}
and identified several low-mass eclipsing binaries 
\citep{kraus15a,david15a,lodieu15a,david15b,libralato16a,david16a,kraus17b}
and a few transiting planets at different ages
\citep{libralato16b,mann16a,david16b,mann16b,david16c,mann17a,mann18a}.
These new discoveries are key to constrain evolutionary models at different ages and improve
our knowledge of planet formation.

Before the advent of the $Gaia$ mission \citep{Gaia_Prusti2016}, the selection of members of 
open clusters was predominantly limited to ground-based proper motions and multi-band photometry.
The second data release of $Gaia$ \citep{Gaia_Brown2018} offers a unique
opportunity to revise the census of membership of the nearest open clusters thanks to exquisite
parallaxes, proper motions, and positions with additional radial velocity measurements for a
limited sample of (mainly solar-type) stars. With this new dataset in hand we can revise several
key questions yet to be solved such as the evolution of rotation as a function of mass and age
\citep[see review by][]{bouvier14a}, the possible age spreads within a given cluster 
\citep[e.g.][]{mayne08,parmentier14a,niederhofer15}, the dynamical masses and ages of low-mass
stars to calibrate model predictions \citep[e.g.][]{kraus15a}, among others.

In this manuscript, we present an astrometric selection of cluster member candidates in three 
of the nearest open clusters to the Sun: \APer{}, the Pleiades, and Praesepe exploiting the
second data release of $Gaia$ \citep{Gaia_Brown2018} and based on our previous
work in the Hyades \citep{lodieu19a}. We provide a revised stellar census in these three regions
and a 3D view of the distribution of their members.
In Section \ref{clustersGaia:GaiaDR2_sample} we present the input catalogue of $Gaia$.
In Section \ref{clustersGaia:Known_Members} we describe the three clusters under study and
compile lists of previously-known members to compare with the $Gaia$ catalogues.
In Section \ref{clustersGaia:select_members} we identify a complete sample of member 
candidates in \APer{}, the Pleiades, and Praesepe, addressing the completeness of previous
studies and cleanness of our selection.
In Section \ref{clustersGaia:WDs} we investigate white dwarfs previously reported as members
of these clusters to revise their membership and estimate their ages.
In Section \ref{clustersGaia:LFplusMF} we derive the stellar luminosity and mass functions
and put them into context with previous studies.
In Section \ref{clustersGaia:3D_view} we discuss the spatial distribution of the highest
probability stellar member candidates and present the first 3D map of the clusters.

%
%
\section{The $Gaia$ DR2 sample}
\label{clustersGaia:GaiaDR2_sample}

We made use of the $Gaia$ DR2 data \citep{Gaia_Lindegren2018,Gaia_Hambly2018,Gaia_Riello2018,Gaia_Evans2018}.
Our selection criteria were designed to start as inclusive as possible and to be 
more selective later in the process. We limited our original catalogues to parallaxes
larger than 1 mas, which is the sole criterion applied to the full $Gaia$ samples.
We checked the re-normalised unit weight error (RUWE) defined by \citet{lindegren18b}
for the full files of more than 2 million sources. We find that 5.1\%, 4.4\%, and 4.6\% of 
the sources have RUWE values greater than 1.4 in \APer{}, the Pleiades, and Praesepe, respectively, 
limit set for quality of the sample. 

We cross-matched a series of well-known large-scale surveys to complement the $Gaia$ catalogue,
keeping all $Gaia$ DR2 sources without any counterpart in those surveys either due to coverage 
or brightness/faintness reasons. We considered the following catalogues:
the Two Micron All-Sky Survey \citep[hereafter 2MASS;][]{cutri03,skrutskie06},
the Sloan Digital Sky Survey Data Release 12 \citep[SDSS;][]{abolfathi18a},
the UKIRT Infrared Deep Sky Survey Galactic Clusters Survey \cite[UKIDSS GCS;][]{lawrence07},
the Wide-field Infrared Survey Explorer \citep[AllWISE;][]{wright10,cutri14}, and
the first data release of the Panoramic Survey Telescope and Rapid Response System 
\cite[PS1;][]{kaiser02,chambers16a}. We cross-matched the catalogues with a matching
radius of 5 arcsec so any unresolved pairs of stars in the large surveys resolved spatially
in $Gaia$ were only matched to the closest $Gaia$ source which is also often the brighter of 
the two possible matches.
All matches were made at the epoch of the catalogue by applying the $Gaia$ proper 
motions to move the possible counterparts in the target catalogue epoch. 

%
%
\begin{figure*}
 \centering
  \includegraphics[width=0.43\linewidth, angle=0]{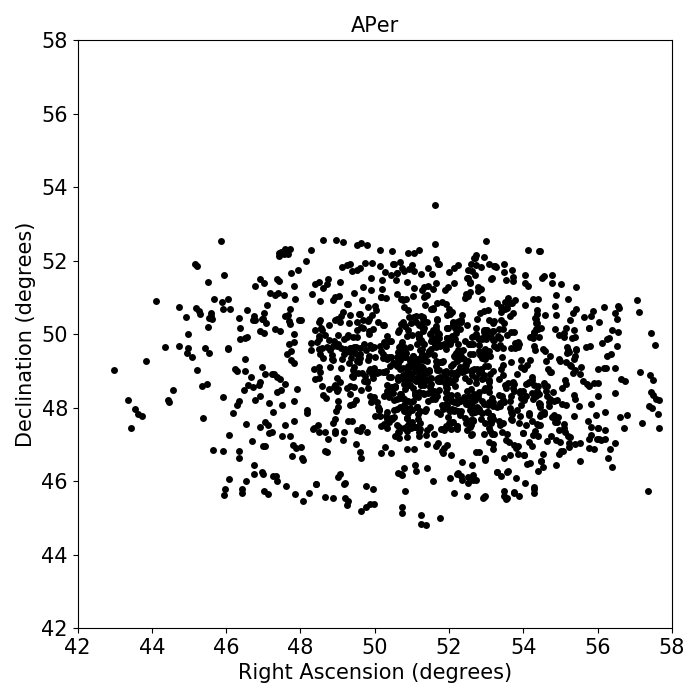}
  \includegraphics[width=0.43\linewidth, angle=0]{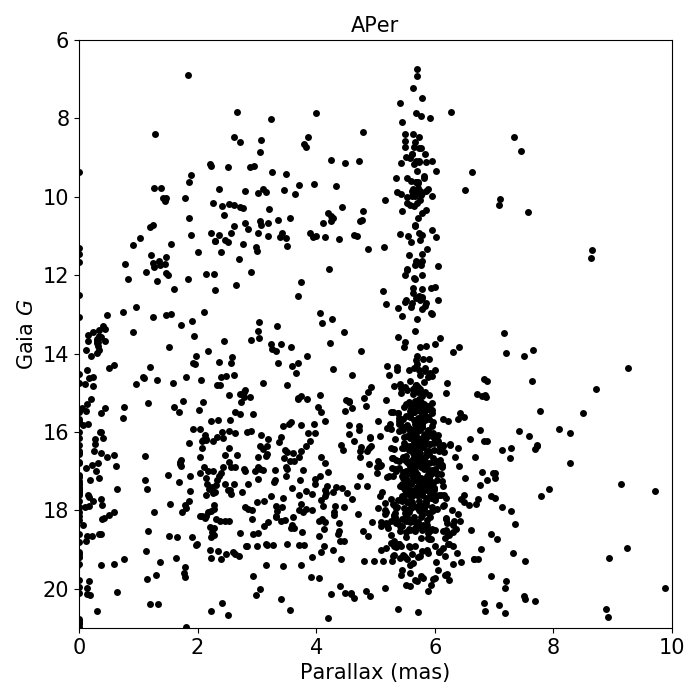}
  \includegraphics[width=0.43\linewidth, angle=0]{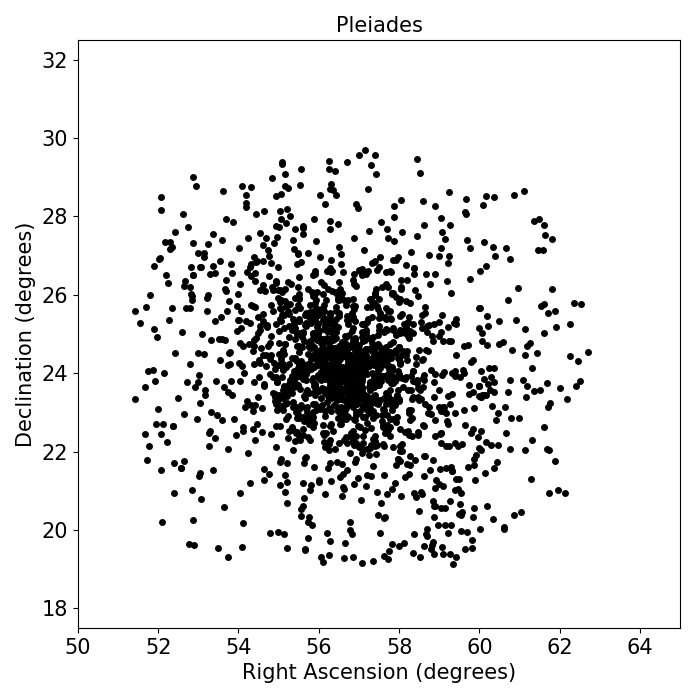}
  \includegraphics[width=0.43\linewidth, angle=0]{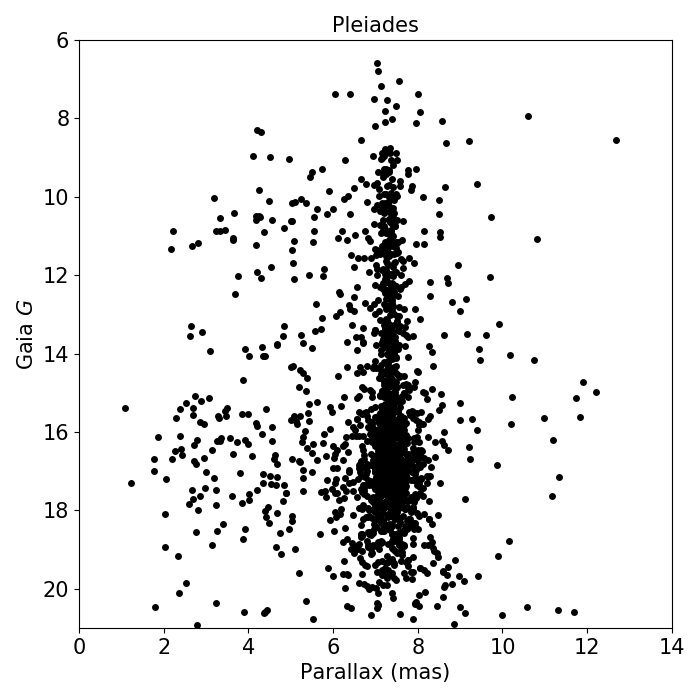}
  \includegraphics[width=0.43\linewidth, angle=0]{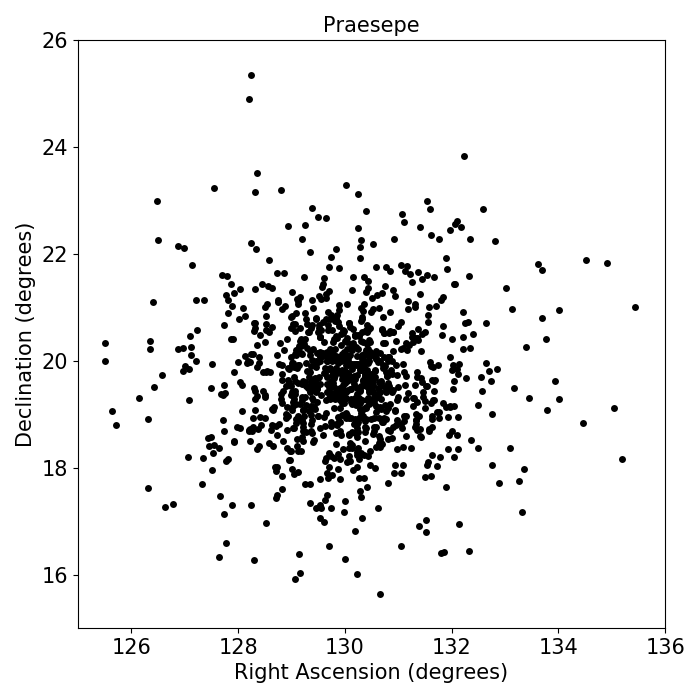}
  \includegraphics[width=0.43\linewidth, angle=0]{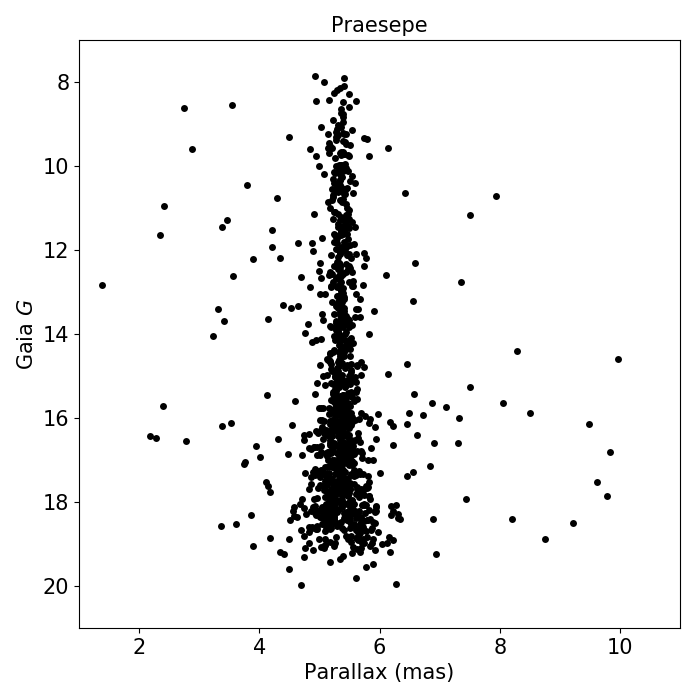}
  \caption{{\it{Left panels:}} Distribution on the sky of selected kinematic member candidates.
  {\it{Right panels:}} $Gaia$ $G$ magnitude as a function of parallax for previously-known
  members with $Gaia$ parallaxes.
  From top to bottom, we plot the figures for the \APer{}, the Pleiades, and Praesepe clusters, respectively.
  }
  \label{fig_clustersGaia:ra_dec_full}
\end{figure*}
%

%
%
\section{Earlier census of stellar and sub-stellar member candidates}
\label{clustersGaia:Known_Members}

In this section, we describe the main characteristics of each cluster and compiled
lists of known members from different proper motion and photometric surveys published 
over the past decades to identify high-mass stars, low-mass stars, and brown dwarfs.

\subsection{The \APer{} cluster}
\label{clustersGaia:Known_Members_APer}

\APer{} (Messier 20) is one of the few open star clusters within 200 pc of the Sun and younger than
200 Myr. The cluster is located to the north-east of the F5V supergiant Alpha Persei at a
distance originally estimated to $\sim$180 pc \citep{pinsonneault98,robichon99} and later
revised to 172.4$\pm$2.7 pc by the re-reduction of the Hipparcos data \citep{vanLeeuwen09}.
The cluster members have solar metallicity \citep{boesgaard90} and the extinction along
the line of sight is estimated to A$_{V}$ = 0.30 mag with a possible differential extinction 
\citep{prosser92}. Its mean proper motion, 
($\mu_{\alpha}\cos{\delta}$,$\mu_{\delta}$)\,=\,($+$22.73,$-$26.51) mas/yr \citep{vanLeeuwen09},
lower than the other clusters studied in this paper, and low galactic latitude ($b$\,=\,$-$7$^{\circ}$), 
makes the membership selection more difficult
in Praesepe and the Pleiades. The combination of larger distance but younger age than the Pleiades 
places the hydrogen-burning limit at similar apparent magnitudes within the reach of $Gaia$.

We collected a total of 3056 sources from 13 surveys in \APer{}:
\citet{heckmann56}; \citet{mitchell60}, \citet{fresneau80}, \citet{stauffer85}, \citet{stauffer89b},
\citet{prosser92}, \citet{prosser94}, \citet{prosser98a}, \citet{prosser98b}, \citet{stauffer99},
\citet{barrado02a}, \citet{deacon04}, \citet{lodieu05a}, 
and the latest sample of 740 members from \citet{babusiaux18}. After removing common sources, 
we are left with a collection of 1711 candidate members.
We did not include the following surveys in our compilation: \citet{kraft66a}, 
\citet{petrie69a}, \citet{petrie70a}, \citet{morgan71a}, and \citet{crawford74a} because 
we could not retrieve the list of members electronically.
We cross-matched this compendium of 1711 members with $Gaia$ DR2, yielding
1429 objects with precise and accurate astrometry. They are distributed in an area of approximately
43--58$^{\circ}$ and 45--53$^{\circ}$ in right ascension and declination, respectively
(Fig.\ \ref{fig_clustersGaia:ra_dec_full}).
They span the $G$-band magnitude from about 6.5 to 21 mag.
In Fig.\ \ref{fig_clustersGaia:APer_plots_general} of Appendix \ref{clustersGaia:Appendix_APer}, 
we plot these objects in the diagram showing the $Gaia$ magnitude as
a function of parallax. We can clearly see the cluster has a mean parallax of about 5.8 mas
with a dispersion of $\sim$0.5 mas, corresponding to 173$\pm$15 pc. We also note that 
a large fraction (about 40\%) of these 1429 previously-known members are most likely 
non-members because of their inconsistent parallaxes.

\subsection{The Pleiades cluster}
\label{clustersGaia:Known_Members_Pleiades}

With the Hyades, the Pleiades (M45) is the best studied northern cluster and qualifies as
a benchmark region \citep{lynga81}. Numerous multi-wavelength surveys have been conducted 
in the region to study in depth the stellar and sub-stellar members. Pleiades members share 
a significant common proper motion compared to neighbouring stars with 
($\mu_\alpha\cos\delta$, $\mu_\delta$)\,$\sim$\,(19.5, $-$45.5) mas/yr
\citep{jones81,robichon99,vanLeeuwen09}. This large mean motion, high galactic latitude
($b$\,$\sim$\,24$^{\circ}$), and low reddening along the line of sight of the cluster
\citep[E($B-V$)\,=\,0.03 mag;][]{Odell94} ease both astrometric and photometric selections. 
The cluster is also nearby, with the Hipparcos distance placing the Pleiades 
at 120.2$\pm$1.9 pc \citep{vanLeeuwen09} while other works suggest a mean distance 
of 134 pc with an uncertainty of 5 pc \citep{johnson57,gatewood00,pinfield00,southworth05}.
The age of the cluster has also been debated in the literature, ranging from 70--80 Myr
from the Zero-Age-Main-Sequence turn-off and isochrone fitting \citep{mermilliod81,vandenberg84} 
up to 120--130$\pm$20 Myr as derived from model fitting \citep{mazzei89,gossage18} 
and the lithium depletion boundary method \citep{basri96,stauffer98,barrado04b} 
with a revision leading a possibly younger age \citep[112$\pm$5 Myr;][]{dahm15}.

Numerous surveys of distinct sizes and depths have looked at the Pleiades
\citep{stauffer91,simons92a,hambly93,stauffer94c,williams96a,cossburn97,zapatero97a,zapatero97b,zapatero97c,stauffer98b,martin98a,bouvier98,festin98,zapatero99b,hambly99,pinfield00,adams01a,moraux01,jameson02,dobbie02b,moraux03,deacon04,bihain06,bihain10a,casewell07,lodieu07c,casewell11,lodieu12a,bouy13,sarro14,zapatero14b,barrado16,babusiaux18,zapatero18}.
\citet{stauffer07} compiled Pleiades members from various sources (see their table 1) while 
\citet{sarro14}, \citet{bouy15b}, and \citet{olivares18a} make use of archival datasets to produce the 
most up-to-date sample of stellar and sub-stellar members. We complemented those two extensive studies 
with the preliminary $Gaia$ catalogue \citep{babusiaux18} to produce a full catalogue of 2010
members after removing duplicates (middle row in Fig.\ \ref{fig_clustersGaia:ra_dec_full}).
We have omitted catalogues of members published before the 1990s, where the first extensive 
astrometric, photometric, and spectroscopic studies of the Pleiades took place 
\citep[e.g.][among others]{trumpler21,hertzsprung47,artyukhina69,eichborn70,jones73,turner79a,jones81,haro82,stauffer82,stauffer84,vanLeeuwen86,jameson89,stauffer89c}.

\subsection{The Praesepe cluster}
\label{clustersGaia:Known_Members_Praesepe}

Praesepe (M44, NGC\,2632) is a rich open cluster also known as the Beehive cluster. 
The cluster is nearby with a mean Hipparcos distance of 181.5$\pm$6.0 pc from the Sun 
\citep{vanLeeuwen09}. Its proper motion is significant compared to field stars
\citep[$-$35.81$\pm$0.29,$-$13.5$\pm$0.24 mas/yr;][]{vanLeeuwen09}. Moreover,
Praesepe might be slightly metal-rich \citep[Fe/H = $+$0.27$\pm$0.10;][]{pace08a}
and has negligible reddening. This is an intermediate-age cluster, much
older than \APer{} and the Pleiades but comparable to the Hyades, with estimated
ages in the 590--660 Myr range \citep{vandenberg84,mermilliod81,delorme11,gossage18}
although older estimates are not discarded 
\citep[700--900 Myr;][]{salaris04,bonatto04a,brandt15a,vandenberg84}.

Praesepe was first investigated by \citet{klein_wassink24} and \citet{klein_wassink27} with 
subsequent studies by \citet{johnson52a}, \citet{artyukhina66}, \citet*{crawford69a}, 
\citet{artyukhina73}, \citet{upgren79}, \citet{weis81a}, \citet{hauck81}, and \citet{jones83} 
before exploring lower mass members \citep{mermilliod90,jones91,gatewood94,williams94a}, 
binarity \citep{bolte91,mermilliod94,bouvier01a,patience02,gillen17a,hillenbrand18}, 
lithium \citep{king96a,cummings17a}, and its luminosity/mass functions \citep{lee97a,kraus07d,boudreault12}.
The advent of photographic plates with several epochs of observations
led to a exhaustive census of stellar members over a large area down to 0.1 M$_{\odot}$ 
\citep{hambly95}. The cluster has also been targeted in X-rays by ROSAT \citep{randich95b}
and XMM-Newton \citep{franciosini03}.
The arrival of large-scale surveys and deep optical pencil-beam
surveys revealed the low-mass and sub-stellar populations 
\citep{adams02,kraus07d,pinfield97,chappelle05,gonzales_garcia06,boudreault10,baker10,wang11,boudreault12,boudreault13,khalaj13,wang14a}.
We compiled all the candidates identified in these surveys and found a total of 6479
sources, 2078 of them being unique among all studies, including the most recent census
of members from the $Gaia$ consortium \citep{babusiaux18}.

%
%
\section{Selection of member candidates}
\label{clustersGaia:select_members}

We implemented the kinematic procedure described by \citet{perryman98} in combination with the
Bayesian procedure proposed by \citet{luri18a} to deal with the proper motions and parallaxes
of the $Gaia$ mission.
This method determines the barycentre of the cluster and identifies potential members 
based on their velocities in space (Table \ref{clustersGaia:tab_cluster_param}). For the details 
of the implementation, we refer the reader to our study of the Hyades cluster \citep{lodieu19a}.

For completeness, we briefly summarise the steps of the procedure here.
We calculate the space velocity of all objects in our initial catalogue following the equations
in \citet{perryman98}. We calculate the cluster barycentre and mean space velocity from the
sample of \citet{babusiaux18} and improve it with our final sample. For the sample of $Gaia$
sources with radial velocity measurements, we compute the velocity vector in space.
The member candidates statisfy a maximum value of a ``degree-of-freedom'' parameter,
which is described by a confidence region based on the barycentric and velocity positions.

%
%
%
\begin{table*}
\centering
\caption{
Positional and kinematics data of the Pleiades, Praesepe, and \APer{} clusters. 
We infer tidal radii of 11.6, 10.7, and 9.5 pc for the Pleiades, Praesepe, and \APer{} clusters.
}
\begin{tabular}{@{\hspace{0mm}}l @{\hspace{2mm}}c @{\hspace{2mm}}c @{\hspace{2mm}}c @{\hspace{2mm}}c @{\hspace{2mm}}c @{\hspace{2mm}}c @{\hspace{2mm}}c @{\hspace{2mm}}c@{\hspace{0mm}}}
\hline
\hline
Cluster   & Distance  & Velocity &  \multicolumn{3}{c}{$\mathbf{b}_c$(pc)}  & \multicolumn{3}{c}{$\mathbf{v}_c$(km s$^{-1}$)} \cr
          &           &          &    $ b_x$ & $b_y$ & $b_z$  &  $v_x$      &  $v_y$      &  $v_z$ \cr
\hline
          &  pc       & km s$^{-1}$ &   pc   &  pc   &  pc    & km s$^{-1}$ & km s$^{-1}$ & km s$^{-1}$ \cr
\hline
\APer{}   & 177.68$\pm$0.84 & 28.70$\pm$0.52 &  $-$148.72$\pm$0.56 & 95.20$\pm$0.60 & $-$19.77$\pm$0.47 & $-$14.22$\pm$0.49 & $-$23.99$\pm$0.31 & $-$6.77$\pm$0.08 \cr
Pleiades  & 135.15$\pm$0.43 & 32.73$\pm$0.18 &  $-$120.25$\pm$0.33 & 28.94$\pm$0.24 & $-$54.49$\pm$0.23 & $-$7.04$\pm$0.27 & $-$28.53$\pm$0.07 & $-$14.40$\pm$0.13 \cr
Praesepe  & 187.35$\pm$3.89 & 49.18$\pm$2.44 &  $-$141.91$\pm$2.59 & $-$69.44$\pm$1.97 & 100.71$\pm$2.23 & $-$43.44$\pm$1.87 & $-$20.73$\pm$1.02 & $-$10.10$\pm$1.74 \cr
\hline
\label{clustersGaia:tab_cluster_param}
\end{tabular}
\end{table*}
\subsection{Member candidates in \APer{}}
\label{clustersGaia:select_members_APer}

We derived a mean distance of 177.68$\pm$0.84 pc (Table \ref{clustersGaia:tab_cluster_param}),
consistent with the original range of distances \citep[175--190 pc;][]{pinsonneault98,robichon99,makarov06}
and slightly further (within 2$\sigma$) than distance from the re-reduction of the Hipparcos data 
\citep[172.4$\pm$2.7 pc]{vanLeeuwen09}. Our value lies within 3$\sigma$ from the $Gaia$ DR2 value 
\citep[174.89$\pm$0.16 pc;][]{babusiaux18}. The brightest members of the cluster with $G$\,=\,6.3--13.7 mag
show a mean line-of-sight velocity of $-$0.47$\pm$3.78 km s$^{-1}$ with a mean dispersion of 6.5 km s$^{-1}$, 
in agreement with earlier studies of bright members \citep{stauffer89b,prosser94,mermilliod08b}. This 
translates into a mean cluster velocity of 28.7$\pm$0.5 km s$^{-1}$ (Table \ref{clustersGaia:tab_cluster_param}).
The kinematic method returned a total of 3123 member candidates with $G$\,=\,3.86--21.0 mag, corresponding
to model-dependent masses of 5.34--0.048 M$_{\odot}$ assuming an age of 90 Myr for \APer{}.

The core of the cluster was estimated to 1.3 degrees in radius \citep{artyukhina72a,kharchenko05a}, 
translating into about 4 pc at the distance of the cluster. The core radius is the radius at which 
the brightness drops to one-half the central value and is equal to 0.64 times the Plummer radius
\citep{plummer15a}. We calculated the density of sources in concentric circles from the centre up
to 30 pc with a sampling of 0.4 pc and dividing by the volume (in 3D space).
We emphasise that we can not reduce the sampling too much because of the small numbers of sources
in the central parsec. We found the best fit of the Plummer model to the density of sources for
a core radius of 2.3$\pm$0.3 pc, significantly smaller than previous estimates.

The tidal radius is usually considered as the limit where cluster candidates are bound to the cluster,
while sources beyond that radius are more affected by the external gravitational field of the galaxy 
than the cluster dynamics.
We calculated the tidal radius of \APer{} using the equation 3 of \citet{roeser11} with values for the
constants from \citet{piskunov06a}\footnote{$M_{c} = xl^{3.} / G \times (4*A*(A-B))$}.
We summed the masses of all kinematic members up to 15 pc per bins of 1 pc and determined the tidal
radius as the radius where the curve derived from equation 3 of \citet{roeser11} crosses the curve
defined by the masses of our candidates. We inferred a tidal radius of $\sim$9.5 pc in agreement
with the 9.7 pc reported by \citet{makarov06}. The half-mass radius is 5.6 pc and contains 247 sources.
We find 21 and 554 candidate members located within the core and tidal radius of the cluster, respectively. 
We consider sources beyond the tidal radius unbound to the cluster, yielding to a much lower statistical 
probability of membership. We limited our study to three times the tidal radius of the cluster, 
i.e.\ 28.5 pc where we have 2041 sources. We find 22 pairs of stars
with separations less than 5 arcsec including four marked as duplicated in $Gaia$ DR2\@.
We have 167 sources with radial velocities from $Gaia$ DR2 (about 5.3\%).

We cross-matched the list of previously-published members with our full sample and the subsample
within the tidal radius with a matching radius of 3 arcsec. The numbers are quoted in
Table \ref{tab_clustersGaia:APer_Xmatch_Summary} and provide a level of reliability for earlier 
studies of the cluster but not strictly completeness due to the variety of areas and depths under study. 

We plot the positions of the \APer{} kinematic member candidates of the cluster within the core, half-mass, 
and tidal radii and up to 28.5 pc with different colours in right ascension and declination in the top 
left panel of Fig.\ \ref{fig_clustersGaia:APer_plots_general} in Appendix \ref{clustersGaia:Appendix_APer}
as well as the other diagrams involving parallax, proper motion, and magnitude from $Gaia$ 
(Fig.\ \ref{fig_clustersGaia:APer_plots_general}). We show various colour-magnitude diagrams
with the $Gaia$ filters and photometry from other public surveys in
Figs.\ \ref{fig_clustersGaia:APer_CMD_Gaia_filters}--\ref{fig_clustersGaia:APer_CMD_nonGaia_filters}
in Appendix \ref{clustersGaia:Appendix_APer}.

We observe large numbers of sources much bluer than the cluster sequences in the two faintest magnitude
bins when the $Gaia$ $\gbp$ magnitude is involved. This behaviour is reduced when the other two $Gaia$
magnitudes are taken into account and disapear when the $Gaia$ $G$ magnitude is combined with infrared
or mid-infrared magnitudes from other public surveys. This peculiarity was already noted in the
case of the Alessi 10 cluster \citep{arenou18a} as well as for nearby faint brown dwarfs and the
Praesepe cluster \citep{smart19a}. We confirm a similar behaviour in all three clusters, pointing 
towards unreliable $\gbp$ magnitudes at the faint end of $Gaia$ when dealing with red sources.

%
%
\begin{table*}
\small
\centering
\caption{Summary of the numbers of \APer{} sources from earlier studies recovered in our full $Gaia$ sample
(i.e.\ up to three times the tidal radius of the cluster).
The final list of kinematic member candidates contains 23, 224, 517, 2069 sources in 2.3 (core radius), 
5.6 (half-mass radius), 9.5 (tidal radius), and 28.5 pc (3$\times$tidal radius), respectively.
}
\begin{tabular}{l c c c l}
\hline
\hline
Method &  in $Gaia$  & Kinematic & in R$_{\rm tidal}$ & Comments \cr
\hline
\citet{fresneau80}  &  51/56   & 11  &   7  & Palomar Schmidt astrometric observations  \cr
\citet{heckmann56}  &  139/140 &  62 &  49  & See also  \citet{mitchell60} \cr
\citet{prosser92}   &  79/148  & 57  &  42  & Astrometric$+$photometric$+$ spectroscopic search  \cr
\citet{prosser94}   &  26/31   & 14  &  10  & Photometric, and spectroscopic observations  \cr
\citet{prosser98a}  &  54/89   & 28  &  23  & Follow-up 73 ROSAT sources \citep{prosser96a} \cr
\citet{prosser98b}  &  45/70   & 23  &  15  & Follow-up 130 ROSAT sources \citep{prosser96a} \cr
\citet{stauffer99}  &  24/28   & 14  &   5  & Lithium depletion boundary  \cr
\citet{barrado02a}  &  59/101  & 42  &  24  & Deep wide-field optical survey  \cr
\citet{deacon04}    &  289/302 & 192 & 111  & Photographic plates \cr
\citet{lodieu05a}   &  1/39    &  1  &   0  & Near-infrared photometric selection  \cr
\citet{lodieu12c}: Table A1: &  367/494 & 241  &  150  & Previous members \cr
\citet{lodieu12c}: Tables B1$+$C1: &  690/728$+$617/685 & 441$+$494  &  240$+$268  & New members \cr
\citet{babusiaux18} &  740/740 & 739  &  467  & $Gaia$ DR2  \cr
\hline
\label{tab_clustersGaia:APer_Xmatch_Summary}
\end{tabular}
\end{table*}
%

%
%
%
\begin{table*}
\centering
\caption{Summary of the numbers of Pleiades sources from earlier studies recovered in our full $Gaia$ sample
(i.e.\ up to three times the tidal radius of the cluster).
We have 106, 495, 1248, and 2195 kinematic member candidates in 3.1, 4.5, 11.6, and 34.8 pc, respectively.
}
\begin{tabular}{l c c c l}
\hline
\hline
Method &  in $Gaia$  & Kinematic & in R$_{\rm tidal}$ & Comments \cr
\hline
\citet{hambly93}    & 429/440   &  385 &  334  & Photographic plates  \cr
\citet{zapatero97a} &   3/10    &    1 &    2  & CCD-based $R,I$ survey \cr
\citet{stauffer98b} &  16/20    &   12 &    9  & Lithium depletion boundary; MHO targets  \cr
\citet{bouvier98}   &  19/26    &   13 &    8  & Proper motions in \citet{moraux01}  \cr
\citet{moraux03}    & 103/109   &   93 &   78  & Brown dwarfs and IMF  \cr
\citet{deacon04}    & 882/916   &  776 &  642  & Astrometry and photometry of photographic plates  \cr
\citet{bihain06}    &  14/34    &   11 &    6  & Deep $I+J$ survey of 1.8 deg$^{2}$  \cr
\citet{stauffer07}  & 1361/1416 & 1129 &   928 & Compilation of earlier studies (their Table 1)  \cr
\citet{casewell07}  &   8/23    &    7 &     2 & Deep CFH12K survey combined with UKIDSS  \cr
\citet{lodieu07c}   &  82/116   &   69 &    46 & UKIDSS GCS DR1 combined with 2MASS  \cr
\citet{lodieu12a}   & 1176/1314 &  863 &   700 & Selection in UKIDSS GCS DR9  \cr
\citet{zapatero14b} &   5/24    &    4 &    2  & See \citet{zapatero14c} for spectroscopy  \cr
\citet{bouy15b}     & 1775/2010 & 1421 &  1070 & See also \citet{bouy13} and \citet{sarro14}  \cr
\citet{olivares18a} & 1960/2152 & 1467 &  1097 & p$\geq$0.75 from Bayesian hierarchical model  \cr
\citet{babusiaux18} & 1326/1326 & 1295 &  1087 & $Gaia$ DR2  \cr
\hline
\label{tab_clustersGaia:Pleiades_Xmatch_Summary}
\end{tabular}
\end{table*}
%

%
%
\begin{table*}
\centering 
\caption{Summary of the numbers of Praesepe sources from earlier studies recovered in our $Gaia$ sample
(i.e.\ up to three times the tidal radius of the cluster).
We have 139, 336, 721, and 1847 kinematic member candidates in 3.5, 5.9, 10.7, and 32.1 pc, respectively.
}
\begin{tabular}{l c c c l}
\hline
\hline
Method &  in $Gaia$  & Kinematic & in R$_{\rm tidal}$ & Comments \cr
\hline
\citet*{jones83}          & 269/282   & 179 &  144 & Palomar photographic plates  \cr
\citet{hambly95}          & 495/515   & 407 &  269 & Photographic plates  \cr
\citet{pinfield97}        &   7/26    &   2 &    1 & Follow-up in \citet{hodgkin99} \& \citet{pinfield03}  \cr
\citet{kraus07d}          & 1083/1130 & 948 &  595 & Census of the cluster  \cr
\citet{gonzales_garcia06} &   2/20    &   0 &    0 & Deep survey focusing on sub-stellar members  \cr
\citet{boudreault10}      &  86/151   &  40 &   25 & Wide-field $I$ + $JK$ survey  \cr
\citet{baker10}           & 137/145   &  97 &   49 & Cross-match of SDSS and UKIDSS DR7  \cr
\citet{wang11}            &   6/59    &   2 &    1 & Deep pencil-beam survey for brown dwarfs  \cr
\citet{boudreault12}      &  980/1116 & 751 &  407 & UKIDSS GCS DR9; see also \citet*{boudreault13}  \cr
\citet{khalaj13}          & 864/893   & 726 &  490 & Cross-match of PPMXL and SDSS  \cr
\citet{wang14a}           &  991/1040 & 825 &  568 & Cross-match of MASS, PPMXL, and Pan-STARRS \cr
\citet{babusiaux18}       &  938/938  & 920 &  644 & $Gaia$ DR2  \cr
\hline
\label{tab_clustersGaia:Praesepe_Xmatch_Summary}
\end{tabular}
\end{table*}
\subsection{Member candidates in the Pleiades}
\label{clustersGaia:select_members_Pleiades}

The distance of the Pleiades has been a subject of debate for the past decades 
\citep{pinsonneault98,soderblom98,vanLeeuwen09}. The distance has been evaluated with different techniques 
(isochrone fitting, orbital modeling, trigonometric parallax, moving cluster) usually agreeing 
on a mean value in the 130--140 pc range \citep{gatewood00,southworth05,soderblom05,melis14,an07a}.
On the contrary, analysis of the Hipparcos data suggested smaller distances close to 120 pc 
\citep{vanLeeuwen97a,robichon99,vanLeeuwen99,vanLeeuwen09}.
\citet{abramson18a} used $Gaia$ DR2 to provide a new estimate of the parallax (7.34$\pm$0.27 mas),
translating into a mean distance of 136.2$\pm$5.0 pc. We infer a distance of 135.15$\pm$0.43 pc 
for the Pleiades from the Bayesian approach of \citet{luri18a} and kinematic members
(Table \ref{clustersGaia:tab_cluster_param}), in agreement with the majority of studies and with 
the distance from $Gaia$ DR1 \citep{Gaia_Prusti2016}.
We find a mean line-of-sight velocity of 5.67$\pm$2.93 km s$^{-1}$ for members of the Pleiades with
a dispersion of 4.80 km s$^{-1}$, yielding a mean cluster motion of 32.73$\pm$0.18 km s$^{-1}$.
Our values are consistent with earlier studies \citep{liu91,mermilliod97} and the recent determination
of \citet{babusiaux18}.

The kinematic method returned a total of 2281 member candidates with $G$\,=\,2.77--21.0 mag, corresponding
to model-dependent masses of 4.78--0.048 M$_{\odot}$ assuming an age of 125 Myr for the Pleiades.
We note that 342 sources have radial velocity measurements from $Gaia$ DR2 ($\sim$15\%).
We inferred the core radius following the procedure described in Section \ref{clustersGaia:select_members_APer},
yielding 2.00$\pm$0.25 pc, similar to the intervals of core radii published so far, see \citet{olivares18a}
for a summary of previous determinations \citep{pinfield98,raboud98a,adams01,converse08,converse10}.
We calculated the tidal radius as in Section \ref{clustersGaia:select_members_APer}, deriving 11.6 pc,
with a total of 1248 sources. The number of sources within the core and half-mass (4.5 pc) radii are 
106 and 495, respectively. 
We limited our search for member candidates up to 3 times the tidal radius (i.e.\ 34.8 pc).
We find 36 pairs with separations less than 5 arcsec, including two triple and one quadruple
systems with components marked as duplicated in Gaia DR2\@.

We cross-matched the list of previously-published members with our full sample and the subsample
within the tidal radius with a matching radius of 3 arcsec. The numbers quoted in 
Table \ref{tab_clustersGaia:Pleiades_Xmatch_Summary} provide an estimate of the level of completeness 
of earlier studies of the Pleiades although we should caution that some surveys were dedicated to
the faintest members which are too faint for $Gaia$.
We should note that we recover 2152 sources out of the most complete list of Pleiades members 
with membership probabilities higher than 75\% \citep{olivares18a}.
Among those, 1960 are in our input file (91\%) and 1467 (68\%) common to our list of members.
We display the members of Pleiades in the same diagrams as for \APer{} in 
Figs.\ \ref{fig_clustersGaia:Pleiades_CMD_Gaia_filters}--\ref{fig_clustersGaia:Pleiades_CMD_nonGaia_filters}
in Appendix \ref{clustersGaia:Appendix_Pleiades}.

\subsection{Member candidates in Praesepe}
\label{clustersGaia:select_members_Praesepe}

We determined a mean distance of 187.35$\pm$3.89 pc for Praesepe (Table \ref{clustersGaia:tab_cluster_param}), 
in agreement with the original estimate
from Hipparcos \citep[188.0$\pm$14 pc;][]{vanLeeuwen99} but 1$\sigma$ higher than the value inferred from the
re-reduction of Hipparcos data \citep[181.5$\pm$6.0 pc;][]{vanLeeuwen09}. Our value also agrees with
the values derived from $Gaia$ \citep[186.18$\pm$0.10;][]{babusiaux18}.
We find a mean line-of-sight velocity of 35.1$\pm$1.7 km s$^{-1}$ with a dispersion of 4.2 km s$^{-1}$,
translating into a cluster mean motion of 49.18$\pm$2.44 km s$^{-1}$ 
(Table \ref{clustersGaia:tab_cluster_param}).
Our estimate is comparable to ground-based radial velocity survey \citep{mermilliod99} and
the $Gaia$ DR2 mean value \citep[35.64$\pm$0.10 km s$^{-1}$;][]{babusiaux18}.

The kinematic method returned a total of 2200 member candidates with $G$\,=\,6.13--20.93 mag, corresponding
to model-dependent masses of 2.6--0.115 M$_{\odot}$ assuming an age of 600 Myr for Praesepe.
We note that 223 out of 2200 candidates have $Gaia$ DR2 line-of-sight velocity (about 10\%).
We derived a core radius of 2.6$\pm$0.2 pc, smaller than the determination of 3.5 pc by \citet{adams02a}.
We inferred a tidal radius of 10.7 pc with 721 member candidates. We derived a half-mass radius of 5.9 pc
with 336 sources, while 77 sources lie within the core radius. 
We limited our study to 32.1 pc or three times the tidal radius with a total of 1847 sources.
We find 23 pairs with separations less than 5 arcsec, including two triples with the brightest
components marked as duplicated in Gaia DR2\@.

We cross-matched the list of previously-published members with our full sample and the subsample
within the tidal radius with a matching radius of 3 arcsec. The numbers are quoted in 
Table \ref{tab_clustersGaia:Praesepe_Xmatch_Summary}
and provide a level of completeness for earlier studies in Praesepe.
We display the members of Praesepe in the same diagrams as for \APer{} in 
Figs.\ \ref{fig_clustersGaia:Praesepe_CMD_Gaia_filters}--\ref{fig_clustersGaia:Praesepe_CMD_nonGaia_filters}
in Appendix \ref{clustersGaia:Appendix_Praesepe}.

%
%
\section{White dwarfs}
\label{clustersGaia:WDs}

When the age of a star cluster is sufficiently old, at least $\sim$50 Myr
for a solar composition, some of its high-mass members may have had enough time 
to evolve down to the white dwarf (WD) cooling sequence.
The presence of a few WDs opens the possibility to determine the age of the 
cluster by evaluating the age of each WD member, as we did in our recent 
article on the Hyades cluster \citep{lodieu19a}.
In this section we use the same method to determine the age of the clusters
under study. Without repeating all the details of this technique
already described in \citet{lodieu19a}, we give a summary of the 
main steps, focusing on a few differences with respect to that paper.

\begin{figure}
\centering
\includegraphics[width=8.6cm,angle=0]{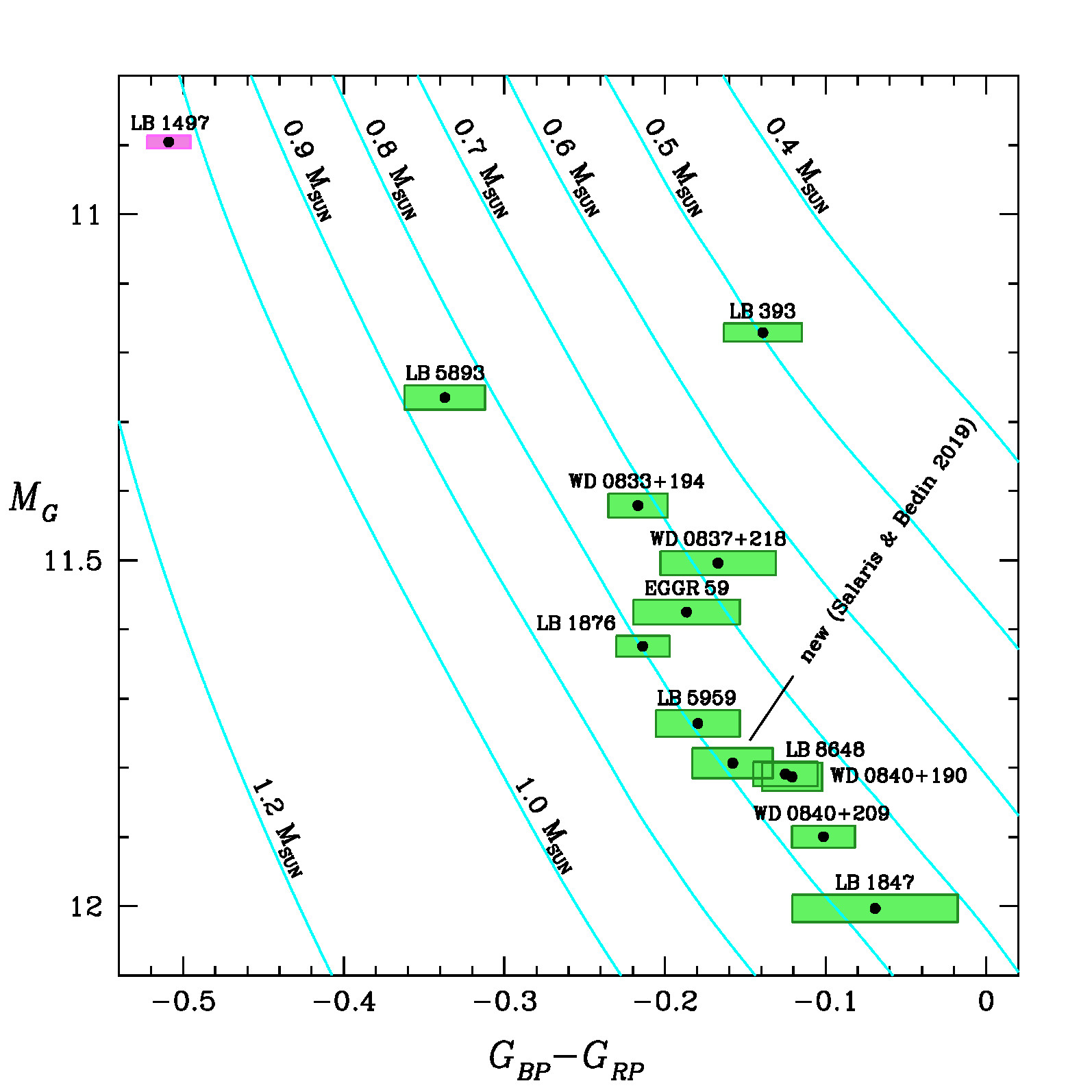}
\caption{Absolute magnitude of the white dwarfs in the Pleiades and Praesepe 
cluster as a function of the $\gbp{} - \grp{}$ colour.
The error boxes (pink for the only known WD in the Pleiades, and green for the
Praesepe WDs) correspond to the $Gaia$ DR2 photometric (+ parallax) errors.
The WD cooling tracks of \citet{bergeron11} for various masses are shown in 
light blue (see text for more details).
}
\label{tab_clustersGaia:Figure_WDs}
\end{figure}

While the selection of the WDs in each single cluster under study is discussed 
in the following sub-sections, we compare the position of the 13 WDs belonging
to the Pleiades and Praesepe clusters in the ($\gbp{} - \grp{}$, $M_G$) HR diagram
(Fig.\ \ref{tab_clustersGaia:Figure_WDs}) with the cooling tracks of single
DA white dwarfs with H-thick envelopes ($M_H$/$M_{\star}$=10$^{-4}$) from 
\citet{bergeron11}, see also \citet{tremblay11b} and references therein.
Fig.\ \ref{tab_clustersGaia:Figure_WDs} is built considering a reddening
$E(B-V)$ equal to 0.045 and 0.027 for the Pleiades and Praesepe, respectively,
using equation 1 of \citet{babusiaux18} to compute the extinction coefficients.
We note that, differently from the relatively bright WDs in the Hyades, 
for which the photometric errors were very small and negligible in the total 
error budget \citep{lodieu19a}, now the photometric errors must be taken into account.
From a cubic spline interpolation of the models, we determine the basic stellar 
parameters of each white dwarf: effective temperature, surface gravity, mass 
and WD cooling age. After having verified that the effective temperatures and 
gravities ($\log$(g)) are in good agreement with the known spectroscopic values 
from the literature, we have used the initial-to-final mass relation (IFMR) 
of \citet{El_Badry18a} to derive the mass of each WD progenitor. 
Then, for each WD progenitor, we have used the Padova evolutionary models for 
massive stars \citep{bressan12,tang14a} with nearly solar 
abundances (Z\,=\,0.017, Y\,=\,0.279) to compute the time needed to evolve from the 
pre-MS to the first thermal pulse in the asymptotic giant branch.

The results of our computations are summarised in Table \ref{tab_clustersGaia:Table_WDs}.
As in \citet{lodieu19a}, we see that the major contribution to the error budget 
comes again from the IFMR uncertainty on the initial mass, which propagates into
the MS evolutionary time. The errors quoted for the initial masses and the 
MS evolutionary times correspond to the 95.4\% probability, as described in 
\citet{El_Badry18a}, see also their figure 3\@.
The errors quoted for the WD cooling times, effective temperatures, gravities, 
and masses are conservative values corresponding to the range of the derived 
parameters from all combinations of the 1-sigma uncertainties plotted in 
Fig.\ \ref{tab_clustersGaia:Figure_WDs}.

Finally it is important to emphasize that the WD ages are very sensititive to the reddening.
As we will see at the end of sections 5.2 and 5.3, small differences in the reddening may 
cause significant differences in age. 
When correcting for reddening, WDs are moved toward bluer colors and brighter magnitudes. 
The strong inclination of the cooling tracks makes the horizontal shift dominant, implying a 
larger mass, a smaller radius, a lower luminosity and a longer cooling time.
This is probably the main limitation of this method, which can produce 
reliable results only when the reddening is well constrained.

\subsection{White dwarfs in \APer{}}
\label{clustersGaia:WDs_APer}

Of the 14 white dwarf (WD) candidates proposed by \citet{casewell15b}
in \APer{}, none were confirmed up to now.
Only two of them (AP\,WD01, AP\,WD09), both classified as DA by \citet{casewell15b}, 
are in our list of kinematic candidates but none lies within the tidal radius 
of the cluster. The closest one, AP\,WD01, is at 12.6 pc from the cluster center, 
while AP\,WD09 is at 27.2 pc. When we compute the total age of these two WDs 
(Main-Sequence (MS) age plus WD cooling age), we obtain 130$^{+30}_{-29}$ Myr and 
734$^{+253}_{-187}$ Myr respectively \citep[assuming $E(B-V)$\,=\,0.090;][]{babusiaux18}.
These values, in particular the second one, are significantly larger than the ages 
derived from isochone fitting of the upper main-sequence \citep[50--80 Myr;][]{meynet93}
and the lithium depletion boundary \citep[85$\pm$10 Myr;][]{stauffer99,barrado04b},
suggesting that at least AP\,WD09 does not belong to the cluster.

%
%

%
%
\begin{table*} 
\centering
\caption[]{Derived parameters for the single WD member of the Pleiades 
(first line) and the 11 WDs of the Praesepe cluster.}
\vspace{2mm}
\begin{tabular}{@{\hspace{0mm}}l @{\hspace{2mm}}c @{\hspace{2mm}}c @{\hspace{2mm}}c @{\hspace{2mm}}c @{\hspace{2mm}}c @{\hspace{2mm}}c @{\hspace{2mm}}c @{\hspace{2mm}}c @{\hspace{2mm}}c @{\hspace{2mm}}c @{\hspace{2mm}}c@{\hspace{0mm}}}

\hline
\hline
Names          &   $Gaia$ DR2 ID    & M$_{G}$ & d$^*$& SC & T$_{\rm eff}$           & $\log$(g)               & M$_{\rm WD}$           & M$_{\rm MS}$           & T$_{\rm WD}$      & T$_{\rm MS}$         & T$_{\rm tot}$ \\
\hline
               &                    &  mag    & pc   &    & K                       & dex                     & M$_{\odot}$            & M$_{\odot}$            & Myr               & Myr                  & Myr         \\
\hline
LB\,1497       & 66697547870378368  & 10.90   & 6.04 & DA & 33001$^{+2033}_{-1595}$ &  8.67$^{+0.06}_{-0.06}$ & 1.05$^{+0.03}_{-0.04}$ & 6.03$^{+1.34}_{-0.71}$ & 58$^{+1}_{-2}$    &   74$^{+26}_{-27}$   &  132$^{+26}_{-27}$ \\
\hline
LB\,5893       & 661270898815358720 & 11.26   & 2.18 & DA & 21334$^{+1332}_{-1201}$ &  8.37$^{+0.08}_{-0.08}$ & 0.85$^{+0.05}_{-0.05}$ & 3.62$^{+0.58}_{-0.44}$ & 126$^{+11}_{-10}$ &  179$^{+81}_{-62}$   &  304$^{+81}_{-62}$   \\ 
LB\,390=EGGR\,59& 664325543977630464& 11.57   & 2.55 & DA & 15428$^{+1087}_{-955}$  &  8.20$^{+0.09}_{-0.09}$ & 0.74$^{+0.06}_{-0.06}$ & 2.96$^{+0.41}_{-0.34}$ & 250$^{+21}_{-21}$ &  418$^{+189}_{-133}$ &  668$^{+189}_{-133}$ \\ 
WD\,0840$+$209 & 661841163095377024 & 11.90   & 4.56 & DA & 13356$^{+515}_{-473}$   &  8.25$^{+0.05}_{-0.05}$ & 0.76$^{+0.03}_{-0.03}$ & 3.18$^{+0.46}_{-0.37}$ & 405$^{+25}_{-23}$ &  365$^{+165}_{-118}$ &  770$^{+165}_{-118}$ \\ 
LB\,8648       & 660178942032517760 & 11.81   & 5.37 & DA & 13913$^{+523}_{-511}$   &  8.23$^{+0.05}_{-0.05}$ & 0.75$^{+0.03}_{-0.03}$ & 3.11$^{+0.45}_{-0.36}$ & 352$^{+21}_{-18}$ &  384$^{+174}_{-123}$ &  736$^{+174}_{-123}$ \\ 
LB\,5959       & 659494049367276544 & 11.74   & 5.72 & DA & 15298$^{+833}_{-714}$   &  8.29$^{+0.07}_{-0.07}$ & 0.79$^{+0.05}_{-0.04}$ & 3.31$^{+0.49}_{-0.39}$ & 297$^{+20}_{-20}$ &  311$^{+138}_{-102}$ &  608$^{+138}_{-102}$ \\ 
WD\,0840+190   & 661010005319096192 & 11.81   & 9.49 & DA & 13809$^{+485}_{-471}$   &  8.23$^{+0.05}_{-0.05}$ & 0.75$^{+0.03}_{-0.03}$ & 3.09$^{+0.44}_{-0.36}$ & 357$^{+22}_{-19}$ &  389$^{+177}_{-125}$ &  746$^{+177}_{-125}$ \\ 
LB\,393        & 661297901272035456 & 11.17   &11.79 & DA & 13876$^{+676}_{-591}$   &  7.79$^{+0.07}_{-0.07}$ & 0.50$^{+0.04}_{-0.04}$ & 0.68$^{+0.01}_{-0.01}$ & 184$^{+13}_{-13}$ & 4559$^{+129}_{-127}$ & 4743$^{+129}_{-127}$ \\ 
LB\,1876       & 661353224747229184 & 11.62   &13.10 & DA & 16332$^{+622}_{-556}$   &  8.30$^{+0.05}_{-0.05}$ & 0.80$^{+0.03}_{-0.03}$ & 3.30$^{+0.49}_{-0.39}$ & 248$^{+13}_{-13}$ &  302$^{+134}_{-99}$  &  550$^{+134}_{-99}$  \\ 
WD\,0833+194   & 662798086105290112 & 11.42   &14.17 & DA & 16349$^{+679}_{-609}$   &  8.17$^{+0.06}_{-0.06}$ & 0.72$^{+0.04}_{-0.04}$ & 2.81$^{+0.38}_{-0.32}$ & 200$^{+12}_{-12}$ &  464$^{+206}_{-146}$ &  664$^{+206}_{-146}$ \\ 
LB\,1847       & 661311267210542080 & 12.00   &25.61 & DA & 12631$^{+1280}_{-943}$  &  8.27$^{+0.10}_{-0.11}$ & 0.77$^{+0.07}_{-0.07}$ & 3.21$^{+0.47}_{-0.38}$ & 482$^{+40}_{-60}$ &  351$^{+158}_{-114}$ &  833$^{+158}_{-114}$ \\ 
WD\,0837+218   & 665139697978259200 & 11.50   &75.69 & DA & 14825$^{+1111}_{-954}$  &  8.11$^{+0.10}_{-0.09}$ & 0.68$^{+0.06}_{-0.06}$ & 2.52$^{+0.32}_{-0.28}$ & 242$^{+22}_{-21}$ &  607$^{+255}_{-189}$ &  849$^{+255}_{-189}$ \\ 
\hline
\multicolumn{10}{l}{$^*$ Distance from the center of the cluster.}\\
\vspace{2mm}
\end{tabular}
\label{tab_clustersGaia:Table_WDs}
\end{table*}
\subsection{White dwarfs in the Pleiades}
\label{clustersGaia:WDs_Pleiades}

In the Pleiades there is only one undisputed white dwarf, LB\,1497 \citep{eggen65a},
at 6.04 pc from the cluster center, within the tidal radius but beyond the half-mass 
radius (Fig.\ \ref{tab_clustersGaia:Figure_WDs}; first line of Table \ref{tab_clustersGaia:Table_WDs}). 
Another candidate, GD\,50, is ruled out based on its $Gaia$ DR2 
distance of 31.21$\pm$0.06 pc, which is not compatible with the Pleiades.

From its unique white dwarf LB\,1497, and assuming $E(B-V)$\,=\,0.045 \citep{babusiaux18},
we derive an age of 132$^{+26}_{-27}$ Myr
for the Pleiades, which is older than the isochrone fitting \citep[77.6 Myr;][]{mermilliod81}
but consistent within 1$\sigma$ to ages from lithium depletion boundary technique 
\citep[130$\pm$20 Myr;][]{stauffer98,barrado04b}, model fitting with enhanced
overshooting \citep[150 Myr;][]{mazzei89}, and models incorporating rotation 
\citep[110--160 Myr;][]{gossage18}.
Just to give an idea of possible systematics due to an incorrect interstellar extinction
coefficient, the age would increase to 174$^{+43}_{-35}$ Myr if we assume zero reddening.

\subsection{White dwarfs in the Praesepe}
\label{clustersGaia:WDs_Praesepe}

There are 11 known WD candidates in the Praesepe cluster published in the literature
\citep{dobbie04,dobbie06a,casewell09}.
\citet{salaris19} confirmed their membership with $Gaia$ DR2 and identified a new member ($Gaia$ DR2 ID 662998983199228032) which is not considered in our analysis given that its spectral class is not known.

For most of the remaining 11 DA WDs, assuming $E(B-V)$=0.027 \citep{babusiaux18}, we obtain a total age 
between 
550 and 850 Myr (Table \ref{tab_clustersGaia:Table_WDs}).
Only one star, LB\,5893, is below this range with 304 Myr.
Another star, LB\,393, returns an age completely different of 4.7 Gyr.
This star lies at 11.8 pc from the cluster center, about 1 pc beyond the tidal radius, implying that it might 
not be bound to Praesepe. 
Moreover, together with LB\,1847, it was excluded from the analysis 
of \citealt{salaris19} because of possible problems with $\gbp{}$ and $\grp{}$ photometry 
\citep[see][for more details]{salaris19}.
And finally another doubt is raised by the low mass of LB\,393, with a mass of 0.50 M$_{\odot}$.
This value is close to the typical mass of a hot subdwarf (0.47 M$_{\odot}$), that is a star with 
an initial mass up to $\sim$3 M$_{\odot}$, which has lost most of its envelope near the tip of 
the red giant branch and, after having exhausted its He fuel during the (extreme) 
horizontal branch, has evolved directly to the WD cooling track, without 
experiencing a second expansion of the envelope in the asymptotic giant branch \citep{han02a}.
If this was the true evolutionary path of LB\,393, its real age could be much less than 4.7 Gyr.
When excluding only LB\,393, we obtain a mean age of 673$^{+55}_{-39}$ Myr 
for Praesepe. When excluding also the ``outlier'' LB\,5893 and LB\,1847 that might have photometric 
problems as stated above, we derive a very similar mean age of 699$^{+65}_{-47}$ Myr. 
Considering only the five WDs within the tidal radius of the cluster (10.7 pc), 
and excluding LB\,5893, we infer a mean age of 706$^{+76}_{-54}$ Myr. 
These estimates are in relatively good agreement with the oldest ages from models incorporating
stellar rotation \citep[$\sim$800 Myr;][]{brandt15b} but also compatible with the range of ages 
quoted for Praesepe by other authors
\citep[590--790 Myr;][]{mermilliod81,bonatto04a,fossati08,delorme11,gossage18}.

With a zero reddening, the previous age estimates 
of 673$^{+55}_{-39}$, 677$^{+65}_{-46}$ and 706$^{+76}_{-54}$ Myr 
would turn into 799$^{+64}_{-47}$, 792$^{+74}_{-53}$ and 815$^{+85}_{-61}$ Myr respectively, 
with a $\sim$17\% increase.

%
%
\section{The luminosity and mass functions}
\label{clustersGaia:LFplusMF}
\subsection{Luminosity functions}
\label{clustersGaia:LFplusMF_LF}

We derive the luminosity functions of the three clusters from our sample of candidates identified 
astrometrically from the second data release of $Gaia$. Our samples contain 517, 1248, and 721
member candidates within the tidal radii of the \APer{}, the Pleiades, and Praesepe clusters.
We did not attempt to correct the system luminosity function for binaries and postpone this analysis 
to later $Gaia$ releases where astrometric parameters of multiple systems will be incorporated. 

\begin{figure}
 \centering
  \includegraphics[width=0.86\linewidth, angle=0]{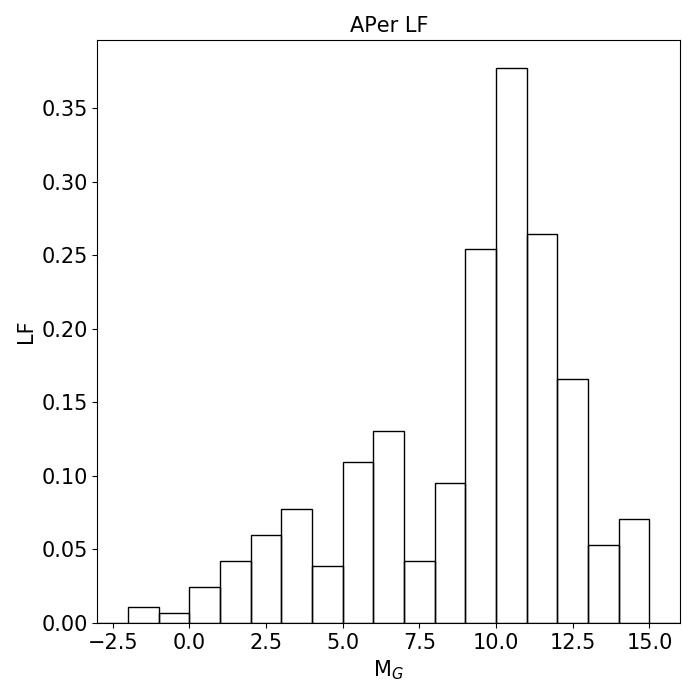}
  \includegraphics[width=0.86\linewidth, angle=0]{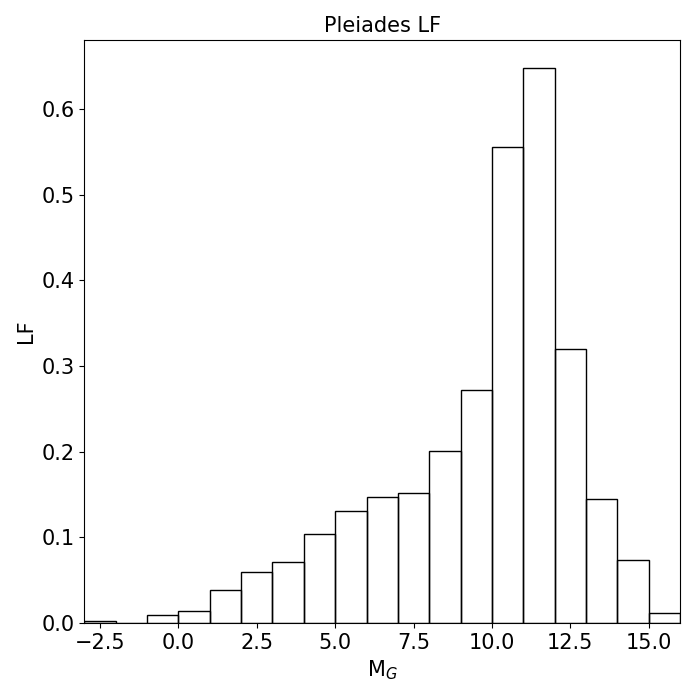}
  \includegraphics[width=0.86\linewidth, angle=0]{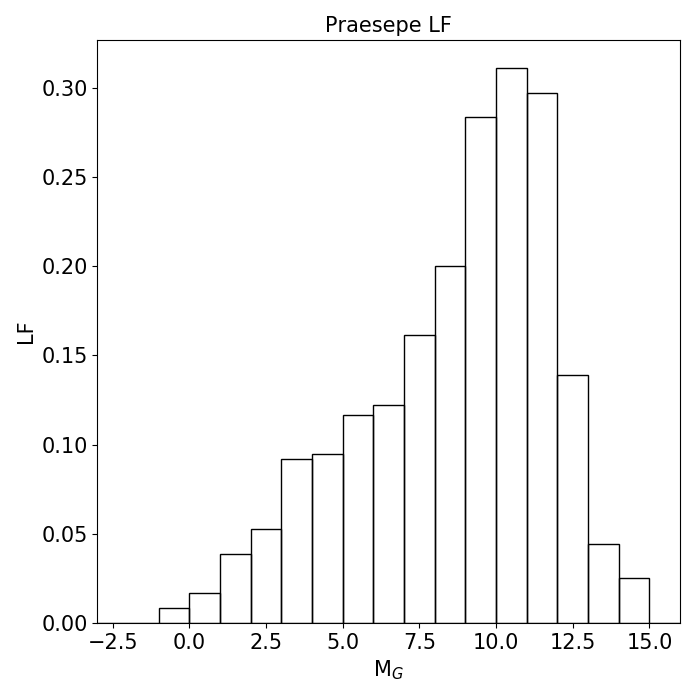}
  \caption{Luminosity functions, i.e.\ numbers of objects as a function of absolute $G$ magnitude,
  within the tidal radius of \APer{} (bottom), the Pleiades (middle), and Praesepe (top panel).
  The number of objects is scaled to a volume of one cubic parsec.
  }
  \label{fig_clustersGaia:LF_clusters}
\end{figure}

We display the system luminosity functions, i.e.\ the number of objects per absolute magnitude 
bins, with bin width of 1 mag scaled to a volume of one cubic parsec in 
Fig.\ \ref{fig_clustersGaia:LF_clusters}. We do not apply any correction to the luminosity function 
because we only consider bound members within the tidal radius of the clusters. Nonetheless, we can
discard incompleteness at the very bright end or contamination at the faint end where $Gaia$
photometry and astrometry is less reliable, especially in the case of \APer{} which lies closer
to the Galactic plane than the Pleiades and Praesepe.

The luminosity function of \APer{}, the Pleiades, and Praesepe show a peak at M$_{G}$\,=\,9--10 mag,
11--12 mag, and 9--12 mag, respectively. The luminosity functions of the Pleiades and Praesepe increase 
steadily until they reach a peak and fall off quickly after, due to a combination of smaller number of 
very-low mass members and $Gaia$ detection limits. The luminosity function of \APer{} displays more up
and down at bright magnitudes. The peak of the luminosity function appears broader with older ages
(Fig.\ \ref{fig_clustersGaia:LF_clusters}).

%
%
%

%
%
\begin{figure}
 \centering
  \includegraphics[width=0.83\linewidth, angle=0]{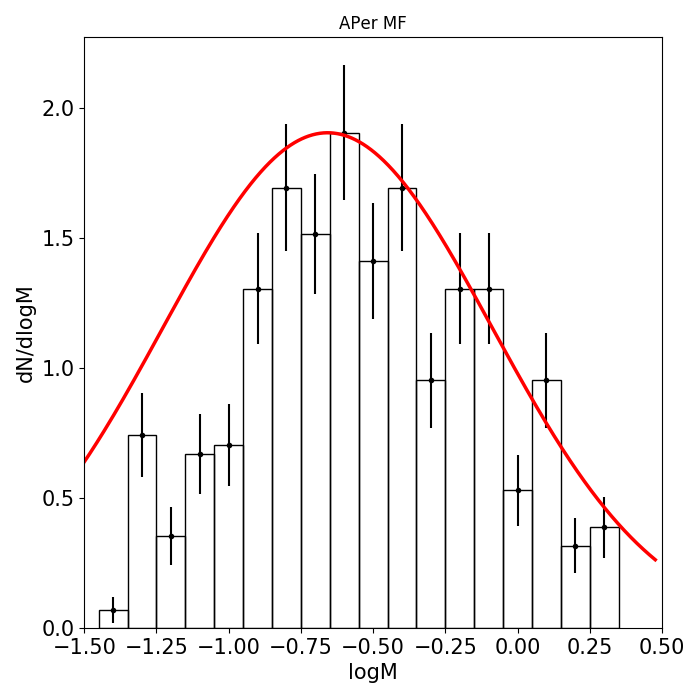}
  \includegraphics[width=0.83\linewidth, angle=0]{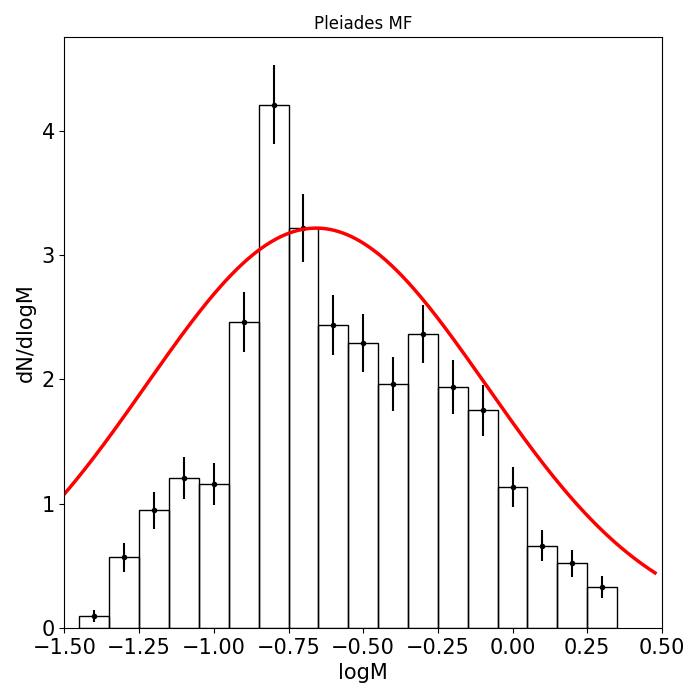}
  \includegraphics[width=0.83\linewidth, angle=0]{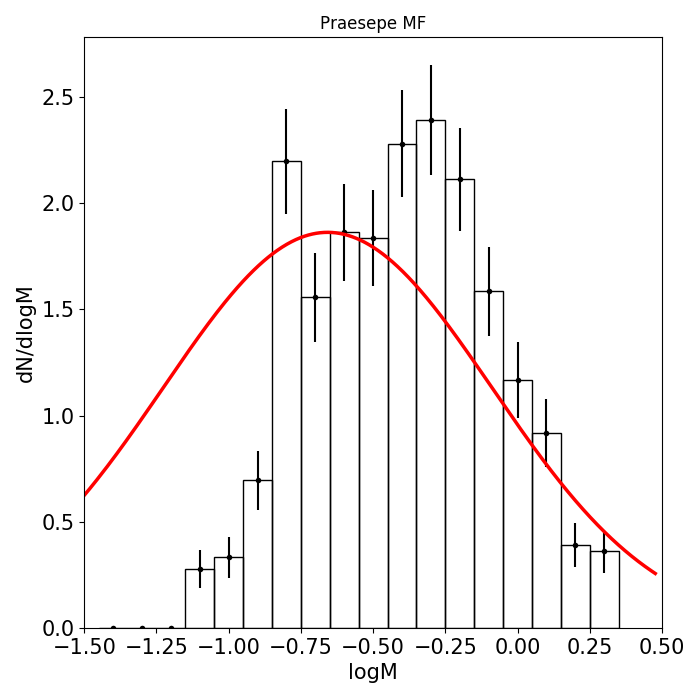}
  \caption{Mass functions of the \APer{} (bottom), the Pleiades (middle), and Praesepe (top)
  clusters scaled to a volume of one cubic parsec and a logMass bin of 0.1 using isochrones
  of 90 Myr, 125 Myr, and 600 Myr, respectively. Error bars represent the Poisson noise.
  The field mass function from \citet{chabrier03} is overplotted as a thick red line and
  normalised to the maximum value of the observed mass function at $\sim$0.25 M$_{\odot}$.
  }
  \label{fig_clustersGaia:MF_clusters}
\end{figure}
\subsection{Mass functions}
\label{clustersGaia:LFplusMF_MF}

To convert magnitudes into masses, we need a mass-luminosity relation over a wide
mass range, from A-type stars down to the hydrogen-burning limit. To derive the most
reliable present-day mass function, we would require eclipsing binaries with accurate 
masses and radii over a wide mass range in all three clusters. Unfortunately, such
binaries spanning a wide range of properties are not available. Therefore, 
we adopt a model-dependent mass-magnitude relation. We should keep in mind
this fact when interpreting the shape of the mass function.  

We combined the Padova isochrones \citep[PARSEC v1.2S + COLIBRI PR16;][]{marigo08,bressan12,marigo13,rosenfield16a,marigo17}\footnote{http://stev.oapd.inaf.it/cgi-bin/cmd} and the BT-Settl models \citep{allard12,baraffe15} to convert observed magnitudes
into masses. We choose ages of 90 Myr, 125 Myr, and 600 Myr for \APer{}, the Pleiades,
and Praesepe as the most common ages quoted in the literature (Section \ref{clustersGaia:Known_Members}).
We opted to merge both models to cover the full range of masses where the luminosities at a given
mass were the closest. In the case of \APer{} we kept the Padova and BT-Settl models above and below 
1.4 M$_{\odot}$ (M$_{G}$\,$\sim$\,3.1 mag), respectively. 

We plot the mass functions in Fig.\ \ref{fig_clustersGaia:MF_clusters} counting the number of
objects per square degree and per bins of 0.1 dex in logarithmic units, assuming the ages quoted above.
We included Poisson error bars to each bin of the mass functions.
We overplotted the field mass function in logarithmic units (thick red line) from \citet{chabrier05a},
normalised to the numbers of objects per per square degree in the bin centered at 0.25 M$_{\odot}$.
The Pleiades mass function increases steadily up to $\log(M)$\,$\sim$\,$-$0.3 dex (0.5 M$_{\odot}$),
reaching a plateau followed by a peak at around $\log(M)$\,$\sim$\,$-$0.8 dex (0.16 M$_{\odot}$)
before falling down quickly at lower masses and in the substellar regime.
In Praesepe, we observe a steep increase from high-mass stars to solar-type stars followed by a 
plateau with less low-mass stars (0.1--0.4 M$_{\odot}$). We note the excess of solar-type stars
(0.5--1.0 M$_{\odot}$ and a clear dearth of very-low-mass stars (brown dwarfs in Praesepe are too 
faint for $Gaia$).
In \APer{}, we observe a slow increase with a broad peak from 0.1 to 0.8 M$_{\odot}$ centered 
on the peak of the field mass function \cite{chabrier03}. We note a large number of sources at
$\log(M)$\,$\sim$\,$-$1.3 dex (0.05 M$_{\odot}$) in \APer{} not present in the other two clusters
(Fig.\ \ref{fig_clustersGaia:MF_clusters}). Some of these sources might be photometric non members
based on their location in colour-magnitude diagrams 
(Figs.\ \ref{fig_clustersGaia:APer_CMD_Gaia_filters}--\ref{fig_clustersGaia:APer_CMD_nonGaia_filters}
in Appendix \ref{clustersGaia:Appendix_APer}) but the excess might be real and not due to small
number statistics because we observe a peak in the luminosity function of \APer{} not detected 
in the Pleiades (Fig.\ \ref{fig_clustersGaia:LF_clusters}).

We also investigated the shape of the mass functions as a function of distance from the centre of
the cluster. We find that stars contribute in a similar way in the core and half-mass radii
with a dearth of low-mass stars. In the concentric circle delineated by the half-mass and tidal 
radii, we observe a decrease of high-mass and intermediate-mass stars as well as an increase in 
the numbers of low-mass stars as expected as a result of mass segregation.

%
%
\section{Discussion: 3D maps of the nearest open clusters}
\label{clustersGaia:3D_view}

\begin{figure*}
 \centering
  \includegraphics[width=0.31\linewidth, angle=0]{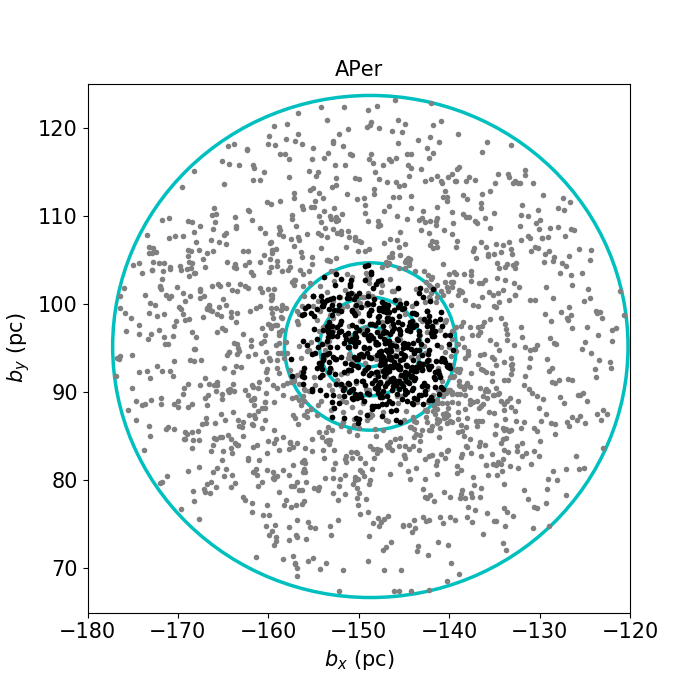}
  \includegraphics[width=0.31\linewidth, angle=0]{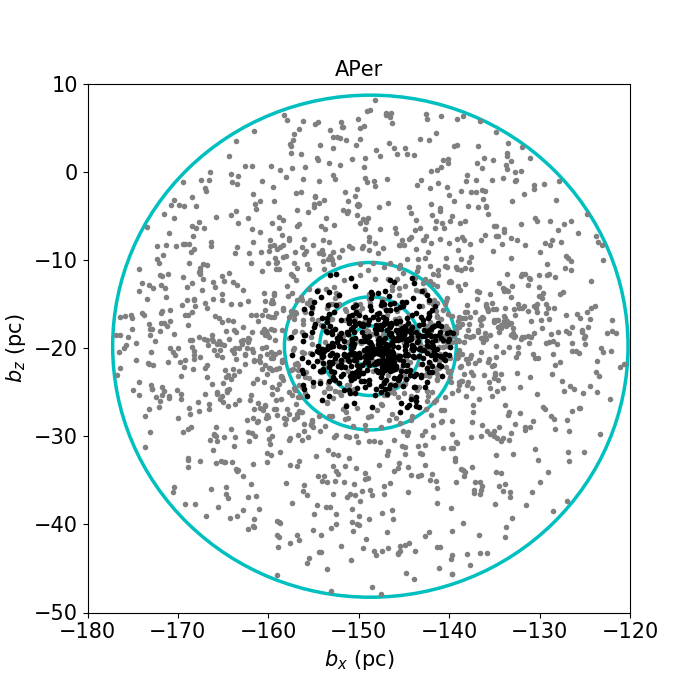}
  \includegraphics[width=0.31\linewidth, angle=0]{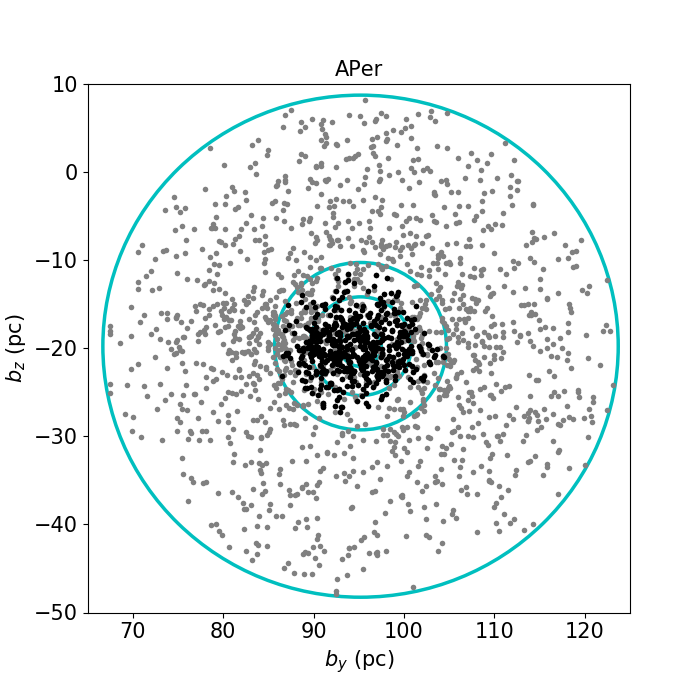}
  \includegraphics[width=0.31\linewidth, angle=0]{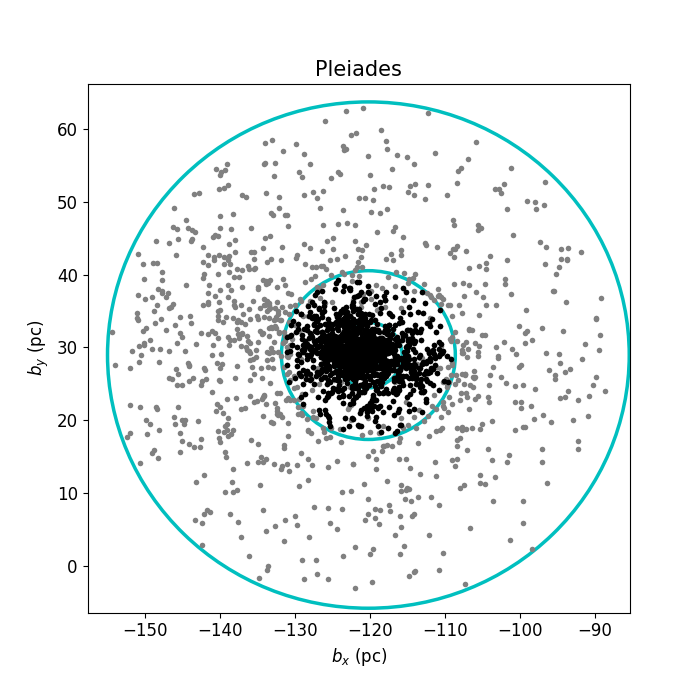}
  \includegraphics[width=0.31\linewidth, angle=0]{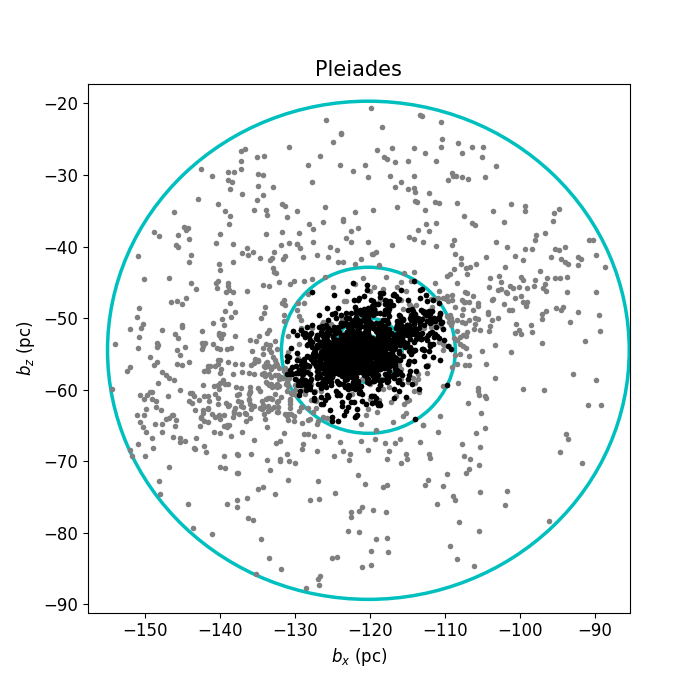}
  \includegraphics[width=0.31\linewidth, angle=0]{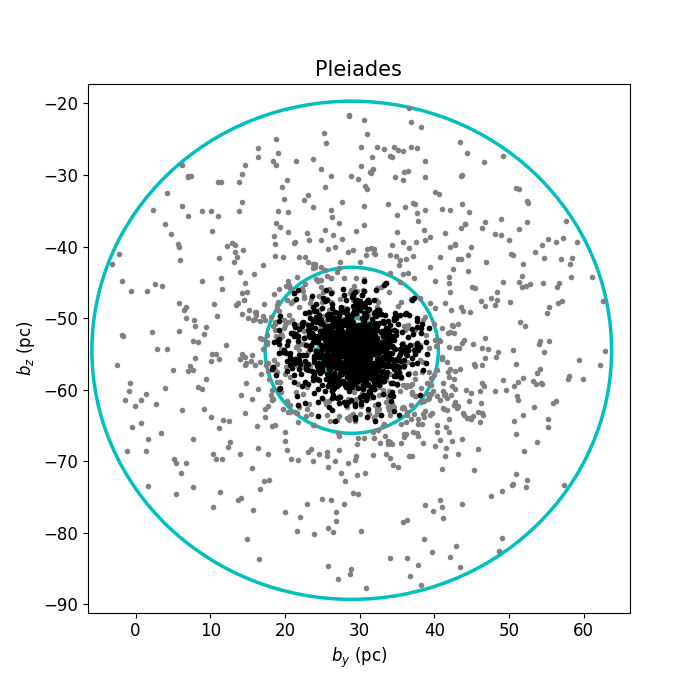}
  \includegraphics[width=0.31\linewidth, angle=0]{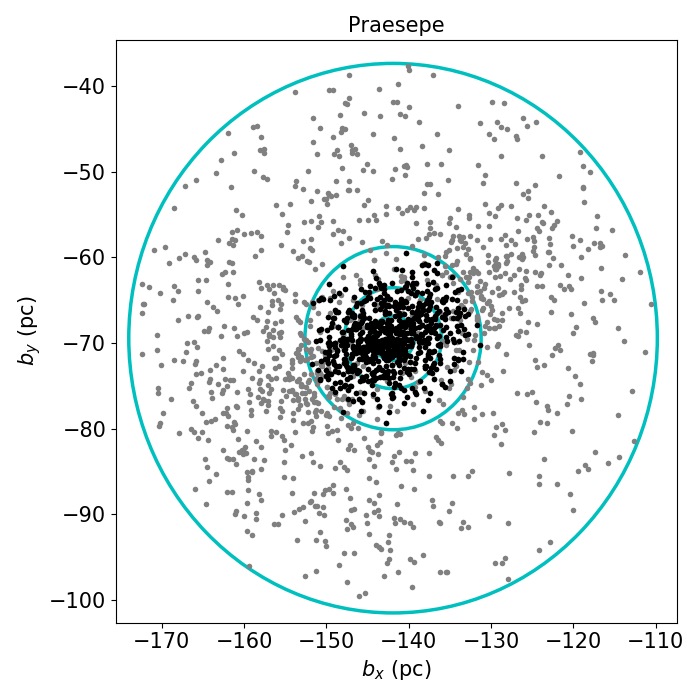}
  \includegraphics[width=0.31\linewidth, angle=0]{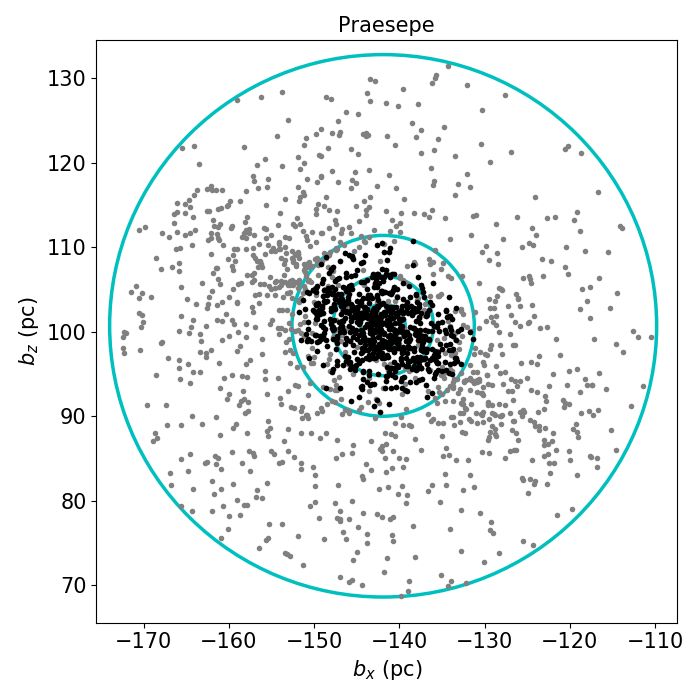}
  \includegraphics[width=0.31\linewidth, angle=0]{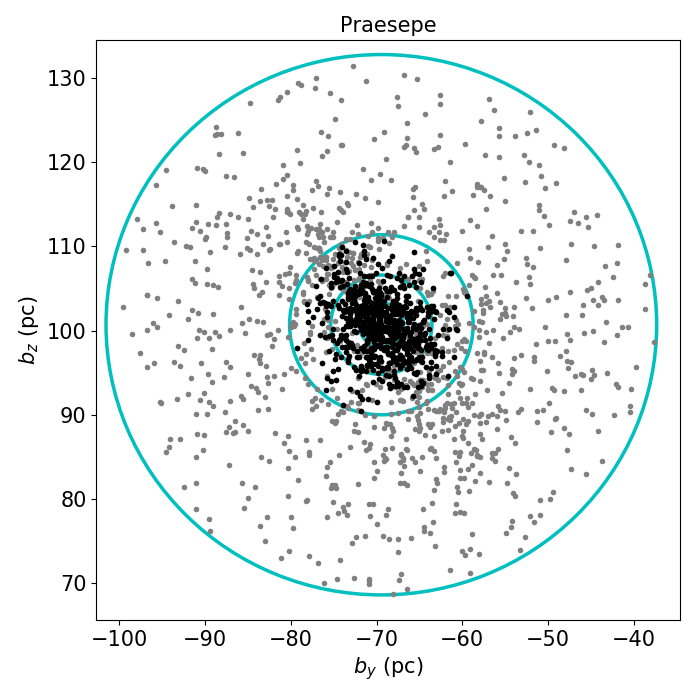}
  \caption{Positions ($\vec{b}_c$ in pc) of the \APer{} (top row),
the Pleiades (middle row), and Praesepe (bottom row) stellar and sub-stellar member candidates
in space. Overplotted in cyan are different annuli from the cluster to appreciate the
distribution and extension of cluster members. Black symbols represent member candidates
candidates inside the tidal radius while grey symbols highlight all candidates in our search
up to three times the tidal radius.
  }
  \label{fig_HyaGaia:APer_3Dmap_XYZ}
\end{figure*}

In Figure \ref{fig_HyaGaia:APer_3Dmap_XYZ} we show the positions in space of all member candidates 
within the tidal radius of the three clusters (black circles) on top of all candidates within three times
the tidal radius (grey dots).
The Galactic system is defined as follows: x is the unit vector towards ($\alpha$,$\delta$)\,=\,(0,0),
y towards (90$^{\circ}$,0), and z towards $\delta$\,=\,90$^{\circ}$.
We observe that Praesepe is the most dispersed of the three clusters most likely due to its older
age. Its members seem to disperse in all three directions in a homogeneous manner in the form
of a tidal-tail structure.
The Pleiades is well concentrated with about half of its members within the half-mass radius.
We detect the presence of a tail-like structure in the (X,Z) direction with an angle of 
about 30 degrees, extending up to several tens of parsecs from the cluster center, suggesting that 
the Pleiades is disintegating and dissipating into the greater stream of the Milky Way
\citep{maedler1846,eggen75a,eggen83a,eggen92,eggen95,dehnen98a,liang17a}.
In the case of \APer{}, most of the sources appear concentrated within about 8 pc.

We investigated further this tail-like structure of the Pleiades, selecting all candidates
within a radius of 50 pc from the centre of the Pleiades. We found 3390 sources satisfying the kinematic 
criteria. We observe that the tail extends up to about 30 pc in the direction (X,Z) and further up 
to 50 pc in the opposite direction ($-$X,$-$Z). This structure might be related to the Pleiades 
stream investigated by many authors but with no significant peculiarity 
detected in terms of abundances and age at the spectral resolution of RAVE
\citep[Radial Velocity Experiment; R\,$\sim$7500;][]{steinmetz06}, suggesting that it might not be solely 
the result of cluster disintegration \citep{famaey05,famaey08,antoja08,zhao09a,antoja12,kushniruk17}.
A better understanding of the origin of the tail-like structures is important to shed light on the 
formation of the Pleiades and Praesepe clusters and more generally the dynamical history of the Milky Way.

%
%
%

%
%
\subsection{Conclusions}
\label{clustersGaia:conclusions}

We presented updated parameters for three nearby open clusters (\APer{}, the
Pleiades, and Praesepe) as well as a revised census of member candidates within
the tidal radius over the full mass range down to the hydrogen-burning limit.

We summarise the main results of our work:
\begin{itemize}
\item [$\bullet$] We derived a mean distance and velocity of 177.68 pc and 28.70 km/s
for \APer{}, 135.15 pc and 32.73 km/s for the Pleiades, and 187.35 pc and 49.18 km/s
for Praesepe. We derived the position of the centre of the cluster and its velocity
in space for all regions.
\item [$\bullet$] We identified 517, 1248, and 721 member candidates inside the tidal radius of
\APer{} (9.5 pc), the Pleiades (11.6 pc), and Praesepe (10.7 pc), respectively. 
\item [$\bullet$] We inferred ages of 132$^{+26}_{-27}$ Myr and 705$^{+76}_{-54}$ Myr from the
WD members within the tidal radius of the Pleiades and Praesepe, respectively, taking into
account reddening towards the clusters. None of the WDs 
previously proposed as members of \APer{} are recovered as kinematic candidates.
\item [$\bullet$] We derived the luminosity and mass functions of the three clusters. We find 
two peaks at $\sim$0.5 and $\sim$0.2 M$_{\odot}$ in the mass function of the Pleiades.
The mass functions of \APer{} and Praesepe appear relatively flat in the 0.3--1.0 M$_{\odot}$.
\item [$\bullet$] We calculated the positions in 3D space and find that the members of these three 
clusters are well concentrated in a circular manner. We observe the presence of a tail-like structure
along the (X,Z) directions in the case of the Pleiades, suggesting that the cluster
might soon dissipate. We also see a tail-like structure in all three directions in the case
of Praesepe.
\item [$\bullet$] We will public release through the Vizier database the catalogues of all sources 
up to three times the tidal radius of the cluster.
\end{itemize}

Our study has ignored multiplicity over the full mass range. The next $Gaia$ release
should include information on binaries with some preliminary orbits to study the
multiplicity as a function of mass and improve the mass determinations for the members 
of these three clusters with different ages to derive a more accurate mass function.
The increment in the total mass of the cluster can reach several tens of percent
depending on the numbers of binaries and mass ratios, as in Praesepe for example
\citep{khalaj13,borodina19a}

%
%
\begin{acknowledgements}
NL and APG were financially supported by the Spanish Ministry of Economy and 
Competitiveness (MINECO) under the grants AYA2015-69350-C3-2-P and 
AYA2015-69350-C3-3-P\@. 
RS thanks Pierre Bergeron for having made available, before publication, 
the WD cooling tracks that include $Gaia$ magnitudes. 
This research has made use of the Simbad and Vizier databases, operated
at the centre de Donn\'ees Astronomiques de Strasbourg (CDS), and
of NASA's Astrophysics Data System Bibliographic Services (ADS).
This research has also made use of some of the tools developed as part of the
Virtual Observatory.
This work has made use of data from the European Space Agency (ESA) mission
{\it Gaia} (https://www.cosmos.esa.int/gaia), processed by the {\it Gaia}
Data Processing and Analysis Consortium (DPAC,
https://www.cosmos.esa.int/web/gaia/dpac/consortium). Funding for the DPAC
has been provided by national institutions, in particular the institutions
participating in the {\it Gaia} Multilateral Agreement.\\
Funding for the Sloan Digital Sky Survey IV has been provided by the Alfred P. Sloan Foundation, the U.S. Department of Energy Office of Science, and the Participating Institutions. SDSS-IV acknowledges
support and resources from the Center for High-Performance Computing at
the University of Utah. The SDSS web site is www.sdss.org.
SDSS-IV is managed by the Astrophysical Research Consortium for the 
Participating Institutions of the SDSS Collaboration including the 
Brazilian Participation Group, the Carnegie Institution for Science, 
Carnegie Mellon University, the Chilean Participation Group, the French Participation Group, 
Harvard-Smithsonian Center for Astrophysics, 
Instituto de Astrof\'isica de Canarias, The Johns Hopkins University, 
Kavli Institute for the Physics and Mathematics of the Universe (IPMU) / 
University of Tokyo, Lawrence Berkeley National Laboratory, 
Leibniz Institut f\"ur Astrophysik Potsdam (AIP),  
Max-Planck-Institut f\"ur Astronomie (MPIA Heidelberg), 
Max-Planck-Institut f\"ur Astrophysik (MPA Garching), 
Max-Planck-Institut f\"ur Extraterrestrische Physik (MPE), 
National Astronomical Observatories of China, New Mexico State University, 
New York University, University of Notre Dame, 
Observat\'ario Nacional / MCTI, The Ohio State University, 
Pennsylvania State University, Shanghai Astronomical Observatory, 
United Kingdom Participation Group,
Universidad Nacional Aut\'onoma de M\'exico, University of Arizona, 
University of Colorado Boulder, University of Oxford, University of Portsmouth, 
University of Utah, University of Virginia, University of Washington, University of Wisconsin, 
Vanderbilt University, and Yale University.
This publication makes use of data products from the Two Micron All Sky Survey, which is a joint 
project of the University of Massachusetts and the Infrared Processing and Analysis 
Center/California Institute of Technology, funded by the National Aeronautics and Space 
Administration and the National Science Foundation.\\
The UKIDSS project is defined in \citet{lawrence07}. UKIDSS uses the UKIRT Wide Field Camera 
\citet[WFCAM;][]{casali07}. The photometric system is described in \citet{hewett06}, and the 
calibration is described in \citet{hodgkin09}. The pipeline processing and science archive are 
described in Irwin et al.\ (2009, in prep) and \citet{hambly08}.\\
This publication makes use of data products from the Wide-field Infrared Survey Explorer, which 
is a joint project of the University of California, Los Angeles, and the Jet Propulsion 
Laboratory/California Institute of Technology, and NEOWISE, which is a project of the Jet 
Propulsion Laboratory/California Institute of Technology. WISE and NEOWISE are funded 
by the National Aeronautics and Space Administration. \\
The Pan-STARRS1 Surveys (PS1) and the PS1 public science archive have been made possible 
through contributions by the Institute for Astronomy, the University of Hawaii, the Pan-STARRS 
Project Office, the Max-Planck Society and its participating institutes, the Max Planck Institute 
for Astronomy, Heidelberg and the Max Planck Institute for Extraterrestrial Physics, Garching, 
The Johns Hopkins University, Durham University, the University of Edinburgh, the Queen's 
University Belfast, the Harvard-Smithsonian Center for Astrophysics, the Las Cumbres Observatory
Global Telescope Network Incorporated, the National Central University of Taiwan, the Space 
Telescope Science Institute, the National Aeronautics and Space Administration under 
Grant No.\ NNX08AR22G issued through the Planetary Science Division of the NASA Science 
Mission Directorate, the National Science Foundation Grant No. AST-1238877, the University 
of Maryland, Eotvos Lorand University (ELTE), the Los Alamos National Laboratory, and the 
Gordon and Betty Moore Foundation. \\
\end{acknowledgements}
%

%
%
\bibliographystyle{aa}
\bibliography{../../AA/mnemonic,../../AA/biblio_old} 

%
%
\begin{appendix}
\newpage
%

%
%
%
\section{Tables and diagrams for \APer{}}
\label{clustersGaia:Appendix_APer}
%

%
%
\begin{figure*}
 \centering
  \includegraphics[width=0.48\linewidth, angle=0]{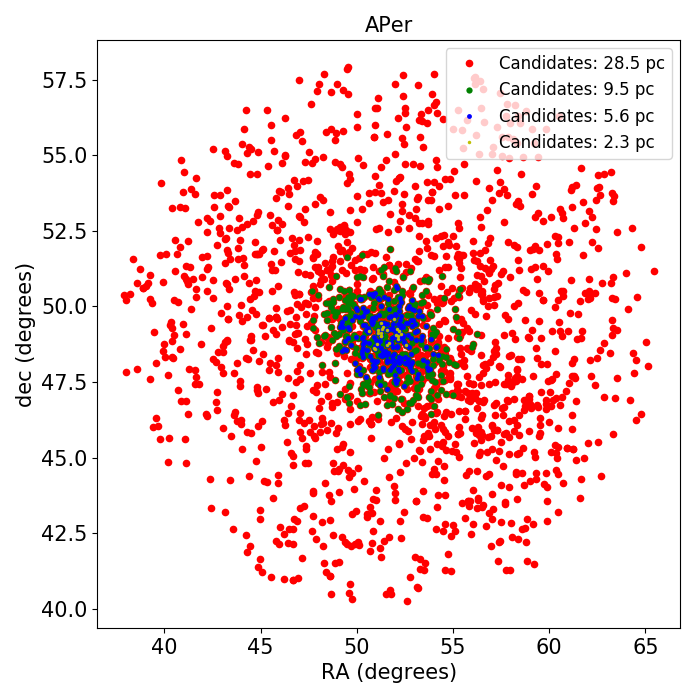}
  \includegraphics[width=0.48\linewidth, angle=0]{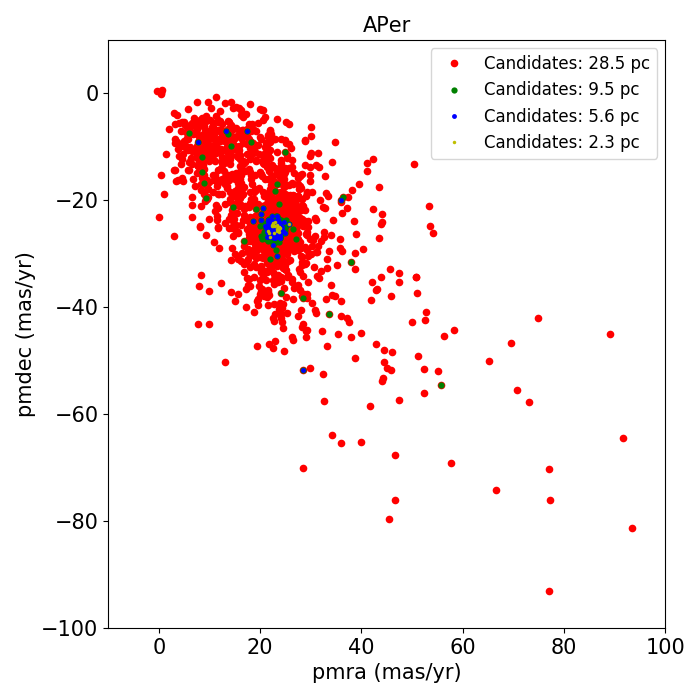}
  \includegraphics[width=0.48\linewidth, angle=0]{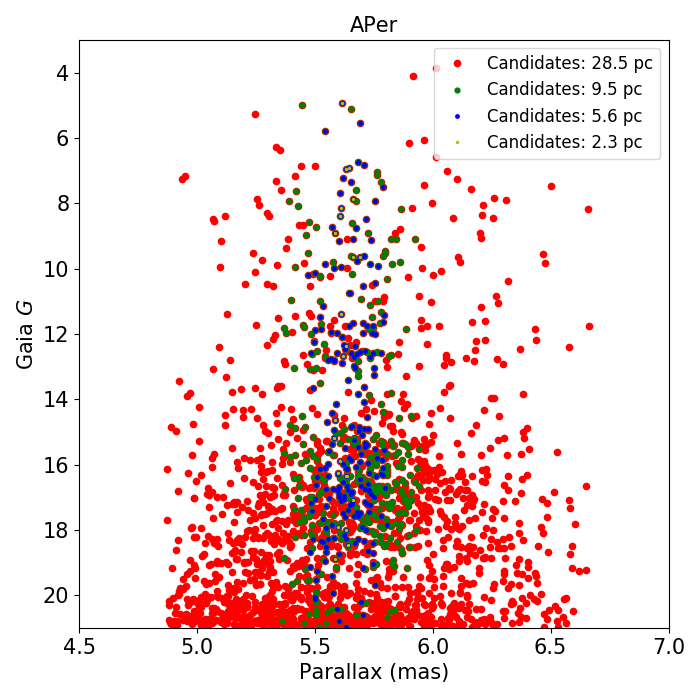}
  \includegraphics[width=0.48\linewidth, angle=0]{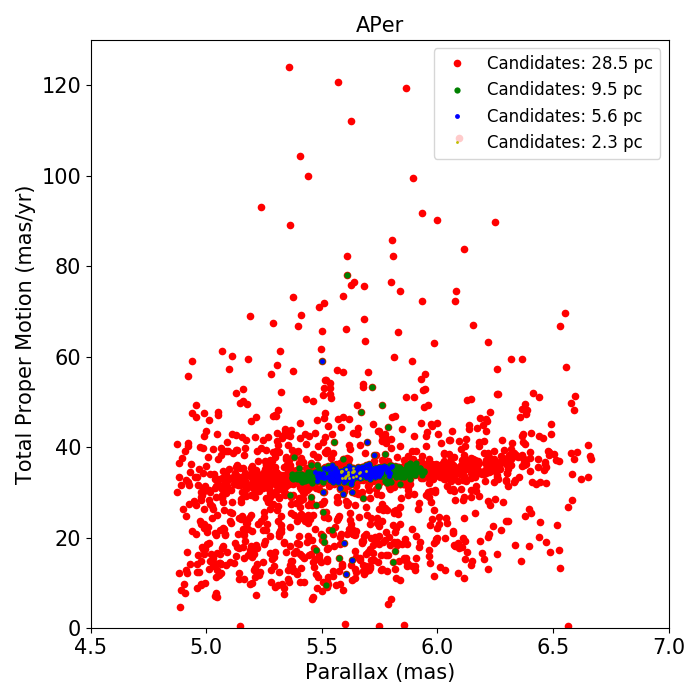}
  \caption{{\it{Top left:}} (RA,dec) diagram showing the position of \APer{} member candidates.
  Samples at different distances from the cluster centre are highlighted with distinct colours and sizes.
  {\it{Top right:}} Vector point diagram.
  {\it{Bottom left:}} $Gaia$ magnitude vs parallax. 
  {\it{Bottom right:}} Total proper motion as a function parallax.
}
  \label{fig_clustersGaia:APer_plots_general}
\end{figure*}
\begin{figure*}
 \centering
  \includegraphics[width=0.43\linewidth, angle=0]{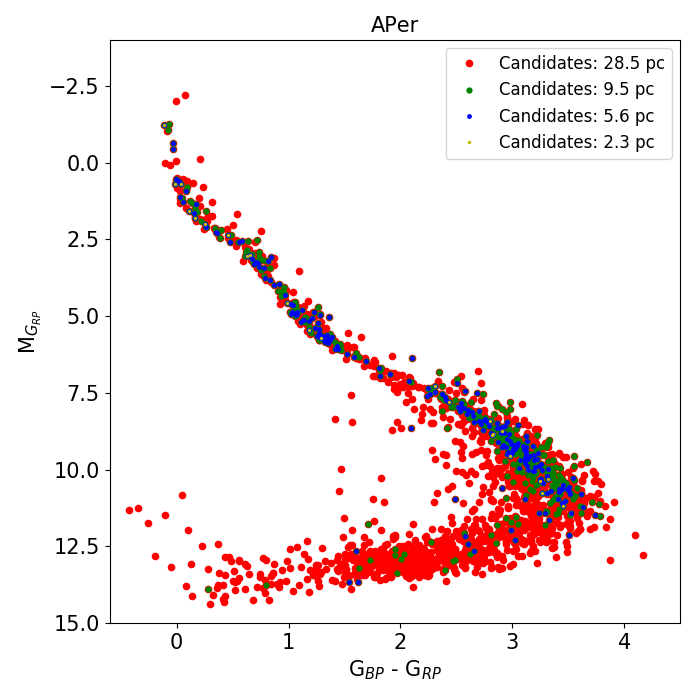}
  \includegraphics[width=0.43\linewidth, angle=0]{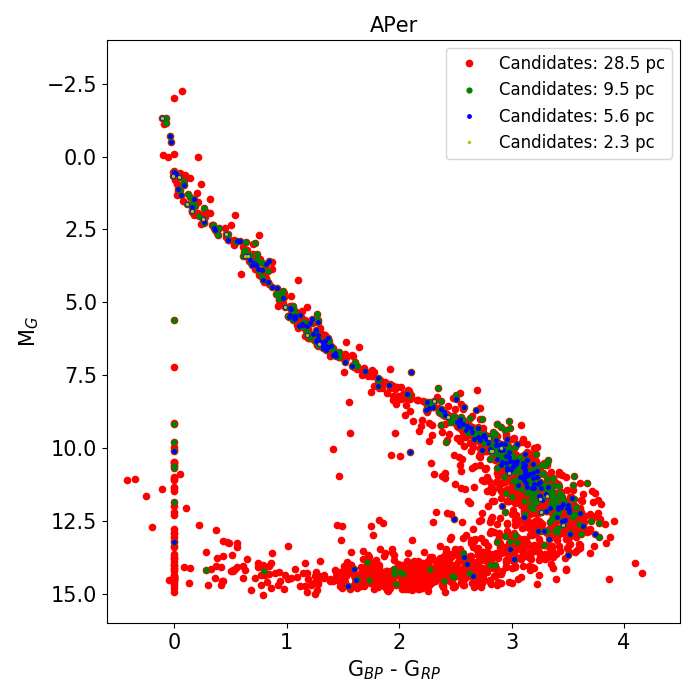}
  \includegraphics[width=0.43\linewidth, angle=0]{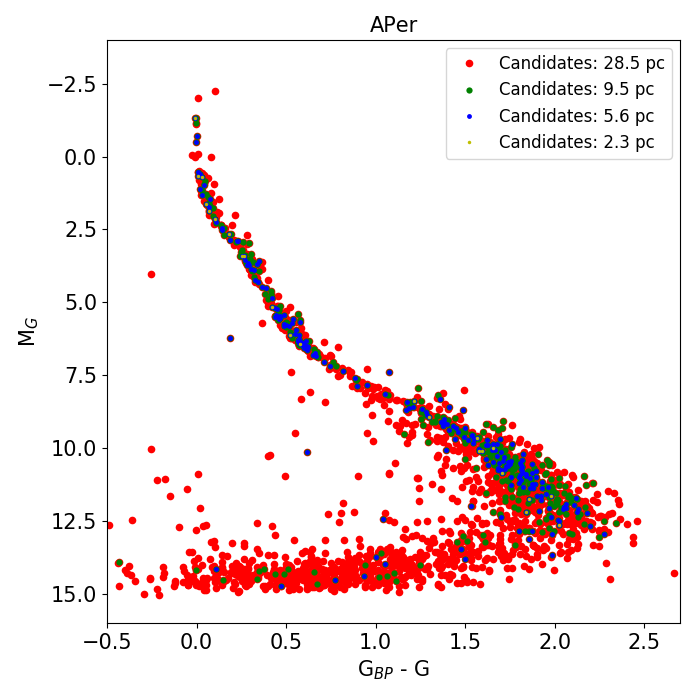}
  \includegraphics[width=0.43\linewidth, angle=0]{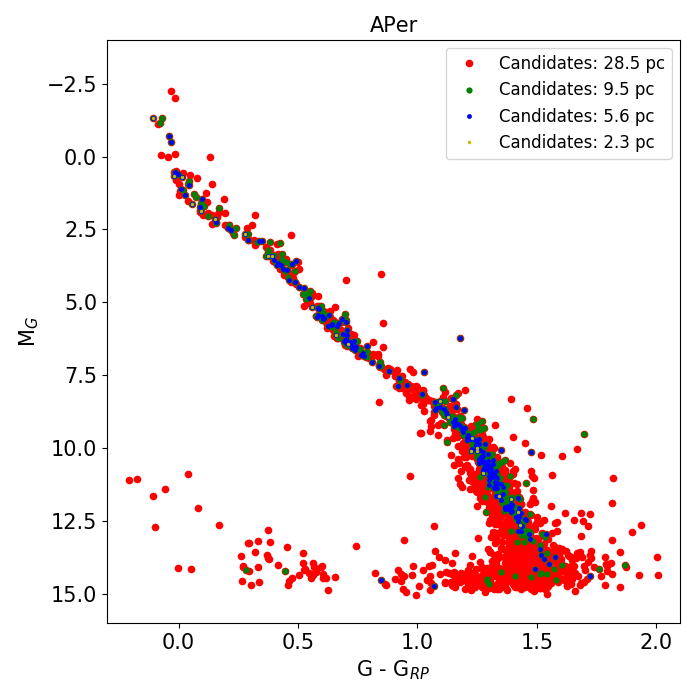}
  \includegraphics[width=0.43\linewidth, angle=0]{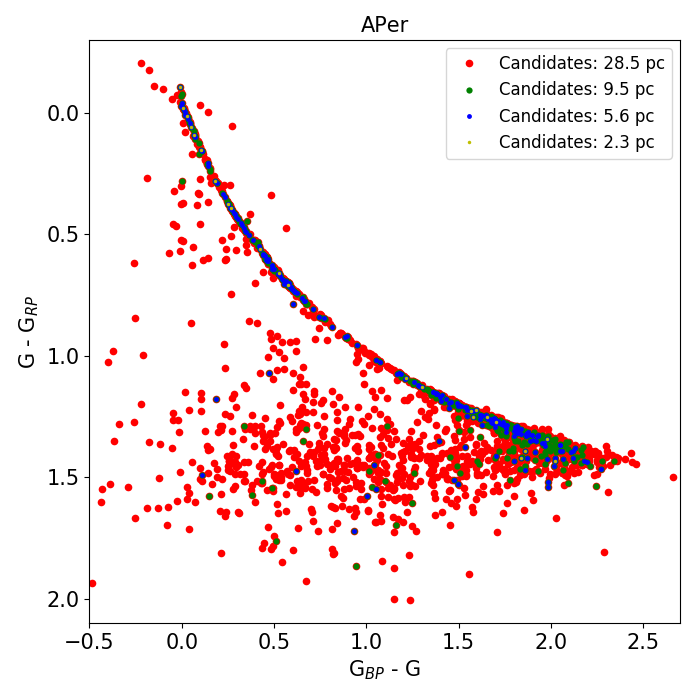}
  \caption{Colour-magnitude (top and middle rows) and colour-colour (bottom panel) diagrams 
  with $Gaia$ photometry only for all \APer{} candidates 
  within a radius up to 28.5 degrees from the cluster centre. 
  We added as small grey dots the full $Gaia$ catalogue in a large region around \APer{}.
}
  \label{fig_clustersGaia:APer_CMD_Gaia_filters}
\end{figure*}
\begin{figure*}
 \centering
  \includegraphics[width=0.48\linewidth, angle=0]{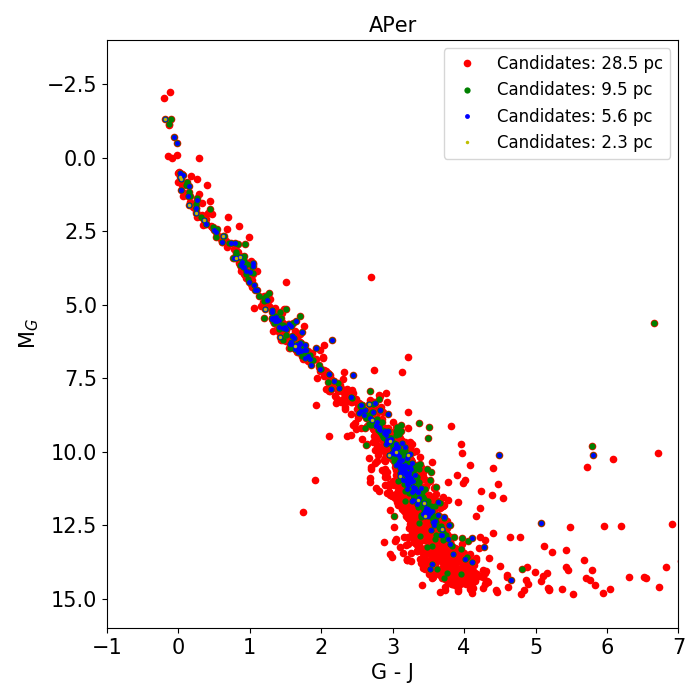}
  \includegraphics[width=0.48\linewidth, angle=0]{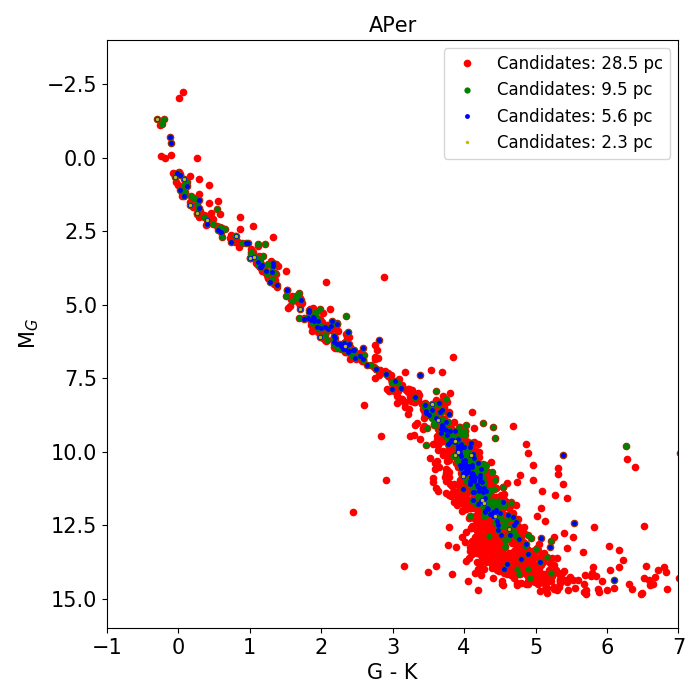}
  \includegraphics[width=0.48\linewidth, angle=0]{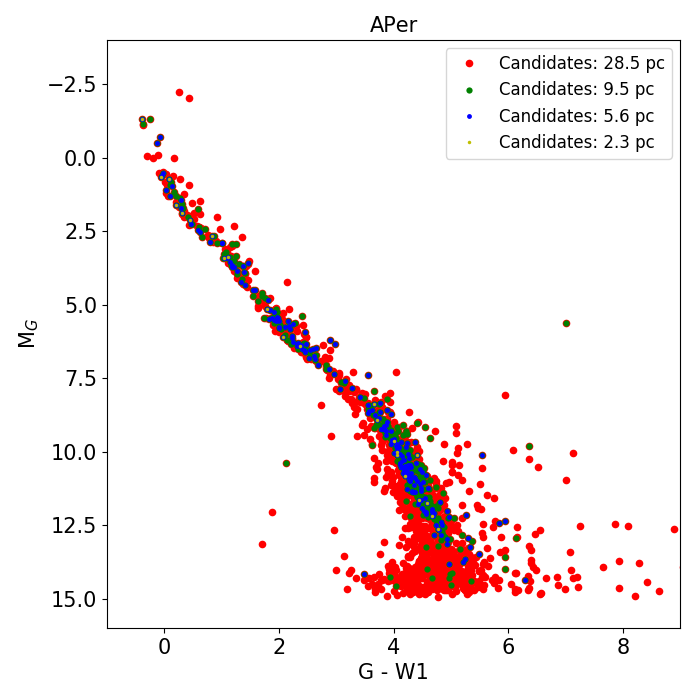}
  \includegraphics[width=0.48\linewidth, angle=0]{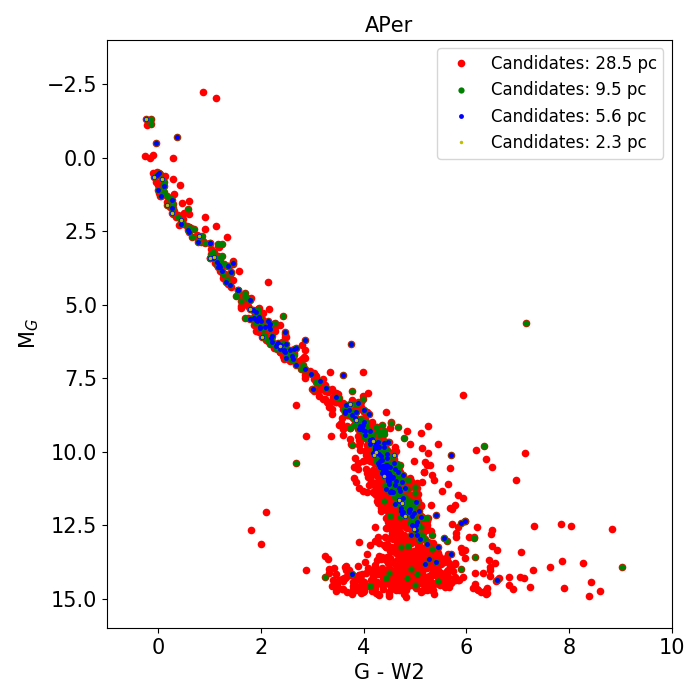}
  \caption{Colour-magnitude diagrams for \APer{} combining the $Gaia$ magnitude with infrared photometry
  from 2MASS ($J+K_{s}$) and AllWISE ($W1+W2$).
  Symbols are as in Fig.\ \ref{fig_clustersGaia:APer_CMD_Gaia_filters}.
  }
  \label{fig_clustersGaia:APer_CMD_G_others}
\end{figure*}
\begin{figure*}
 \centering
  \includegraphics[width=0.48\linewidth, angle=0]{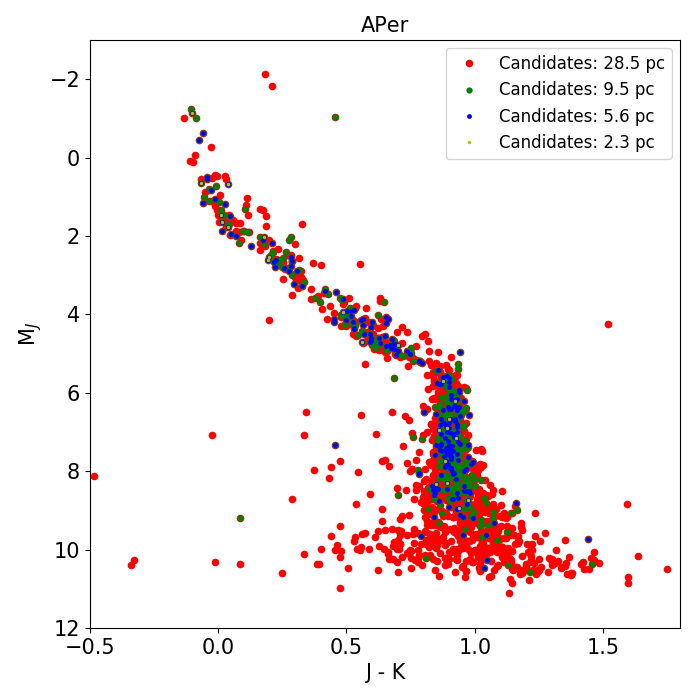}
  \includegraphics[width=0.48\linewidth, angle=0]{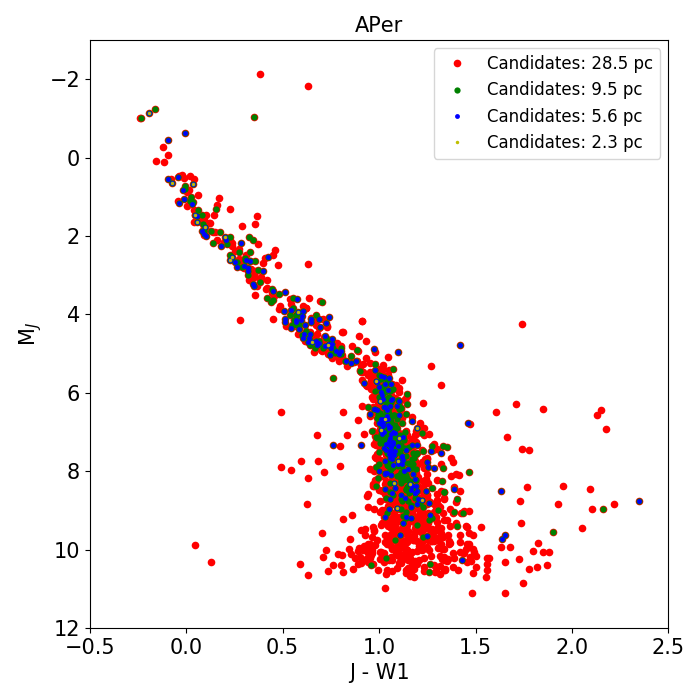}
  \includegraphics[width=0.48\linewidth, angle=0]{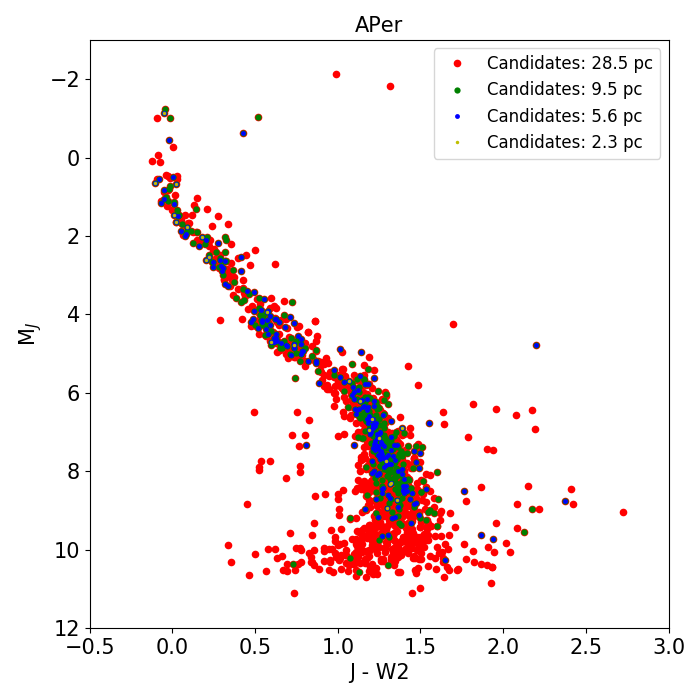}
  \includegraphics[width=0.48\linewidth, angle=0]{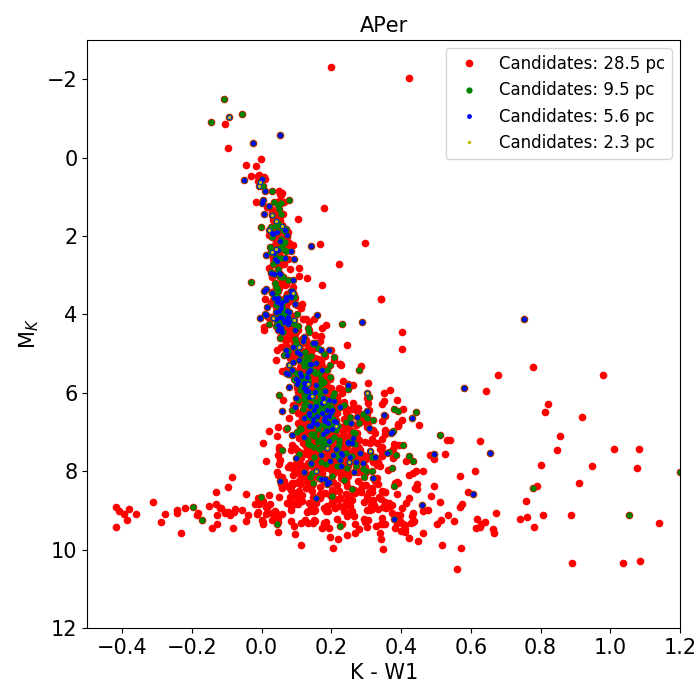}
  \caption{Colour-magnitude diagrams for \APer{} with non-$Gaia$ photometric passbands.
  Symbols are as in Fig.\ \ref{fig_clustersGaia:APer_CMD_Gaia_filters}.
  }
  \label{fig_clustersGaia:APer_CMD_nonGaia_filters}
\end{figure*}
%

%
%
We plan to make public via CDS/Vizier the full table of \APer{} candidate members
after applying the kinematic analysis described in Sect.\ \ref{clustersGaia:select_members}. 
The full catalogue contains 2041 sources within 28.5 pc (three times the size of the tidal radius)
with $Gaia$ DR2 data and photometry from several large-scale survey. Below we show a 
subset with a limited numbers of columns for space reasons, including source identifiers, 
coordinates, proper motions, parallaxes, and $G$ magnitudes.
\begin{table*}
\tiny
\centering
\caption{Catalogue of \APer{} member candidates located within the tidal radius of the cluster.
We list only a subsample of member candidates.
The full table will be made public electronically at the Vizier website.
}
\begin{tabular}{@{\hspace{0mm}}l@{\hspace{2mm}}c @{\hspace{1mm}}c @{\hspace{1mm}}c @{\hspace{1mm}}c @{\hspace{1mm}}c @{\hspace{1mm}}c @{\hspace{1mm}}c @{\hspace{1mm}}c @{\hspace{1mm}}c @{\hspace{1mm}}c @{\hspace{1mm}}c @{\hspace{1mm}}c @{\hspace{1mm}}c @{\hspace{1mm}}c @{\hspace{1mm}}c @{\hspace{1mm}}c @{\hspace{1mm}} c@{\hspace{0mm}}}
\hline
\hline
SourceID & RA & DEC & pmRA & pmDEC & Plx & $G$ & $b_{x}$ & $b_{y}$ & $b_{z}$ & $v_{x}$ & $v_{y}$ & $v_{z}$ & d\_centre & c & Mass & RV  \cr
\hline
         & degrees & degrees & mas/yr & mas/yr & mas & mag & pc & pc & pc &  km s$^{-1}$ & km s$^{-1}$ & km s$^{-1}$  & pc &  &  M$_{\sun}$ & km s$^{-1}$ \cr
\hline
455075360092433920  &  34.813289504406 &  52.493073691625 &   27.646 &  $-$19.901 &  5.846 & 16.533 &  $-$122.27 &   117.16 &   $-$24.09 & --- & --- & --- &  0.387 & 34.644 &    0.286 & --- \cr
455037048980464896  &  35.010252041682 &  52.664085343922 &   18.664 &  $-$20.069 &  5.650 & 20.310 &  $-$126.70 &   121.17 &   $-$24.31 & --- & --- & --- & 13.402 & 34.346 &    0.058 & --- \cr
355242590505237888  &  35.095645199238 &  48.516677205487 &   27.676 &  $-$19.837 &  5.940 & 15.840 &  $-$122.03 &   110.78 &   $-$34.36 & --- & --- & --- &  1.227 & 34.172 &    0.391 & --- \cr
355343947436467200  &  35.202350494471 &  49.437643095842 &   36.234 &  $-$24.237 &  5.819 & 17.689 &  $-$124.45 &   113.99 &   $-$32.46 & --- & --- & --- & 19.035 & 33.215 &    0.168 & --- \cr
355706893648353280  &  35.216075974941 &  50.602272518113 &   15.200 &   $-$8.305 &  5.782 & 19.424 &  $-$124.88 &   115.99 &   $-$29.41 & --- & --- & --- & 13.275 & 33.071 &    0.075 & --- \cr
354025942235087872  &  35.272857635410 &  47.078048854761 &   22.771 &   $-$2.507 &  5.072 & 20.892 &  $-$143.60 &   127.53 &   $-$44.64 & --- & --- & --- & 13.241 & 41.109 &    0.049 & --- \cr
\ldots{}           &  \ldots{}   &   \ldots{} & \ldots{} & \ldots{} & \ldots{} & \ldots{} & \ldots{} & \ldots{} & \ldots{} & \ldots{} & \ldots{} & \ldots{} & \ldots{} \cr
257580321299859968  &  68.322204619080 &  48.021785586470 &   20.927 &  $-$27.332 &  5.885 & 17.712 &  $-$155.67 &    68.12 &     0.25 & --- & --- & --- &  5.921 & 34.384 &    0.167 & --- \cr
258437635428197632  &  68.330791212726 &  48.879728011747 &   22.456 &  $-$24.707 &  5.881 & 20.040 &  $-$155.00 &    69.86 &     2.00 & --- & --- & --- &  6.117 & 33.992 &    0.062 & --- \cr
257268682770968960  &  68.353707068218 &  46.624841369087 &   17.959 &  $-$31.619 &  5.857 & 17.395 &  $-$157.60 &    65.59 &    $-$2.53 & --- & --- & --- &  2.246 & 35.395 &    0.188 & --- \cr
271026936190122112  &  68.428008586545 &  51.372991304681 &    6.104 &  $-$10.112 &  5.522 & 20.742 &  $-$162.55 &    79.45 &     7.62 & --- & --- & --- & 10.021 & 34.489 &       "" & --- \cr
271250446285756928  &  68.529601743168 &  52.049391468516 &   22.941 &  $-$38.325 &  6.016 & 19.207 &  $-$148.56 &    74.08 &     8.46 & --- & --- & --- &  5.695 & 35.251 &    0.082 & --- \cr
272473279313927424  &  69.157470869484 &  51.414405941179 &   12.183 &  $-$32.031 &  5.810 & 16.682 &  $-$154.82 &    74.75 &     8.34 & --- & --- & --- & 19.058 & 35.288 &    0.267 & --- \cr
\hline
\label{tab_clustersGaia:APer_catalogue_AfterCP}
\end{tabular}
\end{table*}
%

%
%
%
\section{Tables and diagrams for Pleiades}
\label{clustersGaia:Appendix_Pleiades}
%
%
%
\begin{figure*}
 \centering
  \includegraphics[width=0.48\linewidth, angle=0]{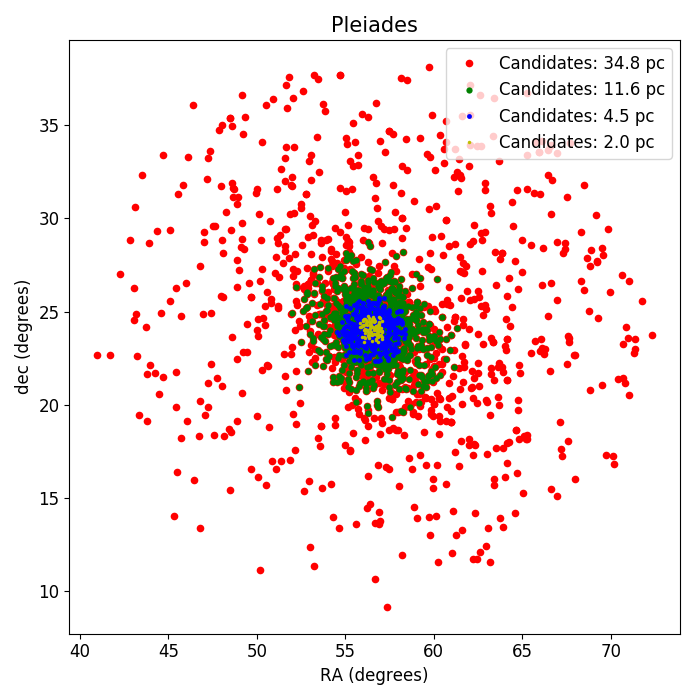}
  \includegraphics[width=0.48\linewidth, angle=0]{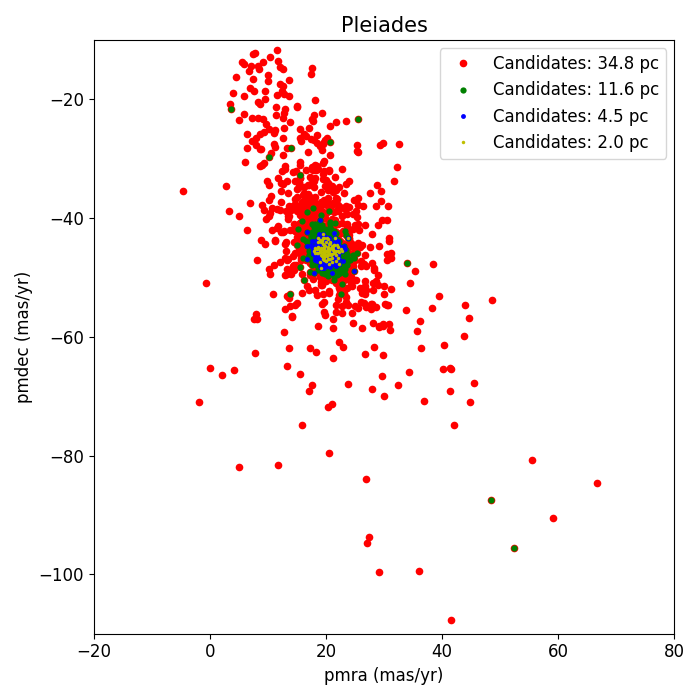}
  \includegraphics[width=0.48\linewidth, angle=0]{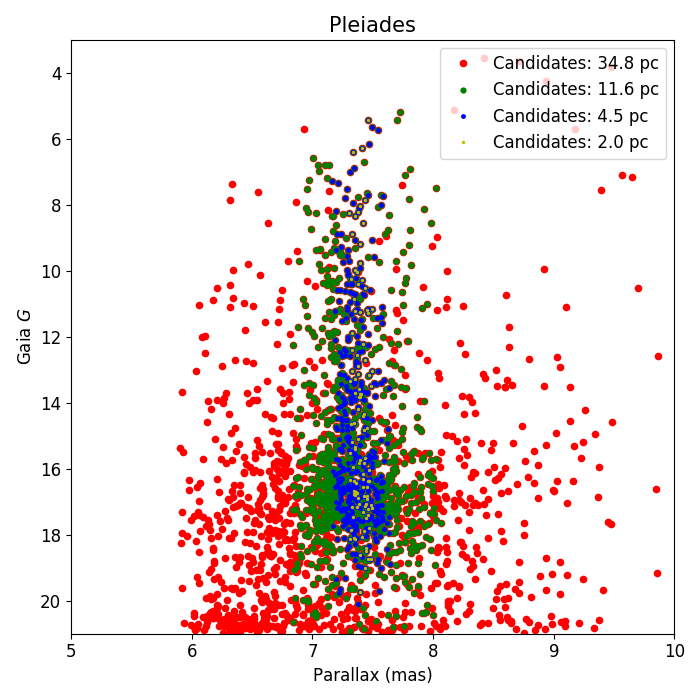}
  \includegraphics[width=0.48\linewidth, angle=0]{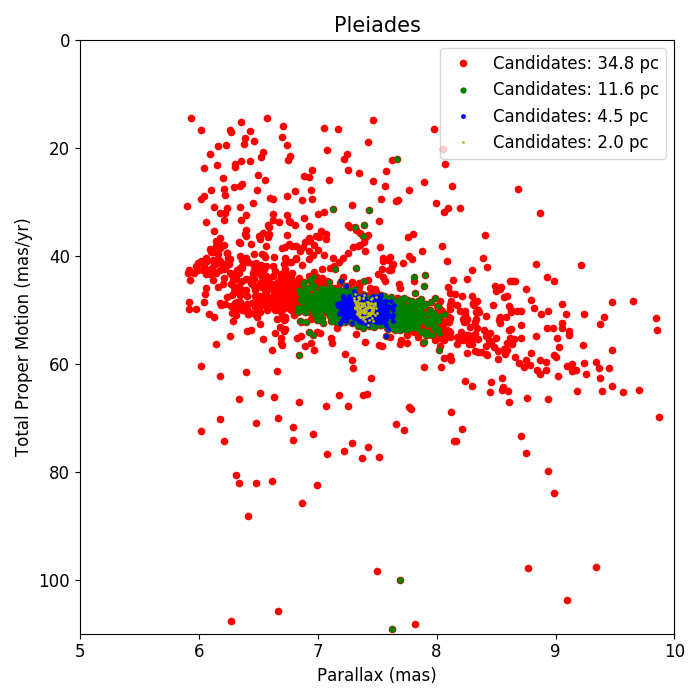}
  \caption{{\it{Top left:}} (RA,dec) diagram showing the position of the Pleiades member candidates.
  Samples at different distances from the cluster centre are highlighted with distinct colours and sizes.
  {\it{Top right:}} Vector point diagram.
  {\it{Bottom left:}} $Gaia$ magnitude vs parallax. 
  {\it{Bottom right:}} Total proper motion as a function parallax.
}
  \label{fig_clustersGaia:Pleiades_plots_general}
\end{figure*}
\begin{figure*}
 \centering
  \includegraphics[width=0.43\linewidth, angle=0]{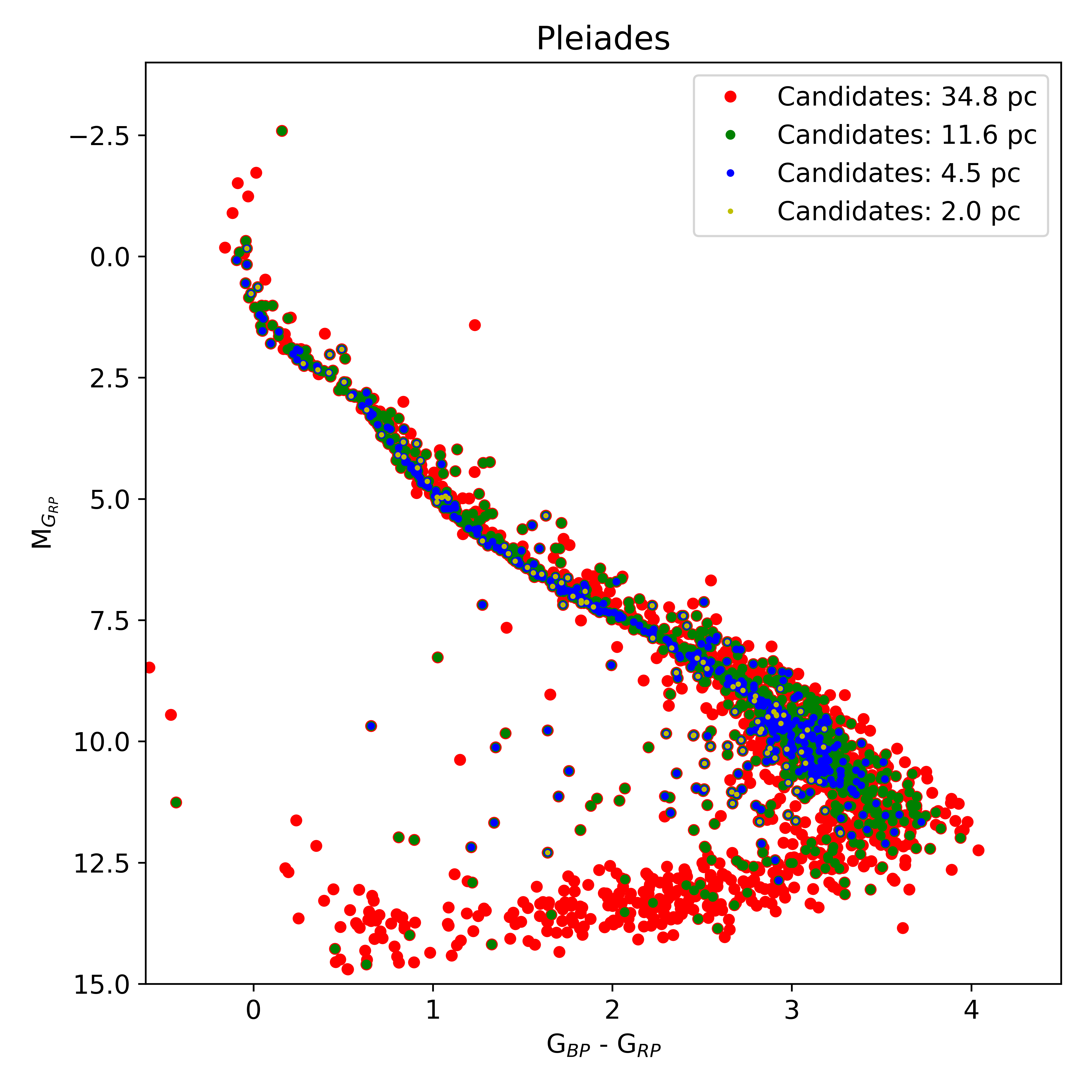}
  \includegraphics[width=0.43\linewidth, angle=0]{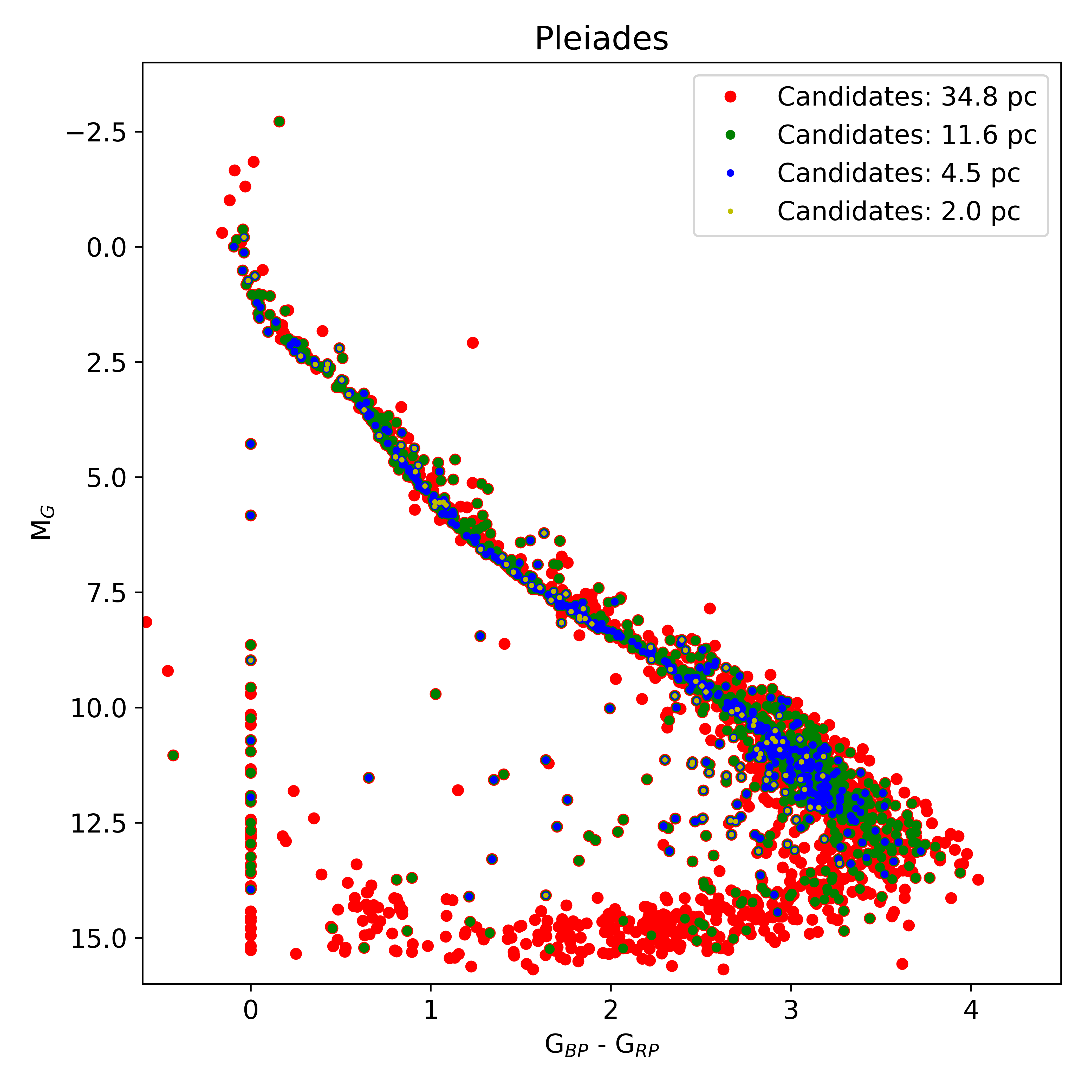}
  \includegraphics[width=0.43\linewidth, angle=0]{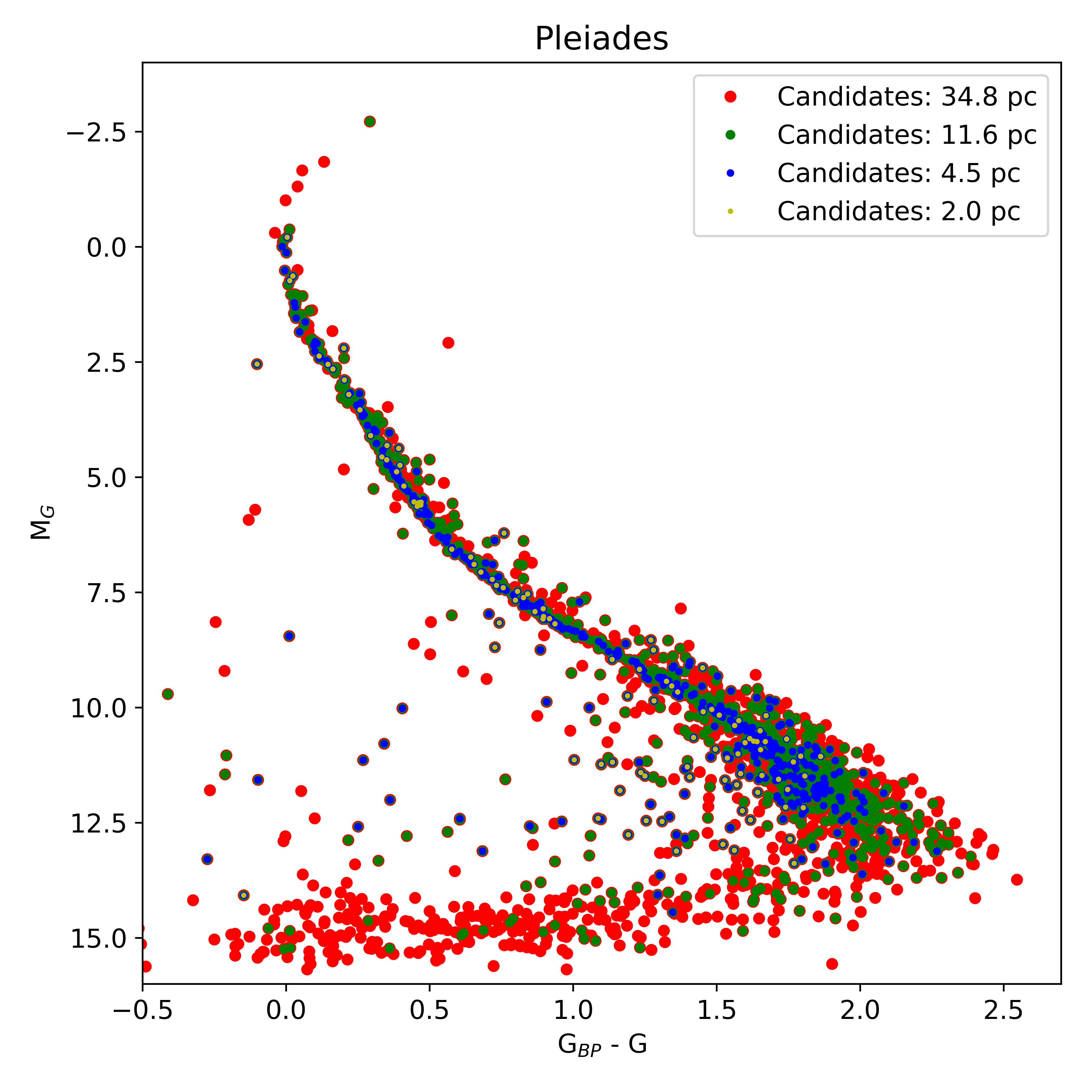}
  \includegraphics[width=0.43\linewidth, angle=0]{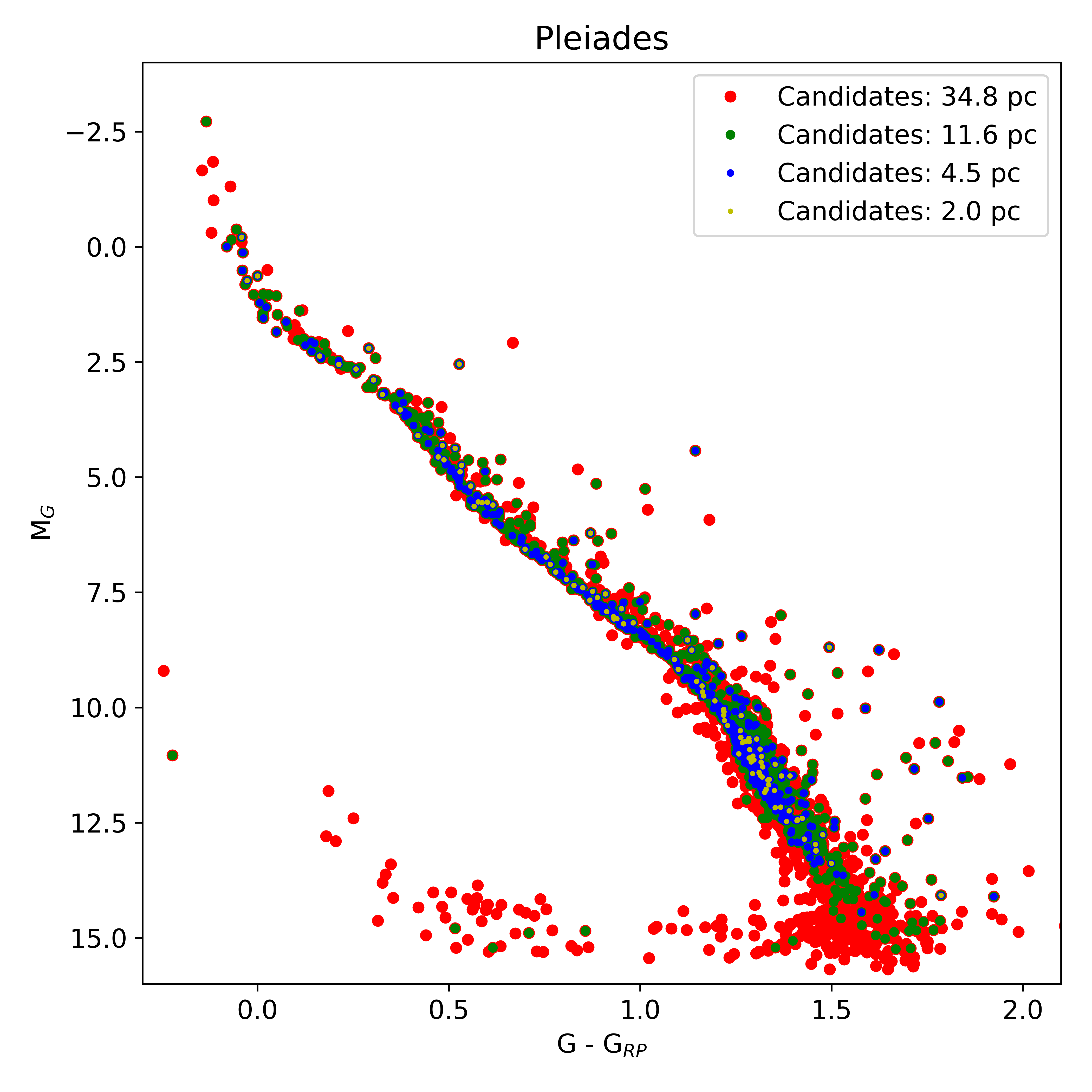}
  \includegraphics[width=0.43\linewidth, angle=0]{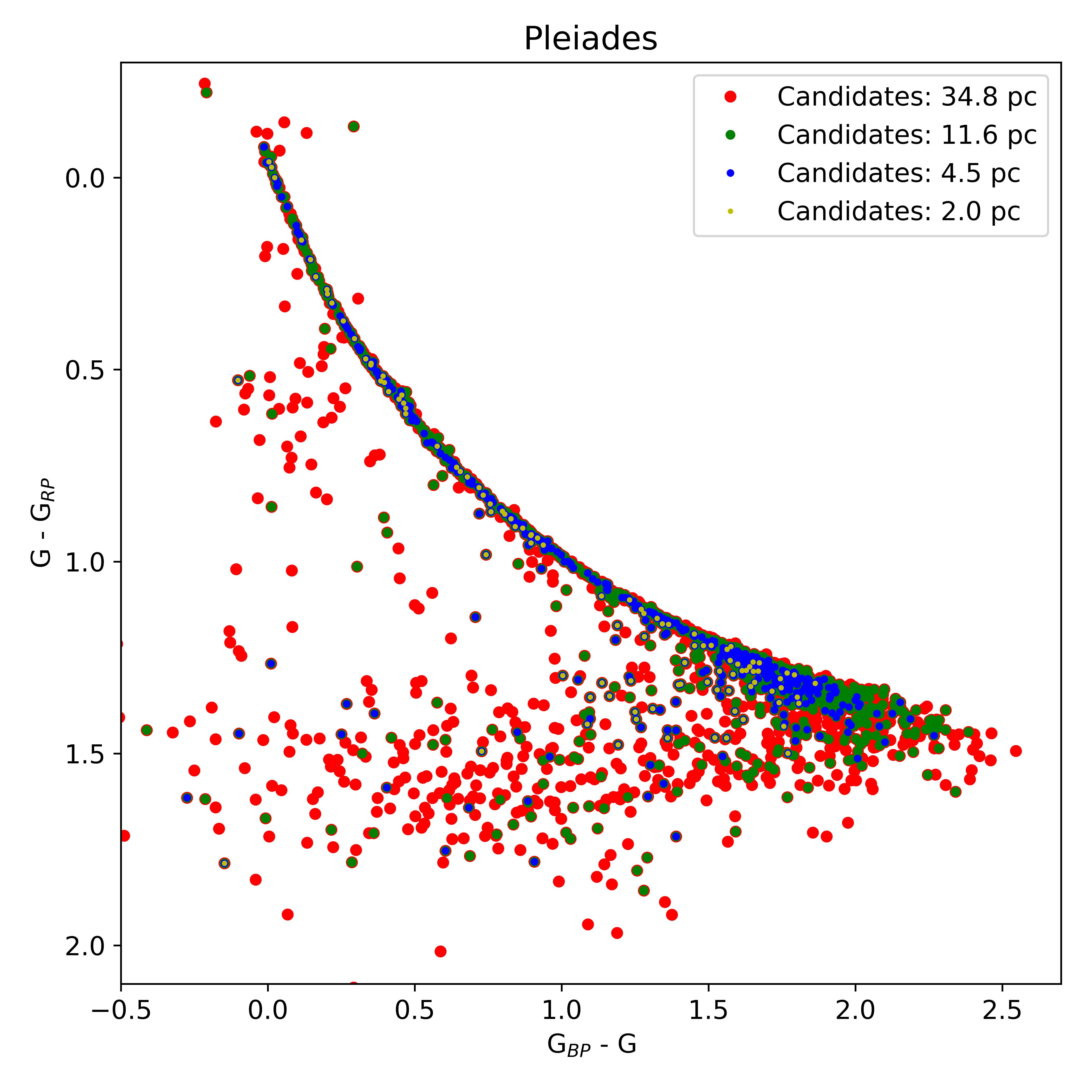}
  \caption{Colour-magnitude (top and middle rows) and colour-colour (bottom panel) diagrams 
  with $Gaia$ photometry only for all Pleiades candidates 
  within a radius up to $\sim$35 degrees from the cluster centre. 
  We added as small grey dots the full $Gaia$ catalogue in a large region around Pleiades.
}
  \label{fig_clustersGaia:Pleiades_CMD_Gaia_filters}
\end{figure*}
\begin{figure*}
 \centering
  \includegraphics[width=0.48\linewidth, angle=0]{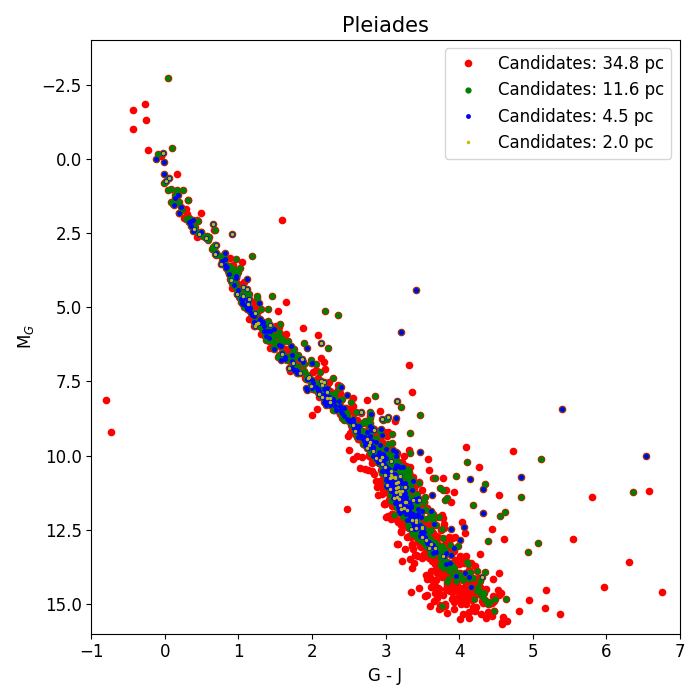}
  \includegraphics[width=0.48\linewidth, angle=0]{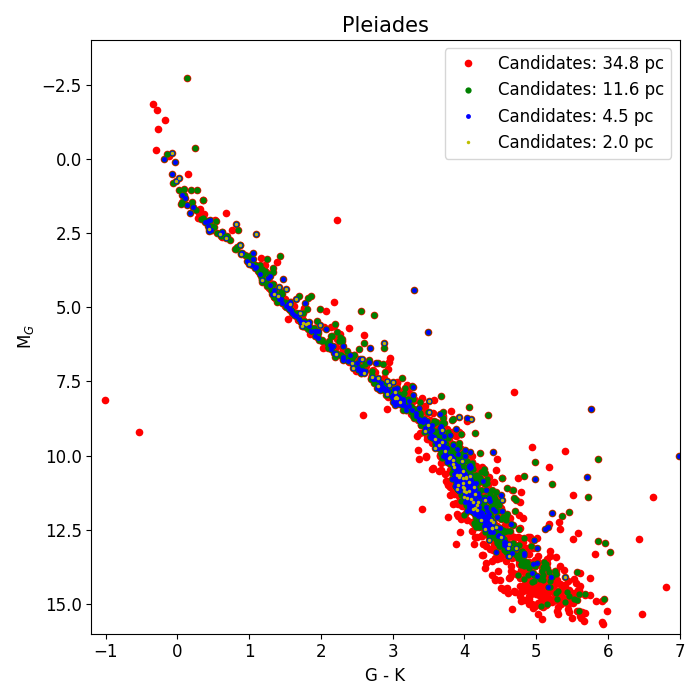}
  \includegraphics[width=0.48\linewidth, angle=0]{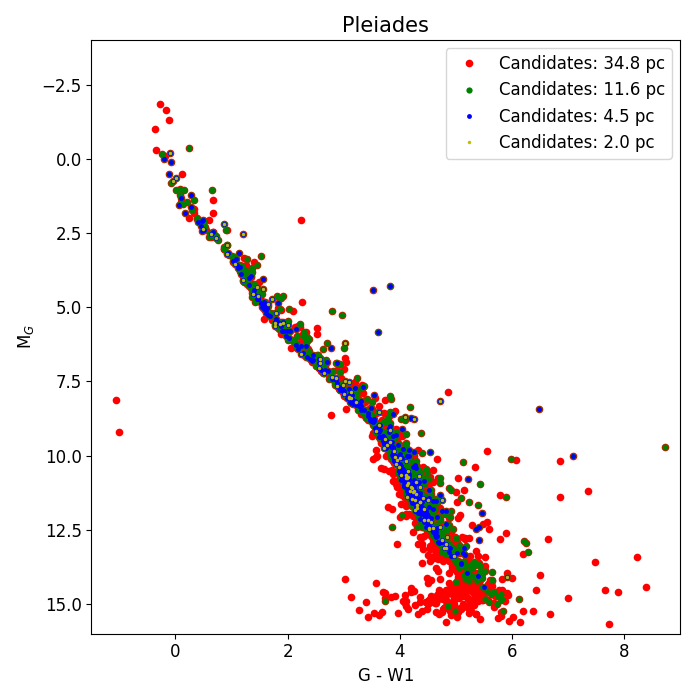}
  \includegraphics[width=0.48\linewidth, angle=0]{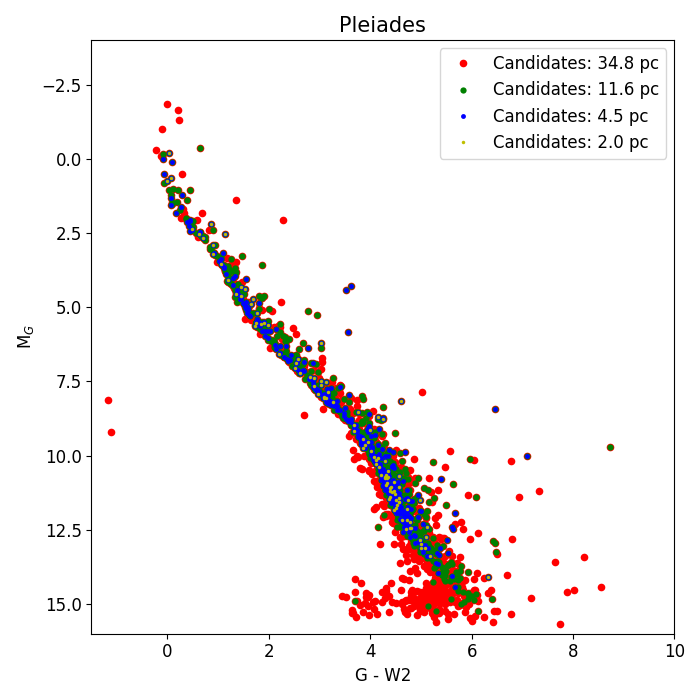}
  \caption{Colour-magnitude diagrams for the Pleiades combining the $Gaia$ magnitude with 
  infrared photometry from 2MASS ($J+K_{s}$) and AllWISE ($W1+W2$).
  Symbols are as in Fig.\ \ref{fig_clustersGaia:Pleiades_CMD_Gaia_filters}.
  }
  \label{fig_clustersGaia:Pleiades_CMD_G_others}
\end{figure*}
\begin{figure*}
 \centering
  \includegraphics[width=0.48\linewidth, angle=0]{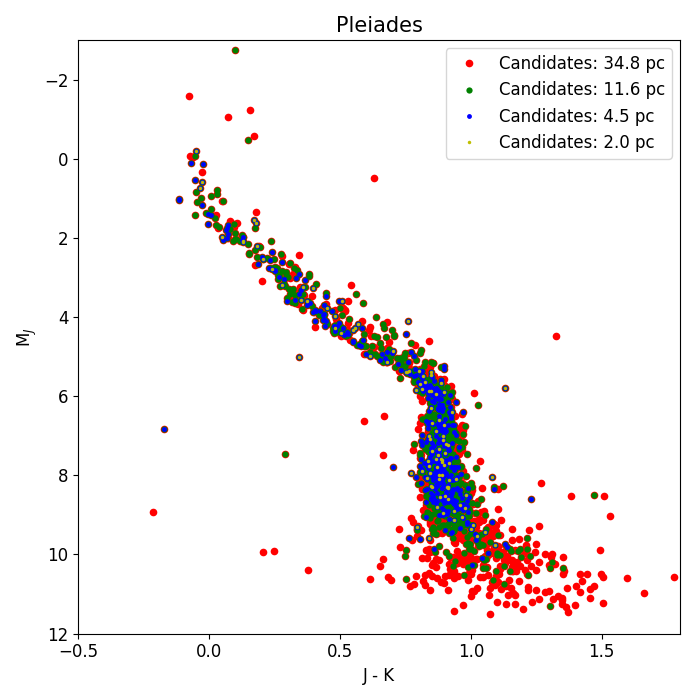}
  \includegraphics[width=0.48\linewidth, angle=0]{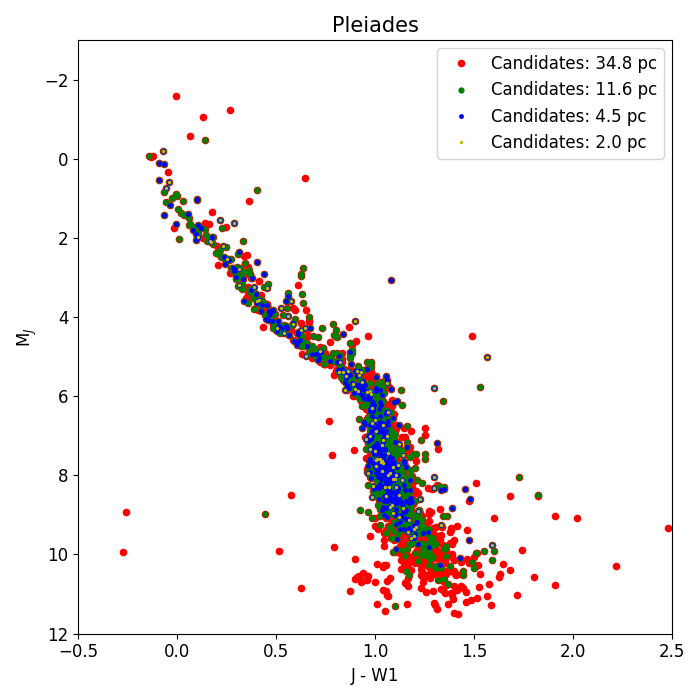}
  \includegraphics[width=0.48\linewidth, angle=0]{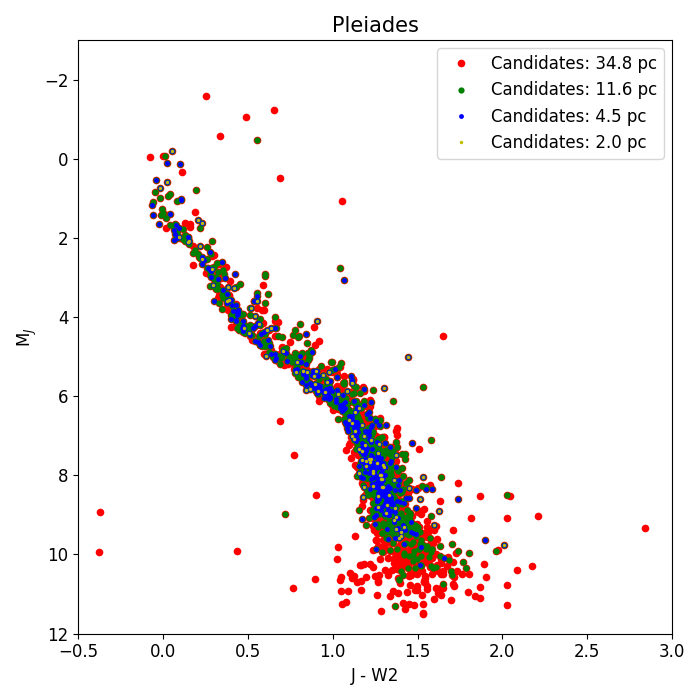}
  \includegraphics[width=0.48\linewidth, angle=0]{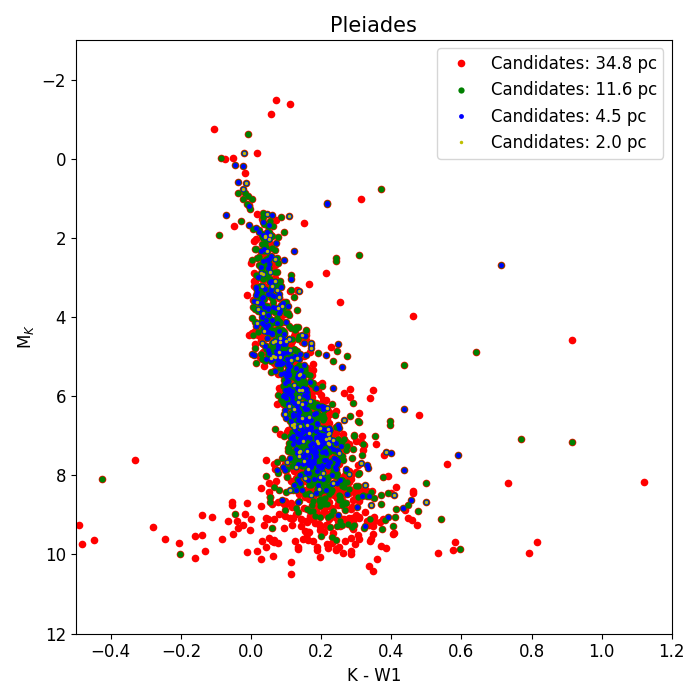}
  \caption{Colour-magnitude diagrams for the Pleiades with non-$Gaia$ photometric passbands.
  Symbols are as in Fig.\ \ref{fig_clustersGaia:Pleiades_CMD_Gaia_filters}.
  }
  \label{fig_clustersGaia:Pleiades_CMD_nonGaia_filters}
\end{figure*}
%

%
%
We plan to make public via CDS/Vizier the full table of Pleiades candidate members
after applying the kinematic analysis described in Sect.\ \ref{clustersGaia:select_members}.
The full catalogue contains 2195 sources located within three times the tidal radius (34.8 pc)
with $Gaia$ DR2 data and photometry from several large-scale survey. Below we show a
subset with a limited numbers of columns for space reasons, including source identifiers,            
coordinates, proper motions, parallaxes, and $G$ magnitudes.
\begin{table*}
\tiny
\centering
\caption{Catalogue of Pleiades member candidates located within the tidal radius of the cluster.
We list only a subsample of member candidates.
The full table will be made public electronically at the Vizier website.
}
\begin{tabular}{@{\hspace{0mm}}l@{\hspace{2mm}}c @{\hspace{1mm}}c @{\hspace{1mm}}c @{\hspace{1mm}}c @{\hspace{1mm}}c @{\hspace{1mm}}c @{\hspace{1mm}}c @{\hspace{1mm}}c @{\hspace{1mm}}c @{\hspace{1mm}}c @{\hspace{1mm}}c @{\hspace{1mm}}c @{\hspace{1mm}}c @{\hspace{1mm}}c @{\hspace{1mm}}c @{\hspace{1mm}}c@{\hspace{0mm}}}
\hline
\hline
SourceID & RA & DEC & pmRA & pmDEC & Plx & $G$ & $b_{x}$ & $b_{y}$ & $b_{z}$ & $v_{x}$ & $v_{y}$ & $v_{z}$ & d\_centre & c & Mass & RV  \cr
\hline
         & degrees & degrees & mas/yr & mas/yr & mas & mag & pc & pc & pc &  km s$^{-1}$ & km s$^{-1}$ & km s$^{-1}$  & pc &  &  M$_{\sun}$ & km s$^{-1}$ \cr
\hline
88861709418842496   &  40.974970833896 &  22.655743940905 &   15.587 &  $-$33.165 &  6.712 & 18.172 &  $-$112.22 &    54.06 &   $-$81.76 & --- & --- & --- & 18.615 & 37.934 &    0.114 & --- \cr
112681288804808064  &  41.714674914238 &  22.655860086477 &   22.806 &  $-$54.808 &  8.078 & 20.067 &   $-$94.15 &    43.94 &   $-$67.30 & --- & --- & --- & 17.527 & 32.715 &    0.059 & --- \cr
114833750319750912  &  42.242873946349 &  27.033836378275 &   20.579 &  $-$32.107 &  8.040 & 18.397 &   $-$96.91 &    49.63 &   $-$60.14 & --- & --- & --- & 21.623 & 31.687 &      0.1 & --- \cr
128604205745014656  &  42.822225147935 &  28.811112669751 &   41.455 &  $-$65.346 &  7.391 & 20.044 &  $-$106.71 &    55.83 &   $-$61.68 & --- & --- & --- &  8.794 & 30.946 &    0.059 & --- \cr
130317932056151680  &  42.928644668868 &  31.943391145743 &   30.722 &  $-$44.277 &  7.494 & 20.243 &  $-$106.07 &    59.40 &   $-$55.02 & --- & --- & --- &  5.057 & 33.595 &    0.056 & --- \cr
114535958762185600  &  43.052790394089 &  26.245931711561 &   36.093 &  $-$99.403 &  6.614 & 20.782 &  $-$118.65 &    57.73 &   $-$73.83 & --- & --- & --- &  7.584 & 34.715 &    0.049 & --- \cr
\ldots{}           &  \ldots{}   &   \ldots{} & \ldots{} & \ldots{} & \ldots{} & \ldots{} & \ldots{} & \ldots{} & \ldots{} & \ldots{} & \ldots{} & \ldots{} & \ldots{} \cr
154342093619417728  &  71.030644785071 &  26.611266953339 &   10.410 &  $-$49.344 &  7.199 & 15.072 &  $-$134.90 &    14.18 &   $-$29.92 & --- & --- & --- &  1.687 & 32.193 &    0.427 & --- \cr
3413122927160114944 &  71.338224175959 &  22.763339165068 &   10.233 &  $-$34.361 &  7.418 & 20.540 &  $-$130.28 &     6.23 &   $-$34.06 & --- & --- & --- &  2.589 & 32.161 &    0.052 & --- \cr
146576410496389120  &  71.371876531125 &  23.546202736402 &   14.408 &  $-$48.123 &  8.025 & 18.367 &  $-$120.63 &     7.06 &   $-$30.40 & --- & --- & --- &  1.414 & 32.545 &    0.102 & --- \cr
3413153850924754816 &  71.392270839100 &  23.005022102866 &    7.543 &  $-$57.013 &  6.847 & 20.868 &  $-$141.25 &     7.16 &   $-$36.43 & --- & --- & --- &  3.988 & 35.245 &    0.049 & --- \cr
147465369941559040  &  71.758353684925 &  25.557751301650 &   17.011 &  $-$41.693 &  7.023 & 15.203 &  $-$138.50 &    11.49 &   $-$31.05 & --- & --- & --- & 13.155 & 34.460 &    0.411 & --- \cr
3413681994462425216 &  72.525370378847 &  24.013526031333 &    8.657 &  $-$68.246 &  7.207 & 19.598 &  $-$135.01 &     7.21 &   $-$31.22 & --- & --- & --- & 14.661 & 35.096 &    0.066 & --- \cr
\hline
\label{tab_clustersGaia:Pleiades_catalogue_AfterCP}
\end{tabular}
\end{table*}
%

%
%
%
\section{Tables and diagrams for Praesepe}
\label{clustersGaia:Appendix_Praesepe}
%
%
%
\begin{figure*}
 \centering
  \includegraphics[width=0.48\linewidth, angle=0]{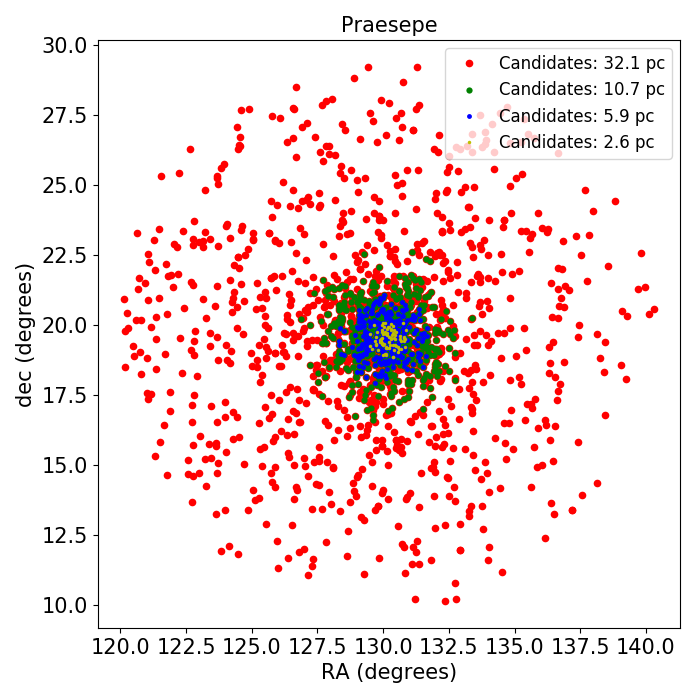}
  \includegraphics[width=0.48\linewidth, angle=0]{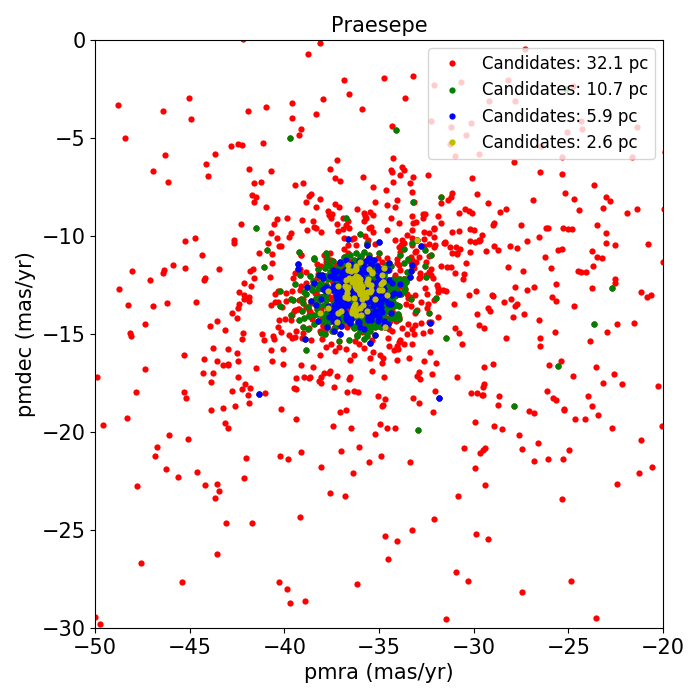}
  \includegraphics[width=0.48\linewidth, angle=0]{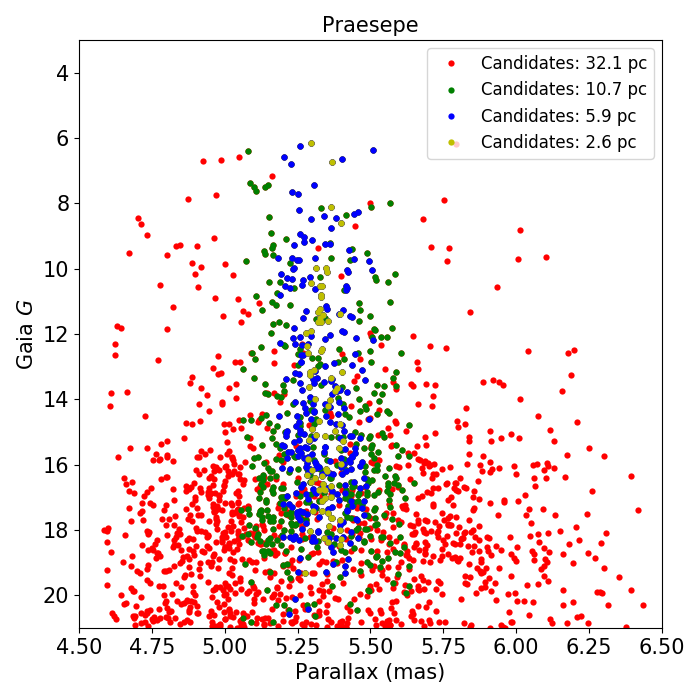}
  \includegraphics[width=0.48\linewidth, angle=0]{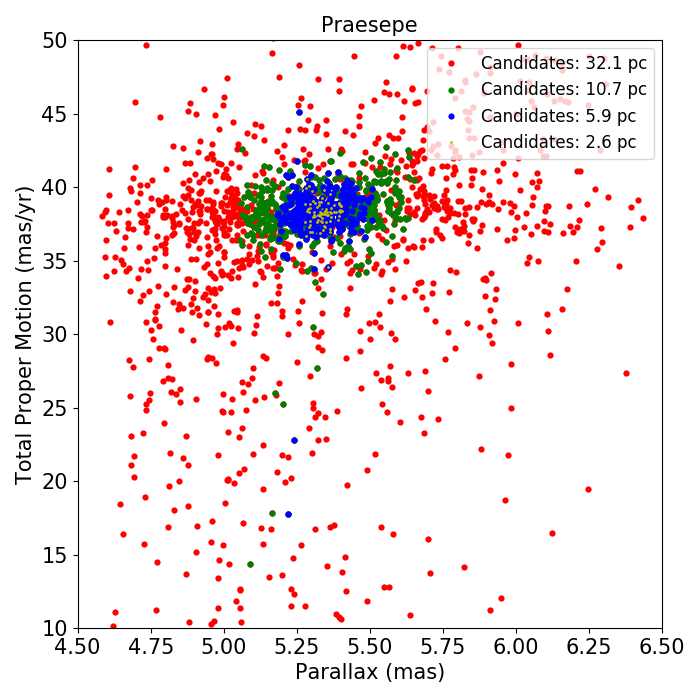}
  \caption{{\it{Top left:}} (RA,dec) diagram showing the position of Praesepe member candidates.
  Samples at different distances from the cluster centre are highlighted with distinct colours and sizes.
  {\it{Top right:}} Vector point diagram.
  {\it{Bottom left:}} $Gaia$ magnitude vs parallax. 
  {\it{Bottom right:}} Total proper motion as a function parallax.
}
  \label{fig_clustersGaia:Praesepe_plots_general}
\end{figure*}
\begin{figure*}
 \centering
  \includegraphics[width=0.43\linewidth, angle=0]{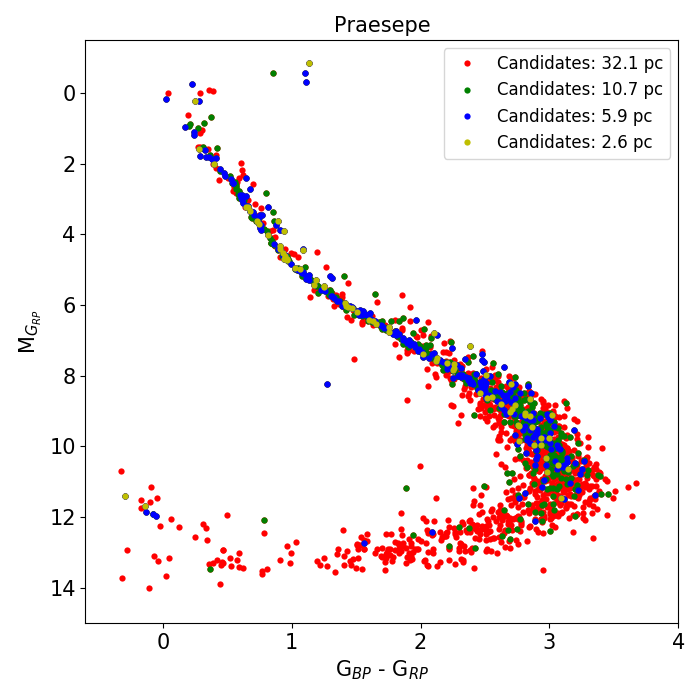}
  \includegraphics[width=0.43\linewidth, angle=0]{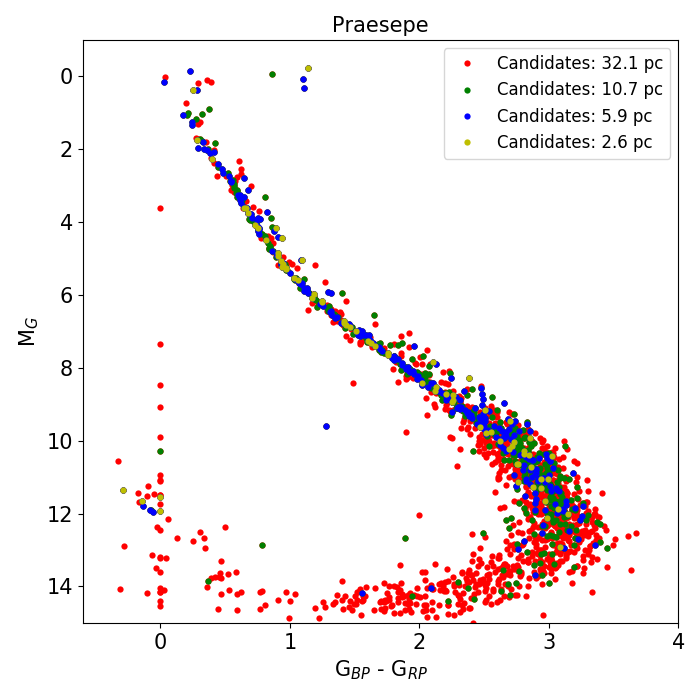}
  \includegraphics[width=0.43\linewidth, angle=0]{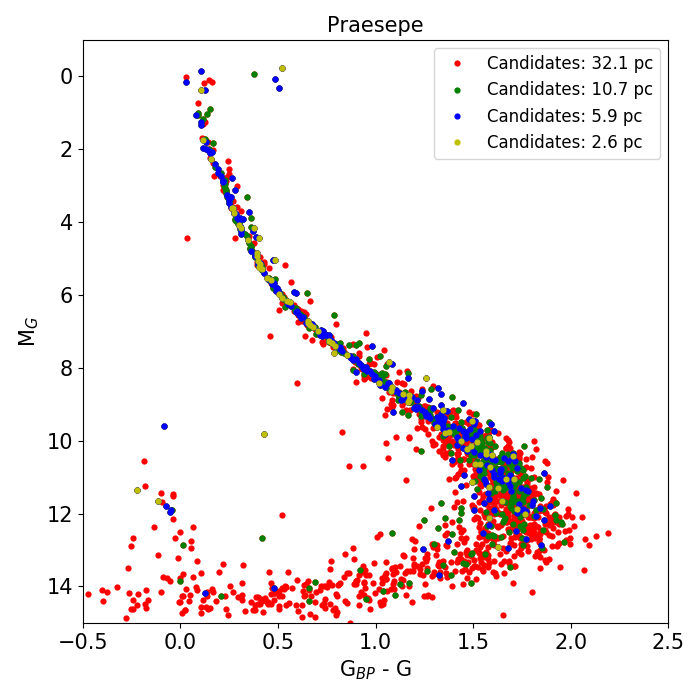}
  \includegraphics[width=0.43\linewidth, angle=0]{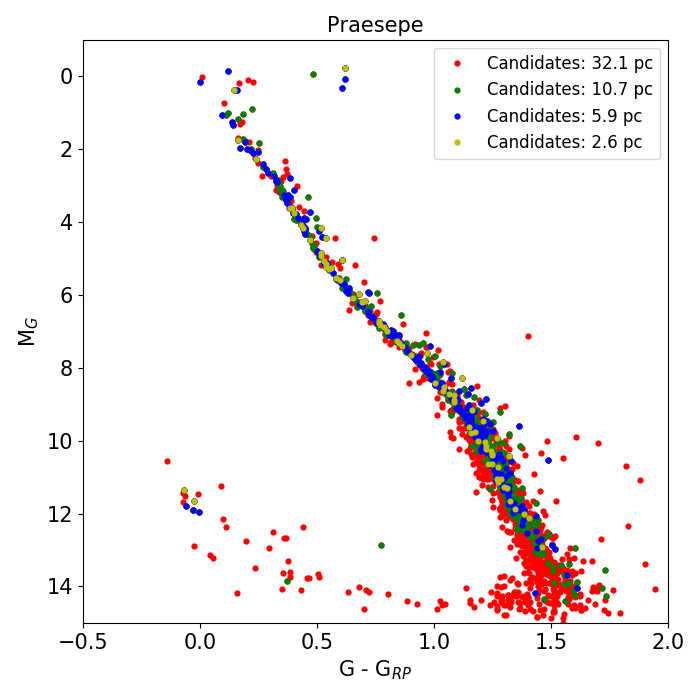}
  \includegraphics[width=0.43\linewidth, angle=0]{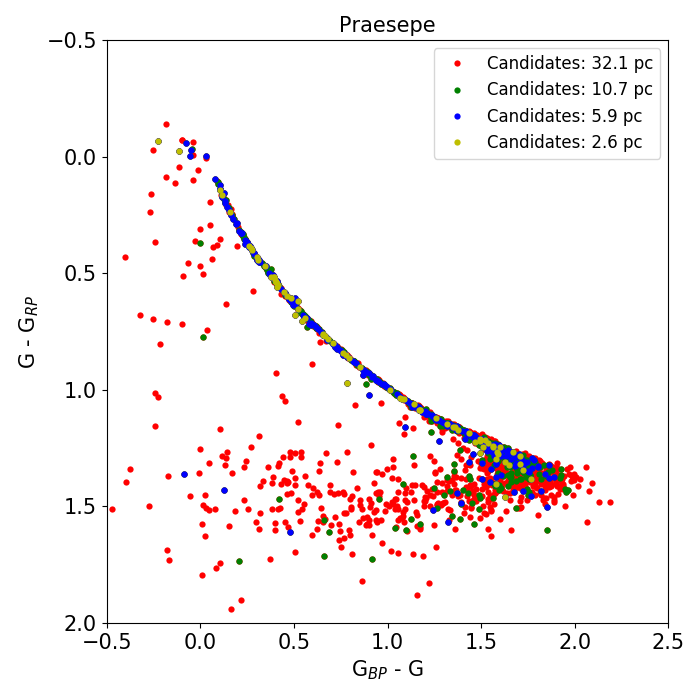}
  \caption{Colour-magnitude (top and middle rows) and colour-colour (bottom panel) diagrams 
  with $Gaia$ photometry only for all Praesepe candidates 
  within a radius up to 32 degrees from the cluster centre. 
  We added as small grey dots the full $Gaia$ catalogue in a large region around Praesepe.
}
  \label{fig_clustersGaia:Praesepe_CMD_Gaia_filters}
\end{figure*}
\begin{figure*}
 \centering
  \includegraphics[width=0.48\linewidth, angle=0]{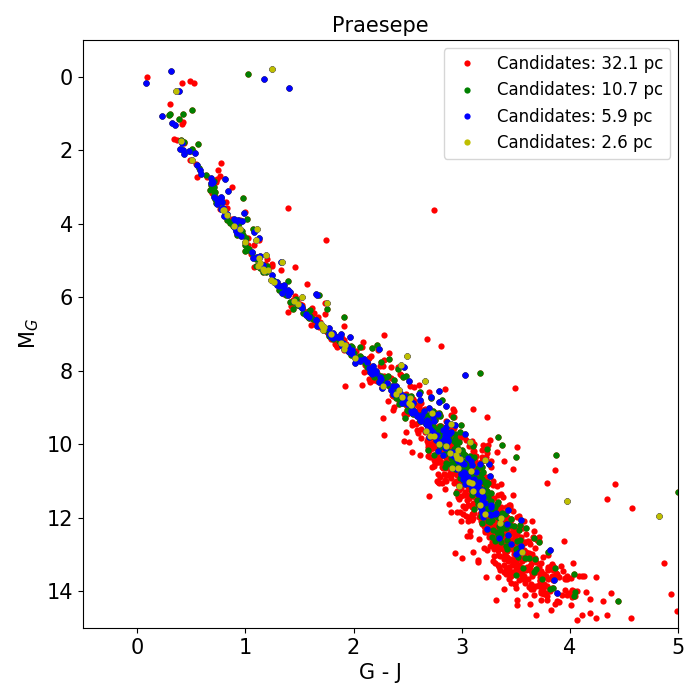}
  \includegraphics[width=0.48\linewidth, angle=0]{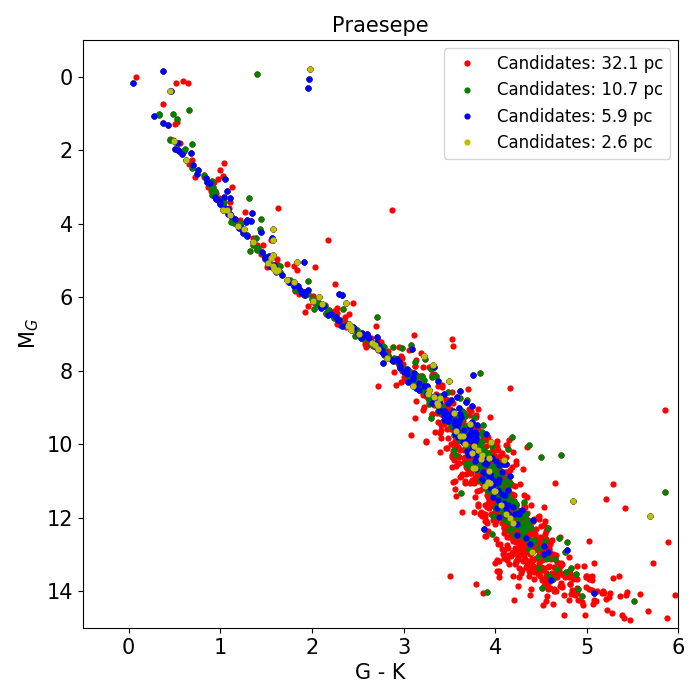}
  \includegraphics[width=0.48\linewidth, angle=0]{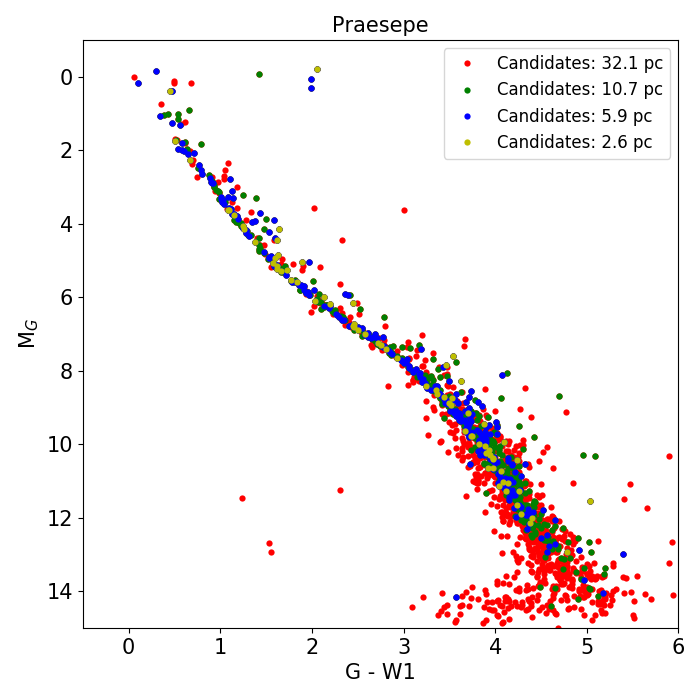}
  \includegraphics[width=0.48\linewidth, angle=0]{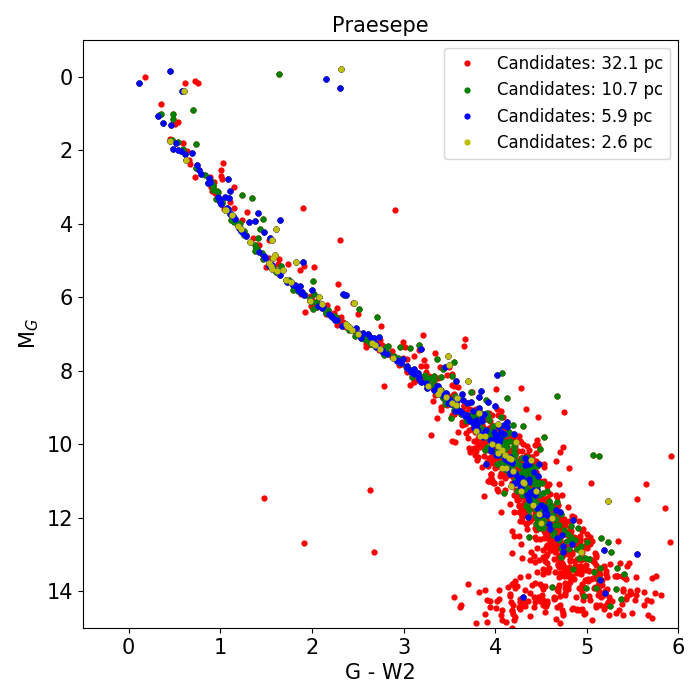}
  \caption{Colour-magnitude diagrams for Praesepe combining the $Gaia$ magnitude with infrared photometry
  from 2MASS ($J+K_{s}$) and AllWISE ($W1+W2$).
  Symbols are as in Fig.\ \ref{fig_clustersGaia:Praesepe_CMD_Gaia_filters}.
  }
  \label{fig_clustersGaia:Praesepe_CMD_G_others}
\end{figure*}
\begin{figure*}
 \centering
  \includegraphics[width=0.48\linewidth, angle=0]{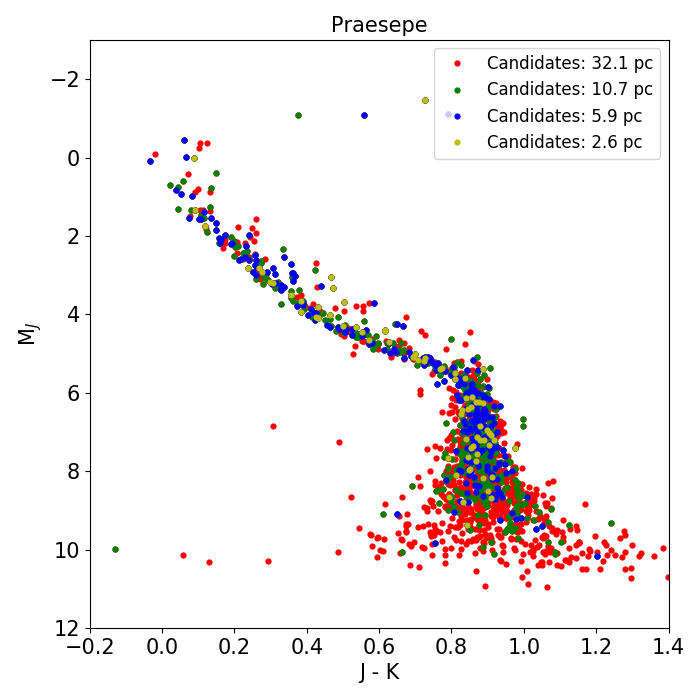}
  \includegraphics[width=0.48\linewidth, angle=0]{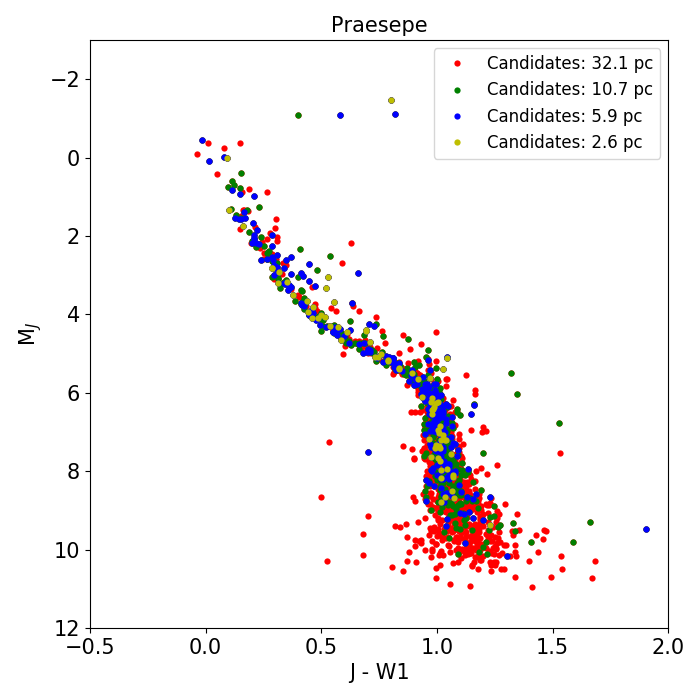}
  \includegraphics[width=0.48\linewidth, angle=0]{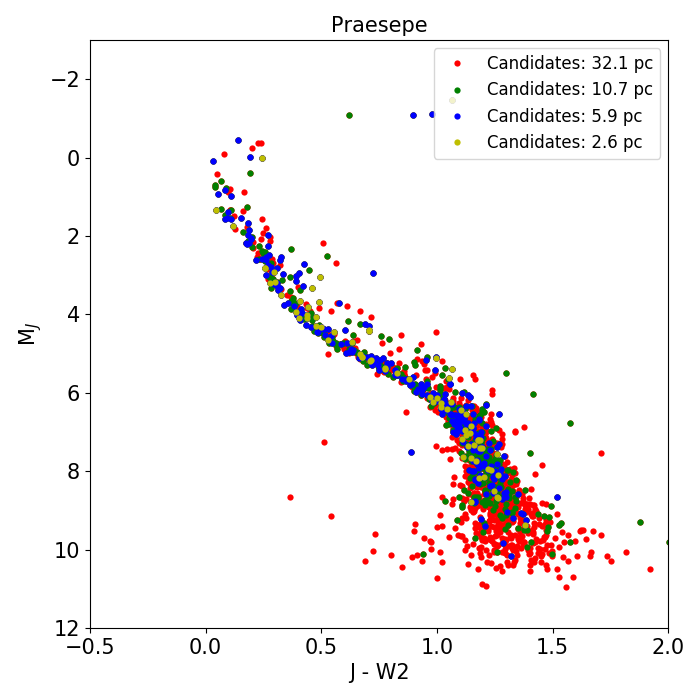}
  \includegraphics[width=0.48\linewidth, angle=0]{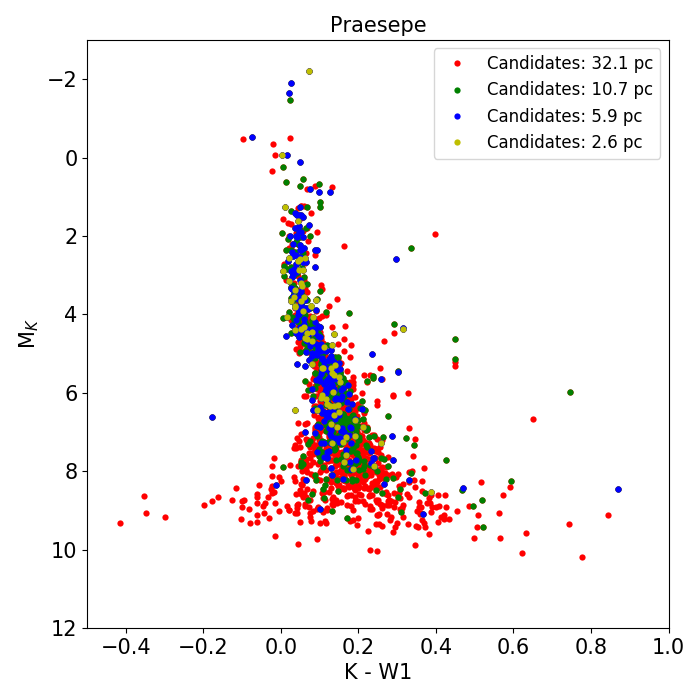}
  \caption{Colour-magnitude diagrams for Praesepe with non-$Gaia$ photometric passbands.
  Symbols are as in Fig.\ \ref{fig_clustersGaia:Praesepe_CMD_Gaia_filters}.
  }
  \label{fig_clustersGaia:Praesepe_CMD_nonGaia_filters}
\end{figure*}
%

%
%
We plan to make public via CDS/Vizier the full table of Praesepe candidate members
after applying the kinematic analysis described in Sect.\ \ref{clustersGaia:select_members}.
The full catalogue contains 1847 sources located within three times the tidal radius
with $Gaia$ DR2 data and photometry from several large-scale survey. Below we show a
subset with a limited numbers of columns for space reasons, including source identifiers,            
coordinates, proper motions, parallaxes, and $G$ magnitudes.
\begin{table*}
\tiny
\centering
\caption{Catalogue of Praesepe member candidates located within the tidal radius of the cluster.
We list only a subsample of member candidates.
The full table will be made public electronically at the Vizier website.
}
\begin{tabular}{@{\hspace{0mm}}l@{\hspace{2mm}}c @{\hspace{1mm}}c @{\hspace{1mm}}c @{\hspace{1mm}}c @{\hspace{1mm}}c @{\hspace{1mm}}c @{\hspace{1mm}}c @{\hspace{1mm}}c @{\hspace{1mm}}c @{\hspace{1mm}}c @{\hspace{1mm}}c @{\hspace{1mm}}c @{\hspace{1mm}}c @{\hspace{1mm}}c @{\hspace{1mm}}c @{\hspace{1mm}}c@{\hspace{0mm}}}
\hline
\hline
SourceID & RA & DEC & pmRA & pmDEC & Plx & $G$ & $b_{x}$ & $b_{y}$ & $b_{z}$ & $v_{x}$ & $v_{y}$ & $v_{z}$ & d\_centre & c & Mass & RV  \cr
\hline
         & degrees & degrees & mas/yr & mas/yr & mas & mag & pc & pc & pc &  km s$^{-1}$ & km s$^{-1}$ & km s$^{-1}$  & pc &  &  M$_{\sun}$ & km s$^{-1}$ \cr
\hline
673332163412736640  & 118.671345250596 &  19.966185887724 &  $-$27.373 &  $-$22.881 &  5.714 & 19.558 &  $-$150.64 &   $-$58.56 &    67.14 & --- & --- & --- &  4.959 & 36.351 &     0.13 & --- \cr
668501115479895808  & 118.849980234877 &  18.734957249821 &  $-$22.459 &  $-$13.183 &  5.645 & 17.424 &  $-$151.43 &   $-$62.86 &    67.07 & --- & --- & --- & 18.919 & 35.568 &    0.244 & --- \cr
670463498920784512  & 119.311702025846 &  20.475050801747 &  $-$19.259 &  $-$17.779 &  5.829 & 18.536 &  $-$147.12 &   $-$56.39 &    67.87 & --- & --- & --- & 19.606 & 35.713 &    0.172 & --- \cr
666720039788305920  & 119.420635803765 &  15.684454972027 &  $-$12.571 &  $-$13.871 &  5.474 & 20.162 &  $-$152.96 &   $-$73.94 &    67.14 & --- & --- & --- & 11.687 & 35.628 &    0.106 & --- \cr
670414330135404416  & 119.431265206749 &  20.070160223743 &  $-$19.887 &   $-$9.106 &  5.233 & 20.119 &  $-$163.43 &   $-$64.13 &    75.46 & --- & --- & --- & 11.391 & 33.594 &       "" & --- \cr
667645008236564864  & 119.448531778971 &  17.921918325649 &   $-$5.426 &   $-$6.718 &  5.219 & 20.816 &  $-$162.27 &   $-$70.86 &    73.21 & --- & --- & --- &  9.975 & 34.246 &       "" & --- \cr
\ldots{}           &  \ldots{}   &   \ldots{} & \ldots{} & \ldots{} & \ldots{} & \ldots{} & \ldots{} & \ldots{} & \ldots{} & \ldots{} & \ldots{} & \ldots{} & \ldots{} \cr
637577896412129024  & 140.294307219569 &  20.557663044063 &  $-$84.731 &  $-$33.850 &  5.501 & 19.828 &  $-$118.67 &   $-$65.39 &   121.19 & --- & --- & --- & 15.877 & 31.236 &       "" & --- \cr
632738395977951488  & 140.474509676173 &  19.042042476738 &  $-$47.127 &  $-$10.999 &  5.838 & 12.636 &  $-$110.19 &   $-$65.80 &   113.43 &   $-$41.03 &   $-$19.46 &    $-$9.11 & 15.909 & 34.370 &    0.787 &    27.84 \cr
632329412012224640  & 140.517789036423 &  18.002924454129 &  $-$49.880 &  $-$14.769 &  5.781 & 17.580 &  $-$110.26 &   $-$69.33 &   113.82 & --- & --- & --- &  5.168 & 34.261 &    0.228 & --- \cr
637763168412025984  & 140.562206859202 &  21.348163316669 &  $-$51.545 &  $-$26.420 &  7.102 & 19.399 &   $-$92.01 &   $-$48.82 &    94.76 & --- & --- & --- & 20.744 & 54.317 &    0.137 & --- \cr
632722869671684096  & 140.706205038359 &  18.944309012550 &  $-$40.915 &  $-$15.227 &  5.998 & 14.759 &  $-$106.72 &   $-$64.29 &   110.79 & --- & --- & --- &  8.066 & 36.963 &    0.563 & --- \cr
632454348315678080  & 141.410495232392 &  18.577579180111 &  $-$40.884 &  $-$10.565 &  5.497 & 16.857 &  $-$114.60 &   $-$71.15 &   122.08 & --- & --- & --- &  5.337 & 34.721 &       "" & --- \cr
\hline
\label{tab_clustersGaia:Praesepe_catalogue_AfterCP}
\end{tabular}
\end{table*}
\end{appendix}

\end{document}